\documentclass[conference]{IEEEtran}
\IEEEoverridecommandlockouts
\usepackage{graphicx}
\usepackage{amsmath}
\usepackage{xcolor}
\usepackage{pdflscape}
\providecommand{\keywords}[1]{\textbf{\textit{Index terms---}} #1}
\usepackage{hyperref}
\usepackage{cite}
\usepackage{amsmath,amssymb,amsfonts}
\usepackage{algorithmic}
\usepackage{graphicx}
\usepackage{textcomp}
\usepackage{xcolor}
\usepackage{graphicx}
\usepackage{dcolumn}
\usepackage{bm}
\usepackage{xcolor}
\usepackage[utf8]{inputenc}
\usepackage[T1]{fontenc}
\usepackage{lscape}
\usepackage{nomencl}
\usepackage{mathptmx}
\usepackage{amsmath}

\newcommand{\ket}[1]{\left|#1\right\rangle}
\newcommand{\bra}[1]{\left\langle#1\right|}

\def\BibTeX{{\rm B\kern-.05em{\sc i\kern-.025em b}\kern-.08em
    T\kern-.1667em\lower.7ex\hbox{E}\kern-.125emX}}
\usepackage{graphicx}
\usepackage{amsmath}
\usepackage{lineno}
\makenomenclature
\begin{document}
\onecolumn
\title{Equivalence between finite state stochastic machine, non-dissipative and dissipative tight-binding and Schroedinger model }
\author{Krzysztof Pomorski$^{1,2}$ \\ \\

1: Cracow University of Technology \\ Faculty of Computer Science and Telecommunications \\ Department of Computer Science \\ \\
2: Quantum Hardware Systems ($www.quantumhardwaresystems.com$) \\ \\ E-mail: $kdvpomorski@gmail.com$ }
\maketitle

\begin{abstract}
The mathematical equivalence between finite state stochastic machine and non-dissipative and dissipative quantum tight-binding and Schroedinger model is derived.  Stochastic Finite state machine is also expressed by classical epidemic model and can reproduce the quantum entanglement emerging in the case of electrostatically coupled qubits described by von-Neumann entropy both in non-dissipative and dissipative case. The obtained results shows that quantum mechanical phenomena might be simulated by classical statistical model as represented by finite state stochastic machine. It includes the quantum like entanglement and superposition of states. Therefore coupled epidemic models expressed by classical systems in terms of classical physics can be the base for possible incorporation of quantum technologies and in particular for quantum like computation and quantum like communication. The classical density matrix is derived and described by the equation of motion in terms of anticommutator. Existence of Rabi like oscillations is pointed in classical epidemic model. Furthermore the existence of Aharonov-Bohm effect in quantum systems \cite{Aharonov} can also be
reproduced by the classical epidemic model or in broader sense by finite state stochastic machine. Every quantum system made from quantum dots and described by simplistic tight-binding model by use of position-based qubits can be effectively described by classical statistical model encoded in finite stochastic state machine with very specific structure of S matrix that has twice bigger size as it is the case of quantum matrix Hamiltonian.
Furthermore the description of linear and non-linear stochastic finite state machine is mapped to tight-binding and Schroedinger model. The concept of N dimensional complex time is incorporated into
tight-binding model, so the description of dissipation in most general case is possible.
Obtained results helps in approximation of non-dissipative or dissipative quantum mechanical phenomena by classical statistical physics expressed by finite state stochastic machine. Furthermore the equivalence of Wannier tight-binding dissipative or non-dissipative formalism to dissipative or non-dissipative Schroedinger formalism and to finite state stochastic machine was shown.
\end{abstract}
\newpage
\tableofcontents
\newpage

\mbox{}

\nomenclature{$c$}{Speed of light in a vacuum inertial frame}
\nomenclature{$h$}{Planck constant}

\printnomenclature

\section{Current status of single-electron technologies}
\begin{figure}[hbt!]
\centering
\includegraphics[scale=0.5]{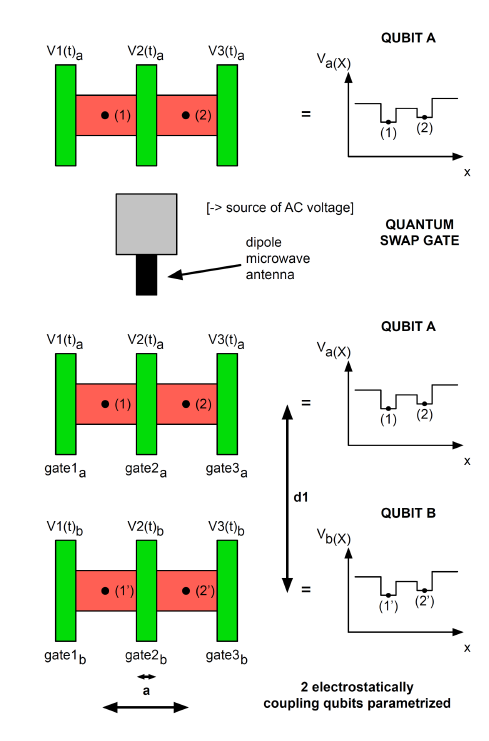}\includegraphics[scale=0.5]{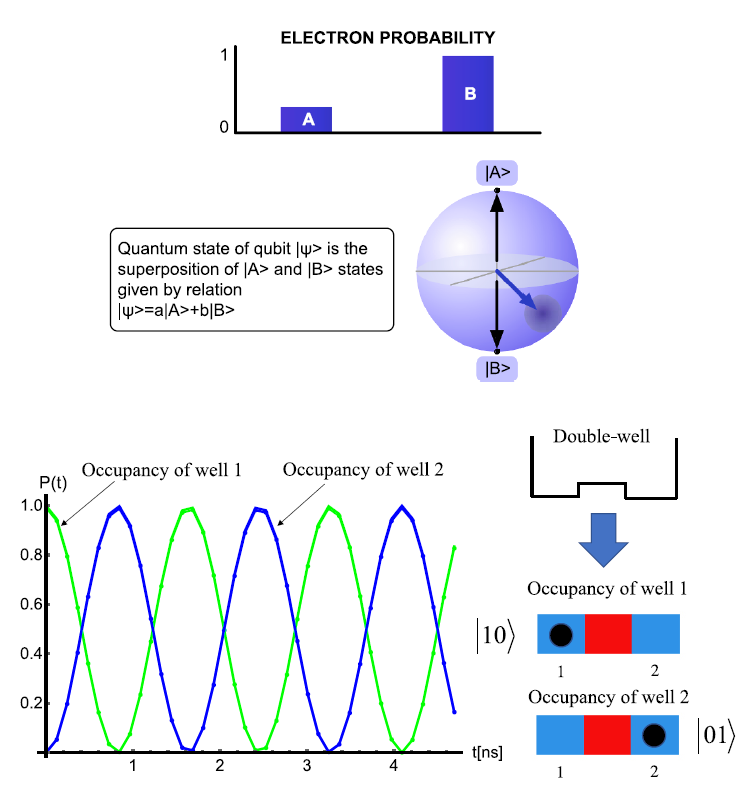}
\caption{Position-based qubit implemented in the chain of coupled quantum dots as given by \cite{Cryogenics} and by \cite{Panos}. }
\label{fig:WannierQ}
\includegraphics[scale=0.5]{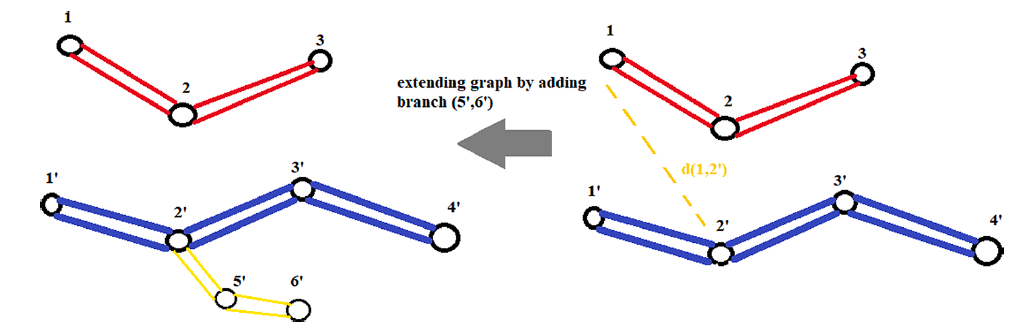}
\caption{Electrostatically controlled graph of coupled quantum dots as by \cite{Cryogenics}. }
\label{fig:QGraph}
\centering
\includegraphics[scale=0.9]{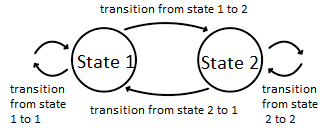}\includegraphics[scale=0.9]{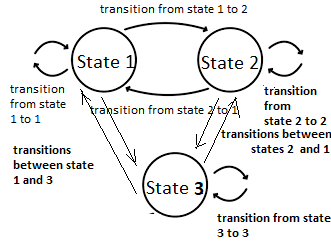}
\caption{Illustration of epidemic model referring to stochastic finite state machine being 2 level system with 2 distinguished states 1 and 2. 4 possible transitions are characterized by 4 time-dependent coefficients $s_{1 \rightarrow 1}(t)=s_{11}(t)$ ,$s_{1 \rightarrow 2}(t)=s_{12}(t)$, $s_{2 \rightarrow 1}(t)=s_{21}(t)$, $s_{2 \rightarrow 2}(t)=s_{22}(t)$ (Left picture). Using induction reasoning we can extend 2 dimensional stochastic finite state machine to 3 state machine and N state machine (Right picture).}
\label{fig:FiniteStates}
\end{figure}

Currently single-electron devices are becoming the more and more dominant trend in implementation of quantum technologies as given by Likharev \cite{Likharev}, Fujisawa \cite{Fujisawa}, Petta \cite{Petta}, Leipold \cite{Dirk}. The theory of operation on single-electron devices was developed in framework of tight-binding and Schroedinger model by Pomorski \cite{Spie},\cite{Cryogenics}, Giounanlis \cite{Panos} and many others.
Essentially one electron is injected into one among N coupled quantum dots and has oscillations of occupancy. Structures with such physical phenomena are shown by Fig.\ref{fig:WannierQ} and in Fig.\ref{fig:QGraph}. Probabilistic nature of this process implies hypothesis that electron occupancy of certain regions can be described by stochastic finite state machine. Indeed we can deal with reconfigurable quantum matter as pointed in Fig.\ref{fig:QGraph}, where we can set quantum dot connectivity in electrical way. Quantum matter has features of superposition of many states at the
same time, entanglement and is subjected to the strong or weak interaction during strong and weak measurement. Furthermore quantum matter is very sensitive towards external noise and decoherence processes. Quantum electrostatic entanglement emerges when we are dealing with 2 single-electron lines as pointed by \cite{2SEL}.
 It turns out that at first we need to
analyze the equation structure of tight-binding model and relate it with classically known epidemic model. Schroedinger equation or tight-binding model deal with square roots of probability occupancy multiplied by exponent of imaginary  phase factor.
 Let us start from epidemic model parametrized by S matrix using derivatives of time with respect to probabilities of 2N states as given by Fig.\ref{fig:FiniteStates}. and smoothly transit to tight-binding model parameterized by N by N Hamiltonian matrix. Quite obviously we can approach continuum of states by setting N to very big values.
 Strong hypothesis will be formulated as saying that any quantum system described by Hubbard model or tight-binding model both in equilibrium and non-equilibrium situations can be mapped to classical epidemic model. The concept of N dimensional complex value time will be discussed and this N dimensional complex value time occurring in quantum mechanics can
 be mapped to real value time occurring in classical epidemic model.
 In that way the uniqueness of quantum mechanics will be waived in favour of classical statistical physics description.
\section{Introduction to classical epidemic model}
Epidemic model can model sickness propagation and various phenomena in sociology, physics and biology \cite{epidemic}. The most basic form of epidemic model susceptible-infectious-susceptible model (abbreviated as SIS) relies on co-dependence of two probabilities of occurrence of state 1 (denoted as S(t)) and 2 (denoted as I(t)) that can be identified at the state of being healthy and sick as it is being depicted in Fig.\ref{fig:FiniteStates}. SIS epidemic model equations gives
\begin{eqnarray}
\begin{pmatrix}
-\lambda & +g \\
-g & +\lambda \\
\end{pmatrix}
\begin{pmatrix}
S_i(t) \\
I_i(t) \\
\end{pmatrix}
= \frac{d}{dt}
\begin{pmatrix}
S_i(t) \\
I_i(t) \\
\end{pmatrix},
\end{eqnarray}
where $S_i$ is number of susceptible individuals (or probability) and  $I_i$ be is the number of infected individuals (or probability).
On another hand the susceptible-infectious-recovered (SIR) model introduced $R_i$ quantity (number or probability of recovered individuals) and is given by equations
\begin{eqnarray}
\frac{d}{dt}
\begin{pmatrix}
S_i(t) \\
I_i(t) \\
R_i(t) \\
\end{pmatrix}
=
\begin{pmatrix}
-\frac{b*I_i(t)}{S_i(t)+I_i(t)+R_i(t)} & 0 & 0 \\
+\frac{b*I_i(t)}{S_i(t)+I_i(t)+R_i(t)} & -\gamma  & 0 \\
0 & \gamma  & 0 \\
\end{pmatrix}
\begin{pmatrix}
S_i(t) \\
I_i(t) \\
R_i(t) \\
\end{pmatrix}
\end{eqnarray}
Formally we can recognize that so called epidemic model is nothing else as stochastic finite state machine expressed by state vector of probabilities (or occurrences of given cases associated with discrete states).
The good example stochasitic (or sometimes more deterministic) finite state machine can be given by observation prizes of stock at the stock market. We can specify the N intervals as by $[price_{1,min}=0,price_{1,max})$, $[price_{1,max},price_{2,max})$ ,$[price_{2,max},price_{3,max})$ , ..,$[price_{k,max},price_{k+1,max})$ ,.. ,$[price_{N-1,max},price_{N,max}=+\infty)$ with N+1 conditions given by $[price_{k,max} < price_{k+1,max})$ for $k \in \{1..N\}$ and with $price_{1,max}>price_{1,min}=0$.
 We have N-dimensional vector $\ket{\psi(t)_{classical}}$ of classical states of stochastic Finite State Machine (or epidemic mode) in Dirac notation \cite{Dirac} and N-dimensional vector $\ket{\psi(t)_{quantum}}$ of quantum states with certain occupancy of N eigenenergies (denoted by probabilities $p_{E_k}(t)$ and phase $\Theta_k(t)$ ) of quantum system given as
\begin{eqnarray}
\ket{\psi(t)_{classical}}=
\begin{pmatrix}
  p_1(t) \\
  p_2(t) \\
  .. \\
  p_n(t)
  \end{pmatrix}
  =p_1(t)
  \begin{pmatrix}
  1 \\
  0 \\
  .. \\
  0
  \end{pmatrix}
  +...+
  p_n(t)
  \begin{pmatrix}
  0 \\
  0 \\
  .. \\
  1
  \end{pmatrix}=
  p_1(t)\ket{1}+..+p_n(t)\ket{N}, \nonumber \\
  \ket{\psi(t)_{quantum}}=
\begin{pmatrix}
  \sqrt{p_{E_1}(t)} \\
  \sqrt{p_{E_2}(t)} \\
  .. \\
  \sqrt{p_{E_N}(t)}
  \end{pmatrix}
  =e^{i\Theta_1(t)}\sqrt{p_{E_1}(t)}
  \begin{pmatrix}
  1 \\
  0 \\
  .. \\
  0
  \end{pmatrix}
  +...+
  e^{i\Theta_N(t)}\sqrt{p_{E_N}(t)}
  \begin{pmatrix}
  0 \\
  0 \\
  .. \\
  1
  \end{pmatrix}=
  e^{i\Theta_1(t)}\sqrt{p_{E1}}(t)\ket{E_1}+..+e^{i\Theta_N(t)}\sqrt{p_{E_N}(t)}\ket{E_N},
\end{eqnarray}
and we obtain projection of k-th state on system state $\ket{\psi_{classical}}$ expressed in so-called bra-ket notation introduced by Dirac \cite{Dirac} as by formula $\bra{k}\ket{\psi}=
\begin{pmatrix}
(0,..,1:\text{where 1 is on k-th position},..,0)\ket{\psi_{classical}}
\end{pmatrix}
=p_k(t)$. We shall recognize one central difference between classical and quantum state .
In case of quantum state we have $\bra{\psi_{quantum}} \ket{\psi_{quantum}}=
\begin{pmatrix}
e^{-i\Theta_1(t)}\sqrt{p_{E_1}(t)} & .. & e^{-i\Theta_N(t)}\sqrt{p_{E_N}(t)}
\end{pmatrix}
\begin{pmatrix}
e^{+i\Theta_1(t)}\sqrt{p_{E_1}(t)} \nonumber \\
 .. \nonumber \\
 e^{+i\Theta_N(t)}\sqrt{p_{E_N}(t)}
\end{pmatrix}
=e^{-i\Theta_1(t)}\sqrt{p_{E_1}(t)}e^{+i\Theta_1(t)}\sqrt{p_{E_1}(t)}+..+e^{-i\Theta_N(t)}\sqrt{p_{E_N}(t)}e^{+i\Theta_N(t)}\sqrt{p_{E_N}(t)}=p_{E_1}(t)+..+p_{E_N}(t)=1$
and in case of classical state
$\bra{\psi_{classical}} \ket{\psi_{classical}}=p_1(t)^2+..+p_N(t)^2<1$, but under condition that $p_1(t)+..+p_n(t)=1$.
One shall recognize the existence of big analogies between classical statistical physics and quantum mechanics as given by \cite{Statistics}, \cite{Statistics1}.
First of all from mathematical point of view Schroedinger equation expressing Quantum Mechanics is diffusion equation (as commonly present in classical statistical physics) with imaginary time. In classical statistical physics that can be expressed symbolically by stochastic Finite State Machine (or epidemic model) to each particle conglomerate is assigned statistical ensemble. In quantum mechanics to each particle is assigned a statistical ensemble.

 It is expressed in the compact way in the following way:
\begin{eqnarray}
(s_{11}(t)\ket{1}\bra{1}+s_{22}(t)\ket{2}\bra{2}+s_{12}(t)\ket{2}\bra{1}+s_{21}(t)\ket{1}\bra{2})(p_{1}(t)\ket{1}+p_{2}(t)\ket{2})=
\frac{d}{dt}(p_{1}\ket{1}+p_{2}\ket{2})
 = \nonumber \\
=\frac{d}{dt}
\begin{pmatrix}
p_{1}(t) \\
p_{2}(t) \\
\end{pmatrix}=
\begin{pmatrix}
s_{11}[p_1(t),p_2(t),t] & s_{12}[p_1(t),p_2(t),t] \\
s_{21}[p_1(t),p_2(t),t] & s_{22}[p_1(t),p_2(t),t] \\
\end{pmatrix}
\begin{pmatrix}
p_{1}(t) \\
p_{2}(t) \\
\end{pmatrix}=
\begin{pmatrix}
s_{11}(t) & s_{12}(t) \\
s_{21}(t) & s_{22}(t) \\
\end{pmatrix}|\psi_{classical}>= \nonumber \\
=\hat{S}_t(p_1(t)\ket{1}+p_2(t)\ket{2})=\frac{d}{dt}\ket{\psi_{classical}}.
\label{eqn:smatrix}
\end{eqnarray}
Quite naturally such system evolves in natural statistical environment before the measurement is done.
Once the measurement is done the statistical system state is changed
from undetermined and spanned by two probabilities into the case of being either with probability $p_1=1$ or $p_1=0$ so it corresponds to two projections:
\begin{eqnarray}
\hat{P}_{\rightarrow 1}=\ket{1}\bra{1}=
\begin{pmatrix}
1 & 0 \\
0 & 0 \\
\end{pmatrix},
\hat{P}_{\rightarrow 2}=\ket{2}\bra{2}=
\begin{pmatrix}
0 & 0 \\
0 & 1 \\
\end{pmatrix},\hat{P}_{\rightarrow 1}+\hat{P}_{\rightarrow 2}=\hat{I}=
\begin{pmatrix}
1 & 0 \\
0 & 1 \\
\end{pmatrix}
 \nonumber \\
\hat{P}_{\rightarrow 1}\ket{\psi}_{classical}= \ket{1}=\ket{\psi_1}_{after},
\begin{pmatrix}
0 & 0 \\
0 & 1 \\
\end{pmatrix},
\hat{P}_{\rightarrow 2}\ket{\psi}_{classical}= \ket{2}=\ket{\psi_2}_{after},
\end{eqnarray}
where occurrence of $\ket{\psi1}_{after}$ and $\ket{\psi2}_{after}$ occurs with probability $p_1(t_{measurement})$ and $p_2(t_{measurement})$.

We notice that matrix $\hat{S}_t$ from equation \ref{eqn:smatrix}
\begin{eqnarray}
\hat{S}_t=
\begin{pmatrix}
s_{11}(t) & s_{12}(t) \\
s_{21}(t) & s_{22}(t) \\
\end{pmatrix}
\end{eqnarray}
has two eigenvalues
\begin{eqnarray}
  E_1(t) &=& \frac{1}{2} \left(-\sqrt{(s_{11}(t)-s_{22}(t))^2+4s_{12}(t)s_{21}(t)}+s_{11}(t)+s_{22}(t)\right) \\
  E_2(t) &=& \frac{1}{2} \left(+\sqrt{(s_{11}(t)-s_{22}(t))^2+4 s_{12}(t)s_{21}(t)}+s_{11}(t)+s_{22}(t)\right).
\end{eqnarray}
 and we have the corresponding two classical eigenstate vectors (eigenstates)
\begin{eqnarray}
\ket{\psi_{E_1}}=\frac{2s_{21}}{2 s_{21}+(-\sqrt{(s_{11}-s_{22})^2+4s_{12}s_{21}}+s_{11}-s_{22})}
\begin{pmatrix}
\frac{-\sqrt{(s_{11}-s_{22})^2+4s_{12}s_{21}}+s_{11}-s_{22}}{2 s_{21}} \\ 1 
\end{pmatrix},   \\
\ket{\psi_{E_2}}=\frac{2s_{21}}{2s_{21}+(+\sqrt{(s_{11}-s_{22})^2+4 s_{12}s_{21}}+s_{11}-s_{22})}
\begin{pmatrix}
\frac{+\sqrt{(s_{11}-s_{22})^2+4 s_{12} s_{21}}+s_{11}-s_{22}}{2 s_{21}} \\
1
\end{pmatrix}.
\end{eqnarray}
We recognize that two states $\ket{\psi_{E_1}}$ and $\ket{\psi_{E_2}}$ are orthogonal, so $\bra{\psi_{E_1}}\ket{\psi_{E_2}}=\bra{\psi_{E_2}}\ket{\psi_{E_1}}=0$.
We also recognize that
\begin{eqnarray}
\bra{\psi_{E_1}}\ket{\psi_{E_1}}=[\frac{-\sqrt{(s_{11}-s_{22})^2+4s_{12}s_{21}}+s_{11}-s_{22}}{2 s_{21}+(-\sqrt{(s_{11}-s_{22})^2+4s_{12}s_{21}}+s_{11}-s_{22})}]^2+[\frac{2s_{21}}{2 s_{21}+(-\sqrt{(s_{11}-s_{22})^2+4s_{12}s_{21}}+s_{11}-s_{22})}]^2= \\
=1-[\frac{4s_{21}(-\sqrt{(s_{11}-s_{22})^2+4s_{12}s_{21}}+s_{11}-s_{22})}{(2 s_{21}+(-\sqrt{(s_{11}-s_{22})^2+4s_{12}s_{21}}+s_{11}-s_{22}))^2}]=n_{E_1}(t), \nonumber \\
\bra{\psi_{E_2}}\ket{\psi_{E_2}}=[\frac{+\sqrt{(s_{11}-s_{22})^2+4 s_{12} s_{21}}+s_{11}-s_{22}}{2s_{21}+(+\sqrt{(s_{11}-s_{22})^2+4 s_{12}s_{21}}+s_{11}-s_{22})}]^2+[\frac{2s_{21}}{2s_{21}+(+\sqrt{(s_{11}-s_{22})^2+4 s_{12}s_{21}}+s_{11}-s_{22})}]^2= \nonumber \\
=1-[\frac{4s_{21}(+\sqrt{(s_{11}-s_{22})^2+4 s_{12}s_{21}}+s_{11}-s_{22})}{(2s_{21}+(+\sqrt{(s_{11}-s_{22})^2+4 s_{12}s_{21}}+s_{11}-s_{22}))^2}]
=n_{E_2}(t).
\end{eqnarray}

It shall be underlined that necessary condition for identification the superposition of two classical eigenstates is expressed by $(-\sqrt{(s_{11}-s_{22})^2+4s_{12}s_{21}}+s_{11}-s_{22})>0$ and $(+\sqrt{(s_{11}-s_{22})^2+4 s_{12}s_{21}}+s_{11}-s_{22})>0$ what preimposes some constrains on
real value functions $s_{11}(t)$, $s_{12}$, $s_{21}$ and $s_{22}$.
The full classical state can be written as the superposition of two ensembles with probabilities $p_I$ and $p_{II}$ expressed by the following classical state
\begin{eqnarray}
\ket{\psi(t)}_{classical}=p_{I}(t)\ket{\psi_{E_1}}+p_{2}(t)\ket{\psi_{E_2}}= \nonumber \\
=p_{I}(t)\frac{2s_{21}}{2 s_{21}+(-\sqrt{(s_{11}-s_{22})^2+4s_{12}s_{21}}+s_{11}-s_{22})}
\begin{pmatrix}
\frac{-\sqrt{(s_{11}-s_{22})^2+4s_{12}s_{21}}+s_{11}-s_{22}}{2 s_{21}} \\ 1 
\end{pmatrix}+ \nonumber  \\
+p_{II}(t)\frac{2s_{21}}{2s_{21}+(+\sqrt{(s_{11}-s_{22})^2+4 s_{12}s_{21}}+s_{11}-s_{22})}
\begin{pmatrix}
\frac{+\sqrt{(s_{11}-s_{22})^2+4 s_{12} s_{21}}+s_{11}-s_{22}}{2 s_{21}} \\
1
\end{pmatrix}= \nonumber \\
=p_{I}(t)
\begin{pmatrix}
\frac{-\sqrt{(s_{11}-s_{22})^2+4s_{12}s_{21}}+s_{11}-s_{22}}{2 s_{21}+(-\sqrt{(s_{11}-s_{22})^2+4s_{12}s_{21}}+s_{11}-s_{22})} \\
\frac{2s_{21}}{2 s_{21}+(-\sqrt{(s_{11}-s_{22})^2+4s_{12}s_{21}}+s_{11}-s_{22})} 
\end{pmatrix}+
p_{II}(t)
\begin{pmatrix}
\frac{+\sqrt{(s_{11}-s_{22})^2+4 s_{12} s_{21}}+s_{11}-s_{22}}{2s_{21}+(+\sqrt{(s_{11}-s_{22})^2+4 s_{12}s_{21}}+s_{11}-s_{22})} \\
\frac{2s_{21}}{2s_{21}+(+\sqrt{(s_{11}-s_{22})^2+4 s_{12}s_{21}}+s_{11}-s_{22})}
\end{pmatrix}=
\nonumber  \\
=
\begin{pmatrix}
p_{I}(t)\frac{-\sqrt{(s_{11}-s_{22})^2+4s_{12}s_{21}}+s_{11}-s_{22}}{2 s_{21}+(-\sqrt{(s_{11}-s_{22})^2+4s_{12}s_{21}}+s_{11}-s_{22})}+p_{II}(t)\frac{+\sqrt{(s_{11}-s_{22})^2+4 s_{12} s_{21}}+s_{11}-s_{22}}{2s_{21}+(+\sqrt{(s_{11}-s_{22})^2+4 s_{12}s_{21}}+s_{11}-s_{22})} \\
p_{I}(t)\frac{2s_{21}}{2 s_{21}+(-\sqrt{(s_{11}-s_{22})^2+4s_{12}s_{21}}+s_{11}-s_{22})}+p_{II}(t)\frac{2s_{21}}{2s_{21}+(+\sqrt{(s_{11}-s_{22})^2+4 s_{12}s_{21}}+s_{11}-s_{22})} 
\end{pmatrix}=
\begin{pmatrix}
p_{1}(t) \\
p_{2}(t)
\end{pmatrix}=\ket{\psi(t)}_{classical}
.
\end{eqnarray}
We have superposition of states with 2 statistical ensembles occurring with probabilities $p_{I}(t)$ and $p_{II}(t)$ that are encoded in probabilities $p_1(t)$ and $p_2(t)$ that are directly observable. We can extract probabilities $p_{I}(t)$ and $p_{II}(t)$ from $\ket{\psi(t)}_{classical}$ in the following way
\begin{eqnarray}
p_{I}(t)=\frac{1}{n_{E_1}(t)}\bra{\psi_{E_1}} \ket{\psi(t)}_{classical}=\Bigg[1-\Bigg[\frac{4s_{21}(-\sqrt{(s_{11}-s_{22})^2+4s_{12}s_{21}}+s_{11}-s_{22})}{(2 s_{21}+(-\sqrt{(s_{11}-s_{22})^2+4s_{12}s_{21}}+s_{11}-s_{22}))^2}\Bigg]\Bigg]\bra{\psi_{E_1}} \ket{\psi(t)}_{classical}= \nonumber \\
=\Bigg[1-\Bigg[\frac{4s_{21}(-\sqrt{(s_{11}-s_{22})^2+4s_{12}s_{21}}+s_{11}-s_{22})}{(2 s_{21}+(-\sqrt{(s_{11}-s_{22})^2+4s_{12}s_{21}}+s_{11}-s_{22}))^2}\Bigg]\Bigg] \times \nonumber \\
\times
\begin{pmatrix}
\frac{-\sqrt{(s_{11}-s_{22})^2+4s_{12}s_{21}}+s_{11}-s_{22}}{2 s_{21}+(-\sqrt{(s_{11}-s_{22})^2+4s_{12}s_{21}}+s_{11}-s_{22})}, & \frac{2s_{21}}{2 s_{21}+(-\sqrt{(s_{11}-s_{22})^2+4s_{12}s_{21}}+s_{11}-s_{22})} 
\end{pmatrix}
\begin{pmatrix}
p_{1}(t), \\
p_{2}(t)
\end{pmatrix} 
\end{eqnarray}
and
\begin{eqnarray}
p_{II}(t)=\frac{1}{n_{E_2}(t)}\bra{\psi_{E_2}} \ket{\psi(t)}_{classical}=\Bigg[1-\Bigg[\frac{4s_{21}(+\sqrt{(s_{11}-s_{22})^2+4 s_{12}s_{21}}+s_{11}-s_{22})}{(2s_{21}+(+\sqrt{(s_{11}-s_{22})^2+4 s_{12}s_{21}}+s_{11}-s_{22}))^2}\Bigg]\Bigg]\bra{\psi_{E_2}} \ket{\psi(t)}_{classical}= \nonumber \\
=\Bigg[1-\Bigg[\frac{4s_{21}(+\sqrt{(s_{11}-s_{22})^2+4s_{12}s_{21}}+s_{11}-s_{22})}{(2 s_{21}+(+\sqrt{(s_{11}-s_{22})^2+4s_{12}s_{21}}+s_{11}-s_{22}))^2}\Bigg]\Bigg] \times \nonumber \\
\times
\begin{pmatrix}
\frac{+\sqrt{(s_{11}-s_{22})^2+4s_{12}s_{21}}+s_{11}-s_{22}}{2 s_{21}+(+\sqrt{(s_{11}-s_{22})^2+4s_{12}s_{21}}+s_{11}-s_{22})}, & \frac{2s_{21}}{2 s_{21}+(+\sqrt{(s_{11}-s_{22})^2+4s_{12}s_{21}}+s_{11}-s_{22})} 
\end{pmatrix}
\begin{pmatrix}
p_{1}(t), \\
p_{2}(t)
\end{pmatrix}= \nonumber \\ =
\Bigg[1-\Bigg[\frac{4s_{21}(+\sqrt{(s_{11}-s_{22})^2+4s_{12}s_{21}}+s_{11}-s_{22})}{(2 s_{21}+(+\sqrt{(s_{11}-s_{22})^2+4s_{12}s_{21}}+s_{11}-s_{22}))^2}\Bigg]\Bigg] \times \nonumber \\
\times
\begin{pmatrix}
\frac{+\sqrt{(s_{11}-s_{22})^2+4s_{12}s_{21}}+s_{11}-s_{22}}{2 s_{21}+(+\sqrt{(s_{11}-s_{22})^2+4s_{12}s_{21}}+s_{11}-s_{22})}p_1(t)+ \frac{2s_{21}}{2 s_{21}+(+\sqrt{(s_{11}-s_{22})^2+4s_{12}s_{21}}+s_{11}-s_{22})}p_2(t) 
\end{pmatrix}. 
\end{eqnarray}
Probabilities $p_{I}(t)$ and $p_{II}(t)$ will describe the occupancy of energy levels $E_1$ and $E_2$ in real time domain of epidemic simplistic model.
We have the same superposition of two eigenergies as in the case of quantum tight-binding model.
The same reasoning can be conducted for N-th state classical epidemic model expressed as
\begin{eqnarray}
(s_{11}(t)\ket{1}\bra{1}+s_{12}(t)\ket{2}\bra{1}+s_{13}(t)\ket{3}\bra{1}+.. + s_{1N}(t)\ket{N}\bra{1}+ \nonumber \\
+s_{21}(t)\ket{2}\bra{1}+s_{22}(t)\ket{2}\bra{2}+s_{23}(t)\ket{3}\bra{2}+.. + s_{2N}(t)\ket{N}\bra{2}+ \nonumber \\
.... + \nonumber \\
+s_{1N}(t)\ket{1}\bra{N}+s_{2N}(t)\ket{2}\bra{N}+s_{3N}(t)\ket{3}\bra{N}+.. + s_{NN}(t)\ket{N}\bra{N})
\nonumber \\
(p_{1}(t)\ket{1}+p_{2}(t)\ket{2}+..+p_{N}(t)\ket{N})=
\frac{d}{dt}(p_{1}\ket{1}+p_{2}\ket{2}+..+p_{N}\ket{N})
 = \nonumber \\
=\frac{d}{dt}
\begin{pmatrix}
p_{1}(t) \\
p_{2}(t) \\
.. \\
p_{N}(t) \\
\end{pmatrix}=
\begin{pmatrix}
s_{11}(t) & s_{12}(t) & .. & s_{1N}(t) \\
s_{21}(t) & s_{22}(t) & .. & s_{1N}(t  \\
.. \\
s_{N1}(t) & s_{N2}(t) & .. & s_{NN}(t  \\
\end{pmatrix}
\begin{pmatrix}
p_{1}(t) \\
p_{2}(t) \\
..
p_{N}(t) \\
\end{pmatrix}=
\begin{pmatrix}
s_{11}(t) & s_{12}(t) & .. & s_{1N}(t) \\
.. \\
s_{1N}(t) & s_{2N}(t)  & .. & s_{NN}(t) \\
\end{pmatrix}|\psi_{classical}>= \nonumber \\
=\hat{S}_t(p_1(t)\ket{1}+p_2(t)\ket{2}+..+p_N(t)\ket{N})=\frac{d}{dt}|\psi_{classical}>.
\end{eqnarray}
In such way N-state stochastic finite state machine can be modeled.
\section{Analytical solutions of simplistic classical epidemic model}
In principle we can also introduce weak measurement procedure that will be partly omitted in this work. In very real way if we have the population of N individuals possibly infected with some diesease we can inspect $N_1$ individuals, where $N_1<N$ and we can introduce some corrections to $p_1(t_{measurement}^{-})\rightarrow \frac{1}{N}[(N-N_1)p_1(t_{measurement}^{-})+N_{1}p_1(t_{test})]=p_1(t_{measurement}^{+})$ and $p_2(t_{measurement}^{-})\rightarrow \frac{1}{N}[(N-N_1)p_2(t_{measurement}^{-})+N_{1}p_2(t_{test})]=p_2(t_{measurement}^{+})$, where $p_1(t_{test}),p_2(t_{test})$ are probabilities obtained by testing $N_1$ individuals what could correspond to weak measurement conducted on assemble of N individuals.  Let us consider the state of the system before measurement and its natural evolution
Such set of equations has two analytical solutions for probabilities $p_1(t)$ and $p_2(t)$ expressed as
\begin{eqnarray}
exp
\begin{pmatrix}
\int_{t0}^{t} s_{11}(t)dt' & \int_{t0}^{t} s_{11}(t)dt' \\
\int_{t0}^{t} s_{21}(t)dt' & \int_{t0}^{t} s_{22}(t)dt' \\
\end{pmatrix}
\begin{pmatrix}
p_{1}(t_0) \\
p_{2}(t_0) \\
\end{pmatrix}
=exp
\begin{pmatrix}
S_{11}(t,t_0) & S_{12}(t,t_0) \\
S_{21}(t,t_0) & S_{22}(t,t_0) \\
\end{pmatrix}
\begin{pmatrix}
p_{1}(t_0) \\
p_{2}(t_0) \\
\end{pmatrix}
= \nonumber \\
\begin{pmatrix}
U_{11}(t,t_0) & U_{12}(t,t_0) \\
U_{21}(t,t_0) & U_{22}(t,t_0) \\
\end{pmatrix}
\begin{pmatrix}
p_{1}(t_0) \\
p_{2}(t_0) \\
\end{pmatrix}
= \hat{U}(t,t_0)
\begin{pmatrix}
p_{1}(t_0) \\
p_{2}(t_0) \\
\end{pmatrix}
=
\begin{pmatrix}
p_{1}(t) \\
p_{2}(t) \\
\end{pmatrix}
\end{eqnarray}
with
\begin{eqnarray}
S_{11}(t,t_0)=\int_{t0}^{t} s_{11}(t)dt', S_{12}(t,t_0)=\int_{t0}^{t} s_{12}(t)dt', S_{22}(t,t_0)=\int_{t0}^{t} s_{22}(t)dt', S_{21}(t,t_0)=\int_{t0}^{t} s_{21}(t)dt',
\end{eqnarray}
and
\begin{eqnarray}
U_{1,1}(t,t_0)=e^{\frac{S_{11}(t,t_0)+S_{22}(t,t_0)}{2}}\Bigg[+\frac{(S_{11}(t,t_0)-S_{22}(t,t_0)) \sinh \left(\frac{1}{2} \sqrt{(S_{11}(t,t_0)-S_{22}(t,t_0))^2+4 S_{12}(t,t_0)
   S_{21}(t,t_0)}\right)}{\sqrt{(S_{11}(t,t_0)-S_{22}(t,t_0))^2+4 S_{12}(t,t_0) S_{21}(t,t_0)}}+\nonumber \\ +\cosh \left(\frac{1}{2} \sqrt{(S_{11}(t,t_0)-S_{22}(t,t_0))^2+4
   S_{12}(t,t_0)S_{21}(t,t_0)}\right)\Bigg]
\end{eqnarray}
\begin{eqnarray}
U_{2,2}(t,t_0)=e^{\frac{S_{11}(t,t_0)+S_{22}(t,t_0)}{2}} \Bigg[-\frac{(S_{11}(t,t_0)-S_{22}(t,t_0)) \sinh \left(\frac{1}{2} \sqrt{(S_{11}(t,t_0)-S_{22}(t,t_0))^2+4 S_{12}(t,t_0)
   S_{21}(t,t_0)}\right)}{\sqrt{(S_{11}(t,t_0)-S_{22}(t,t_0))^2+4 S_{12}(t,t_0)S_{21}(t,t_0)}}+ \nonumber \\ + \cosh \left(\frac{1}{2} \sqrt{(S_{11}(t,t_0)-S_{22}(t,t_0))^2+4
   S_{12}(t,t_0) S_{21}(t,t_0)}\right)\Bigg]
\end{eqnarray}
\begin{eqnarray}
U_{1,2}(t,t_0)=\frac{2 S_{12}(t,t_0) e^{\frac{S_{11}(t,t_0)+S_{22}(t,t_0)}{2}} \sinh \left(\frac{1}{2} \sqrt{(S_{11}(t,t_0)-S_{22}(t,t_0))^2+4 S_{12}(t,t_0)
   S_{21}(t,t_0)}\right)}{\sqrt{(S_{11}(t,t_0)-S_{22}(t,t_0))^2+4 S_{12}(t,t_0) S_{21}(t,t_0)}},
\end{eqnarray}
\begin{eqnarray}
U_{2,1}(t,t_0)=\frac{2 S_{21}(t,t_0) e^{\frac{S_{11}(t,t_0)+S_{22}(t,t_0)}{2}} \sinh \left(\frac{1}{2} \sqrt{(S_{11}(t,t_0)-S_{22}(t,t_0))^2+4S_{12}(t,t_0)
   S_{21}(t,t_0)}\right)}{\sqrt{(S_{11}(t,t_0)-S_{22}(t,t_0))^2+4S_{12}(t,t_0)S_{21}(t,t_0)}}
\end{eqnarray}
We obtain explicit formula for probabilities
\begin{eqnarray}
p_1(t)=e^{\frac{S_{11}(t,t_0)+S_{22}(t,t_0)}{2}}\Bigg[\Bigg[+\frac{(S_{11}(t,t_0)-S_{22}(t,t_0)) \sinh \left(\frac{1}{2} \sqrt{(S_{11}(t,t_0)-S_{22}(t,t_0))^2+4 S_{12}(t,t_0)
   S_{21}(t,t_0)}\right)}{\sqrt{(S_{11}(t,t_0)-S_{22}(t,t_0))^2+4 S_{12}(t,t_0) S_{21}(t,t_0)}}+\nonumber \\ +\cosh \left(\frac{1}{2} \sqrt{(S_{11}(t,t_0)-S_{22}(t,t_0))^2+4
   S_{12}(t,t_0)S_{21}(t,t_0)}\right)\Bigg]p_1(t_0)+ \nonumber \\ +\Bigg[\frac{2 S_{12}(t,t_0)\sinh \left(\frac{1}{2} \sqrt{(S_{11}(t,t_0)-S_{22}(t,t_0))^2+4 S_{12}(t,t_0)
   S_{21}(t,t_0)}\right)}{\sqrt{(S_{11}(t,t_0)-S_{22}(t,t_0))^2+4 S_{12}(t,t_0) S_{21}(t,t_0)}}\Bigg] p_2(t_0)\Bigg],
\end{eqnarray}
\begin{eqnarray}
p_2(t)=e^{\frac{S_{11}(t,t_0)+S_{22}(t,t_0)}{2}}\Bigg[\Bigg[\frac{2 S_{21}(t,t_0)\sinh \left(\frac{1}{2} \sqrt{(S_{11}(t,t_0)-S_{22}(t,t_0))^2+4S_{12}(t,t_0)
   S_{21}(t,t_0)}\right)}{\sqrt{(S_{11}(t,t_0)-S_{22}(t,t_0))^2+4S_{12}(t,t_0)S_{21}(t,t_0)}}\Bigg]p_1(t_0)+ \nonumber \\
   +\Bigg[-\frac{(S_{11}(t,t_0)-S_{22}(t,t_0)) \sinh \left(\frac{1}{2} \sqrt{(S_{11}(t,t_0)-S_{22}(t,t_0))^2+4 S_{12}(t,t_0)
   S_{21}(t,t_0)}\right)}{\sqrt{(S_{11}(t,t_0)-S_{22}(t,t_0))^2+4 S_{12}(t,t_0)S_{21}(t,t_0)}}+ \nonumber \\ + \cosh \left(\frac{1}{2} \sqrt{(S_{11}(t,t_0)-S_{22}(t,t_0))^2+4
   S_{12}(t,t_0) S_{21}(t,t_0)}\right)\Bigg]p_2(t_0)\Bigg].
\end{eqnarray}
It is useful to express the ratio of probabilities $p_1(t)$ and $p_2(t)$ in the following analytical way as
\begin{eqnarray}
r_{12}(t)=\frac{p_1(t)}{p_2(t)}=\nonumber \\
=\Bigg[[(S_{11}(t,t_0)-S_{22}(t,t_0))p_1(t_0)+2 S_{21}(t,t_0)p_2(t_0)] \tanh \left(\frac{1}{2} \sqrt{(S_{11}(t,t_0)-S_{22}(t,t_0))^2+4 S_{12}(t,t_0)
   S_{21}(t,t_0)}\right) +\nonumber \\+ [p_1(t_0)\sqrt{(S_{11}(t,t_0)-S_{22}(t,t_0))^2+4
   S_{12}(t,t_0)S_{21}(t,t_0)}]\Bigg]/ \nonumber \\ \Bigg[ -[(S_{11}(t,t_0)-S_{22}(t,t_0))p_2(t_0)+2 S_{21}(t,t_0)p_1(t_0)] \tanh \left(\frac{1}{2} \sqrt{(S_{11}(t,t_0)-S_{22}(t,t_0))^2+4 S_{12}(t,t_0)
   S_{21}(t,t_0)}\right) +\nonumber \\+ [p_2(t_0)\sqrt{(S_{11}(t,t_0)-S_{22}(t,t_0))^2+4
   S_{12}(t,t_0)S_{21}(t,t_0)}] \Bigg] \nonumber \\
\end{eqnarray}
Although in general case of classical epidemic model probabilities are not normalized their true physical or technical meaning is not lost since the ratio between
probabilities in capturing the essence of their meaning.
\section{Equations of motion for classical epidemic model in projector representation in case of time independent matrix $S$}
Let us consider the equations of motion for the case of time-independent $s_{11}$, $s_{12}$, $s_{21}$ and $s_{22}$. In Dirac notation it can be written in the following way:
\begin{eqnarray}
(E_1\ket{\psi_{E_1}}\frac{1}{n_{E_1}}\bra{\psi_{E_1}}+E_2\ket{\psi_{E_2}}\frac{1}{n_{E_2}}\bra{\psi_{E_2}})(p_{I}\ket{\psi_{E_1}}+p_{II}\ket{\psi_{E_2}}).
=\frac{d}{dt}(p_{I}\ket{\psi_{E_1}}+p_{II}\ket{\psi_{E_2}})=((\frac{d}{dt}p_{I})\ket{\psi_{E_1}}+(\frac{d}{dt})p_{II}\ket{\psi_{E_2}}),
\end{eqnarray}
since $E_1$, $E_2$, $\ket{\psi_{E_1}}$ and $\ket{\psi_{E_2}}$ are time independent. By applying $\ket{\psi_{E1}}$ and $\ket{\psi_{E1}}$ on the left side we obtain
and using orthogonality relation between $\ket{\psi_{E_1}}$ and $\ket{\psi_{E_2}}$ we obtain set of equations
\begin{equation}
\frac{E_1}{n_{E_1}}p_{I}=\frac{d}{dt}p_{I}, \frac{E_2}{n_{E_2}}p_{II}=\frac{d}{dt}p_{II},
\end{equation}
that has the solutions
\begin{equation}
e^{\frac{E_1}{n_{E_1}}(t-t_0)}p_{I}(t_0)=p_{I}(t), e^{\frac{E_2}{n_{E_2}}(t-t_0)}p_{II}(t_0)=p_{II}(t).
\end{equation}
Sum of probabilities is not normalized. However physical significance between ratio $p_1(t)$ and $p_2$ that is expressed by ratio
\begin{equation}
r_{12}(t)=\frac{p_1(t)}{p_2(t)}=\frac{p_{I}(t_0)}{p_{II}(t_0)}exp((\frac{E_{1}}{n_{E_1}}-\frac{E_2}{n_{E_2}})(t-t_0)).
\end{equation}
It means that Rabi oscillations or more precisely change of occupancy among levels is naturally build in classical epidemic model.
Still superposition of two states is mainatianed so the analogy of classical epidemic model to quantum tight-binding model is deep.
\section{Case of constant occupacy of 2 eigenergy levels in classical epidemic model}
We consider the case of $p_{I}(t)=constant_1$ and $p_{II}(t)=constant_{II}$. We have time-dependent parameters $s_{11}$, $s_{22}$, $s_{12}$ and $s_{21}$ and we obtain the following equations of motion
\begin{eqnarray}
(E_1(t)\ket{\psi_{E_1}}_t\bra{\psi_{E_1}}_t+E_2(t)\ket{\psi_{E_2}}_t\bra{\psi_{E_2}}_t)(p_{I}\ket{\psi_{E_1}}_t+p_{II}\ket{\psi_{E_2}}_t).
=\frac{d}{dt}(p_{I}\ket{\psi_{E_1}}_t+p_{II}\ket{\psi_{E_2}}_t)=(p_{I}\frac{d}{dt})\ket{\psi_{E_1}}_t+(p_{II}\frac{d}{dt})\ket{\psi_{E_2}}_t).
\end{eqnarray}
We obtain the set of 2 equations
\begin{eqnarray}
E_1(t) p_I=(p_{I}\bra{\psi_{E_1}}\frac{d}{dt})\ket{\psi_{E_1}}+(p_{II}\bra{\psi_{E_1}}\frac{d}{dt})\ket{\psi_{E_2}}),  \\
E_2(t) p_{II}=(p_{I}\bra{\psi_{E_2}}\frac{d}{dt})\ket{\psi_{E_1}}+(p_{II}\bra{\psi_{E_2}}\frac{d}{dt})\ket{\psi_{E_2}}).
\end{eqnarray}
Consequently we obtain
\begin{eqnarray}
\frac{p_I}{p_{II}}=\frac{\bra{\psi_{E_1}}\frac{d}{dt}\ket{\psi_{E_2}}}{E_1(t)-\bra{\psi_{E_1}}\frac{d}{dt}\ket{\psi_{E_1}}}, \nonumber \\ 
\frac{p_I}{p_{II}}=\frac{(E_2(t)-\bra{\psi_{E_2}}\frac{d}{dt}\ket{\psi_{E_2}})}{\bra{\psi_{E_2}}\frac{d}{dt}\ket{\psi_{E_1}}} 
\end{eqnarray}
and it implies
\begin{eqnarray}
\frac{\bra{\psi_{E_1}}\frac{d}{dt}\ket{\psi_{E_2}}}{(E_1(t)-\bra{\psi_{E_1}}\frac{d}{dt}\ket{\psi_{E_1}})}=\frac{(E_2(t)-\bra{\psi_{E_2}}\frac{d}{dt}\ket{\psi_{E_2}})}{\bra{\psi_{E_2}}\frac{d}{dt}\ket{\psi_{E_1}}}. 
\end{eqnarray}
The last equation guarantees that ratio $\frac{p_I}{p_{II}}$ is constant.
\section{Equations of motion for classical epidemic model in projector representation (Dirac notation) and Rabi oscillations in classical epidemic model}
Let us consider the equations of motion in the following way:
\begin{eqnarray}
(E_1(t)\ket{\psi_{E1}}\frac{1}{n_{E_1}}\bra{\psi_{E1}}+E_2(t)\ket{\psi_{E2}}\bra{\psi_{E2}}+e_{12}(t)\ket{\psi_{E2}}\bra{\psi_{E1}}+e_{21}(t)\ket{\psi_{E1}}\bra{\psi_{E2}})(p_{I}\ket{\psi_{E1}}+p_{II}\ket{\psi_{E2}})=\nonumber \\
=\frac{d}{dt}(p_{I}\ket{\psi_{E1}}+p_{II}\ket{\psi_{E2}}).
\end{eqnarray}
This equation is equivalent to the set of 2 coupled ordinary differential equations given as
\begin{eqnarray}
E_1(t)p_{I}(t)+e_{21}(t)p_{II}(t)=\bra{\psi_{E1}(t)}\frac{d}{dt}(p_{I}(t)\ket{\psi_{E1}(t)})+\bra{\psi_{E1(t)}}\frac{d}{dt}(p_{II}(t)\ket{\psi_{E2}(t)}), \nonumber \\ 
E_2(t)p_{II}(t)+e_{12}(t)p_{I}(t)=\bra{\psi_{E2}(t)}\frac{d}{dt}(p_{I}(t)\ket{\psi_{E1}(t)})+\bra{\psi_{E2}(t)}\frac{d}{dt}(p_{II}(t)\ket{\psi_{E2}(t)}).
\end{eqnarray}
and can be rewritten to be as
\begin{eqnarray}
E_1(t)p_{I}(t)+e_{21}(t)p_{II}(t)=\frac{d}{dt}p_{I}(t)+p_{I}(t)(\bra{\psi_{E1}(t)}\frac{d}{dt}\ket{\psi_{E1}(t)}) 
+p_{II}(t)\bra{\psi_{E1(t)}}\frac{d}{dt}(\ket{\psi_{E2}(t)}), \nonumber \\ 
+e_{12}(t)p_{I}(t)+E_2(t)p_{II}(t)=\frac{d}{dt}p_{II}(t)+p_{II}(t)(\bra{\psi_{E2}(t)}\frac{d}{dt}\ket{\psi_{E2}(t)}) 
+p_{I}(t)\bra{\psi_{E2(t)}}\frac{d}{dt}(\ket{\psi_{E1}(t)}).
\end{eqnarray}
It will lead to further simplification that can be written as
\begin{eqnarray}
+[E_1(t)-(\bra{\psi_{E1}(t)}\frac{d}{dt}\ket{\psi_{E1}(t)})]p_{I}(t)+[e_{21}(t)-p_{II}(t)\bra{\psi_{E1(t)}}\frac{d}{dt}(\ket{\psi_{E2}(t)})]p_{II}(t)=\frac{d}{dt}p_{I}(t) 
\nonumber \\ 
+[e_{12}(t)-\bra{\psi_{E2}(t)}\frac{d}{dt}(\ket{\psi_{E1}(t)})]p_{I}(t)+[E_2(t)-(\bra{\psi_{E2}(t)}\frac{d}{dt}\ket{\psi_{E2}(t)})]p_{II}(t)=\frac{d}{dt}p_{II}(t). 
\end{eqnarray}
We can write it in the compact form as
\begin{eqnarray}
\begin{pmatrix}
[E_1(t)-(\bra{\psi_{E1}(t)}\frac{d}{dt}\ket{\psi_{E1}(t)})] & [e_{21}(t)-\bra{\psi_{E1(t)}}\frac{d}{dt}(\ket{\psi_{E2}(t)})] \\
[e_{12}(t)-\bra{\psi_{E2}(t)}\frac{d}{dt}(\ket{\psi_{E1}(t)})] & [E_2(t)-(\bra{\psi_{E2}(t)}\frac{d}{dt}\ket{\psi_{E2}(t)})] \\
\end{pmatrix}
\begin{pmatrix}
p_{I}(t) \\
p_{II}(t) \\
\end{pmatrix}=
\frac{d}{dt}
\begin{pmatrix}
p_{I}(t) \\
p_{II}(t) \\
\end{pmatrix}.
\end{eqnarray}
The solution is analytical given as
\begin{eqnarray}
exp
\begin{pmatrix}
\int_{t_0}^{t}dt'[E_1(t')-(\bra{\psi_{E1}(t')}\frac{d}{dt'}\ket{\psi_{E1}(t')})] & \int_{t_0}^{t}dt'[e_{21}(t')-\bra{\psi_{E1(t')}}\frac{d}{dt'}(\ket{\psi_{E2}(t')})] \\
\int_{t_0}^{t}dt'[e_{12}(t')-\bra{\psi_{E2}(t')}\frac{d}{dt'}(\ket{\psi_{E1}(t')})] & \int_{t_0}^{t}dt'[E_2(t')-(\bra{\psi_{E2}(t')}\frac{d}{dt'}\ket{\psi_{E2}(t')})] \\
\end{pmatrix}
\begin{pmatrix}
p_{I}(t_0) \\
p_{II}(t_0) \\
\end{pmatrix}=
\begin{pmatrix}
p_{I}(t) \\
p_{II}(t) \\
\end{pmatrix}
\end{eqnarray}
and can be written as
\begin{eqnarray}
\hat{G}(t,t_0)
\begin{pmatrix}
p_{I}(t_0) \\
p_{II}(t_0) \\
\end{pmatrix}=exp
\begin{pmatrix}
g_{1,1}(t,t_0)  & g_{1,2}(t,t_0) \\
g_{2,1})(t,t_0) & g_{2,1}(t,t_0) \\
\end{pmatrix}
\begin{pmatrix}
p_{I}(t_0) \\
p_{II}(t_0) \\
\end{pmatrix}=
\begin{pmatrix}
p_{I}(t) \\
p_{II}(t) \\
\end{pmatrix}
=
\begin{pmatrix}
G_{1,1}(t,t_0) & G_{1,2}(t,t_0) \\
G_{2,1})(t,t_0) & G_{2,1}(t,t_0) \\
\end{pmatrix}
\begin{pmatrix}
p_{I}(t_0) \\
p_{II}(t_0) \\
\end{pmatrix}=
\begin{pmatrix}
p_{I}(t) \\
p_{II}(t) \\
\end{pmatrix},
\end{eqnarray}
where
\begin{eqnarray}
g_{1,1}(t,t_0)=\int_{t_0}^{t}dt'[E_1(t')-(\bra{\psi_{E1}(t')}\frac{d}{dt'}\ket{\psi_{E1}(t')})] ,  \\
g_{1,2}(t,t_0)=\int_{t_0}^{t}dt'[e_{21}(t')-\bra{\psi_{E1}(t')}\frac{d}{dt'}(\ket{\psi_{E2}(t')})] ,  \\
g_{2,1}(t,t_0)=\int_{t_0}^{t}dt'[e_{12}(t')-\bra{\psi_{E2}(t')}\frac{d}{dt'}(\ket{\psi_{E1}(t')})] ,  \\
g_{2,2}(t,t_0)=\int_{t_0}^{t}dt'[E_2(t')-(\bra{\psi_{E2}(t')}\frac{d}{dt'}\ket{\psi_{E2}(t')})].
\end{eqnarray}
 and we have the corresponding classical eigenstates
\begin{eqnarray}
\bra{\psi_{E_1}(t')}(\frac{d}{dt'}\ket{\psi_{E_1}(t')})=\nonumber \\
\Bigg[\frac{1}{2 s_{21}+(-\sqrt{(s_{11}-s_{22})^2+4s_{12}s_{21}}+s_{11}-s_{22})}
\begin{pmatrix}
-\sqrt{(s_{11}-s_{22})^2+4s_{12}s_{21}}+s_{11}-s_{22} & 2 s_{21} 
\end{pmatrix}\Bigg] \times \nonumber \\ \times
\frac{d}{dt}\Bigg[\frac{1}{2 s_{21}+(-\sqrt{(s_{11}-s_{22})^2+4s_{12}s_{21}}+s_{11}-s_{22})}
\begin{pmatrix}
-\sqrt{(s_{11}-s_{22})^2+4s_{12}s_{21}}+s_{11}-s_{22} \\ 2 s_{21} 
\end{pmatrix}\Bigg], \nonumber \\
\bra{\psi_{E_1}(t')}(\frac{d}{dt'}\ket{\psi_{E_2}(t')})=\nonumber \\
\Bigg[\frac{1}{2 s_{21}+(-\sqrt{(s_{11}-s_{22})^2+4s_{12}s_{21}}+s_{11}-s_{22})}
\begin{pmatrix}
-\sqrt{(s_{11}-s_{22})^2+4s_{12}s_{21}}+s_{11}-s_{22} & 2 s_{21} 
\end{pmatrix}\Bigg] \times \nonumber \\ \times
\frac{d}{dt'}\Bigg[\frac{1}{2 s_{21}+(+\sqrt{(s_{11}-s_{22})^2+4s_{12}s_{21}}+s_{11}-s_{22})}
\begin{pmatrix}
+\sqrt{(s_{11}-s_{22})^2+4s_{12}s_{21}}+s_{11}-s_{22} \\ 2 s_{21} 
\end{pmatrix}\Bigg],
\end{eqnarray}
\begin{eqnarray}
\bra{\psi_{E_1}(t')}(\frac{d}{dt'}\ket{\psi_{E_2}(t')})=\nonumber \\
\Bigg[\frac{1}{2 s_{21}+(-\sqrt{(s_{11}-s_{22})^2+4s_{12}s_{21}}+s_{11}-s_{22})}
\begin{pmatrix}
-\sqrt{(s_{11}-s_{22})^2+4s_{12}s_{21}}+s_{11}-s_{22} & 2 s_{21} 
\end{pmatrix}\Bigg] \times \nonumber \\ \times
\frac{d}{dt'}\Bigg[\frac{1}{2 s_{21}+(+\sqrt{(s_{11}-s_{22})^2+4s_{12}s_{21}}+s_{11}-s_{22})}
\begin{pmatrix}
+\sqrt{(s_{11}-s_{22})^2+4s_{12}s_{21}}+s_{11}-s_{22} \\ 2 s_{21} 
\end{pmatrix}\Bigg], \nonumber \\
\bra{\psi_{E_2}(t')}(\frac{d}{dt'}\ket{\psi_{E_2}(t')})=\nonumber \\
\Bigg[\frac{1}{2 s_{21}+(+\sqrt{(s_{11}-s_{22})^2+4s_{12}s_{21}}+s_{11}-s_{22})}
\begin{pmatrix}
+\sqrt{(s_{11}-s_{22})^2+4s_{12}s_{21}}+s_{11}-s_{22} & 2 s_{21} 
\end{pmatrix}\Bigg] \times \nonumber \\ \times
\frac{d}{dt'}\Bigg[\frac{1}{2 s_{21}+(+\sqrt{(s_{11}-s_{22})^2+4s_{12}s_{21}}+s_{11}-s_{22})}
\begin{pmatrix}
+\sqrt{(s_{11}-s_{22})^2+4s_{12}s_{21}}+s_{11}-s_{22} \\ 2 s_{21} 
\end{pmatrix}\Bigg],
\end{eqnarray}

and
\begin{eqnarray}
  \int_{t_0}^{t}dt'E_1(t') &=& \int_{t_0}^{t}dt'\frac{1}{2} \left(-\sqrt{(s_{11}(t')-s_{22}(t'))^2+4s_{12}(t')s_{21}(t')}+s_{11}(t')+s_{22}(t')\right) \\
  \int_{t_0}^{t}dt'E_2(t') &=& \int_{t_0}^{t}dt'\frac{1}{2} \left(+\sqrt{(s_{11}(t')-s_{22}(t'))^2+4s_{12}(t')s_{21}(t')}+s_{11}(t')+s_{22}(t')\right).
\end{eqnarray}
and
\begin{eqnarray}
G_{1,1}(t,t_0)=e^{\frac{g_{11}(t,t_0)+g_{22}(t,t_0)}{2}}\Bigg[+\frac{(g_{11}(t,t_0)-g_{22}(t,t_0)) \sinh \left(\frac{1}{2} \sqrt{(g_{11}(t,t_0)-g_{22}(t,t_0))^2+4 g_{12}(t,t_0)
   g_{21}(t,t_0)}\right)}{\sqrt{(g_{11}(t,t_0)-g_{22}(t,t_0))^2+4 g_{12}(t,t_0) g_{21}(t,t_0)}}+\nonumber \\ +\cosh \left(\frac{1}{2} \sqrt{(g_{11}(t,t_0)-g_{22}(t,t_0))^2+4
   g_{12}(t,t_0)g_{21}(t,t_0)}\right)\Bigg]
\end{eqnarray}
\begin{eqnarray}
G_{2,2}(t,t_0)=e^{\frac{g_{11}(t,t_0)+g_{22}(t,t_0)}{2}} \Bigg[-\frac{(g_{11}(t,t_0)-g_{22}(t,t_0)) \sinh \left(\frac{1}{2} \sqrt{(g_{11}(t,t_0)-g_{22}(t,t_0))^2+4 g_{12}(t,t_0)
   g_{21}(t,t_0)}\right)}{\sqrt{(g_{11}(t,t_0)-g_{22}(t,t_0))^2+4 g_{12}(t,t_0)g_{21}(t,t_0)}}+ \nonumber \\ + \cosh \left(\frac{1}{2} \sqrt{(g_{11}(t,t_0)-g_{22}(t,t_0))^2+4
   g_{12}(t,t_0) g_{21}(t,t_0)}\right)\Bigg]
\end{eqnarray}
\begin{eqnarray}
G_{1,2}(t,t_0)=\frac{2 g_{12}(t,t_0) e^{\frac{g_{11}(t,t_0)+g_{22}(t,t_0)}{2}} \sinh \left(\frac{1}{2} \sqrt{(g_{11}(t,t_0)-g_{22}(t,t_0))^2+4 g_{12}(t,t_0)
   g_{21}(t,t_0)}\right)}{\sqrt{(g_{11}(t,t_0)-g_{22}(t,t_0))^2+4 g_{12}(t,t_0) g_{21}(t,t_0)}},
\end{eqnarray}
\begin{eqnarray}
G_{2,1}(t,t_0)=\frac{2 g_{21}(t,t_0) e^{\frac{g_{11}(t,t_0)+g_{22}(t,t_0)}{2}} \sinh \left(\frac{1}{2} \sqrt{(g_{11}(t,t_0)-g_{22}(t,t_0))^2+4g_{12}(t,t_0)
   g_{21}(t,t_0)}\right)}{\sqrt{(g_{11}(t,t_0)-g_{22}(t,t_0))^2+4g_{12}(t,t_0)g_{21}(t,t_0)}}
\end{eqnarray}

\section{Analogy between quantum entanglement and classical epidemic model}
 It is now easy to generalize our considerations for 2 coupled statistical ensembles as corresponding to cities with flight traffic.
 We have
 \begin{eqnarray}
\begin{pmatrix}
s_{11}(t)_A & s_{12}(t)_A & 0 & s_{1A2B}(t) \\
s_{21}(t)_A & s_{22}(t)_A & s_{2A2B}(t) & 0 \\
0 & s_{2A1B}(t) & s_{11}(t)_B & s_{12}(t)_B \\
s_{2A1B}(t) & 0 & s_{21}(t)_B & s_{22}(t)_B \\
\end{pmatrix}
\begin{pmatrix}
p_{1}(t)_A \\
p_{2}(t)_A \\
p_{1}(t)_B \\
p_{2}(t)_B \\
\end{pmatrix}=
\hat{S}|\psi_{classical}>
=\frac{d}{dt}
\begin{pmatrix}
p_{1}(t)_A \\
p_{2}(t)_A \\
p_{1}(t)_B \\
p_{2}(t)_B \\
\end{pmatrix}=\nonumber \\
=\hat{S}_t(p_{1A}(t)\ket{1}_A|\ket{1}_B+p_{2A}(t)\ket{1}_A\ket{2}_B+p_{1B}(t)\ket{2}_A|\ket{1}_B+p_{2B}(t)\ket{2}_A\ket{2}_B)=\frac{d}{dt}|\psi_{classical}>
\end{eqnarray}
We make the analytic simplifications by assumption of two symmetric systems A and B interacting interacting in asymmetric way as
 \begin{eqnarray}
\begin{pmatrix}
s_{11}(t) & s_{12}(t) & 0 & s(t) \\
s_{21}(t) & s_{22}(t) & s(t) & 0 \\
0 & s(t) & s_{11}(t) & s_{12}(t) \\
s(t) & 0 & s_{21}(t) & s_{22}(t) \\
\end{pmatrix}
\begin{pmatrix}
p_{1}(t)_A \\
p_{2}(t)_A \\
p_{1}(t)_B \\
p_{2}(t)_B \\
\end{pmatrix}=\frac{d}{dt}
\begin{pmatrix}
p_{1}(t)_A \\
p_{2}(t)_A \\
p_{1}(t)_B \\
p_{2}(t)_B \\
\end{pmatrix}
.
\end{eqnarray}
We have 4 eigenstates
\begin{eqnarray}
\ket{V_1(t)}=
\begin{pmatrix}
-\frac{\sqrt{4 (S-\text{S12}) (S-\text{S21})+(\text{S11}-\text{S22})^2}-\text{S11}+\text{S22}}{2 (S-\text{S21})}, \\
-1, \\
\frac{\sqrt{4(S-\text{S12}) (S-\text{S21})+(\text{S11}-\text{S22})^2}-\text{S11}+\text{S22}}{2 (S-\text{S21})}, \\
   1
\end{pmatrix}
\end{eqnarray}

\begin{eqnarray}
\ket{V_2(t)}=
\begin{pmatrix}
\frac{\sqrt{4 (S-\text{S12}) (S-\text{S21})+(\text{S11}-\text{S22})^2}+\text{S11}-\text{S22}}{2 (S-\text{S21})}, \\
-1, \\
-\frac{\sqrt{4(S-\text{S12}) (S-\text{S21})+(\text{S11}-\text{S22})^2}+\text{S11}-\text{S22}}{2 (S-\text{S21})}, \\
   1
\end{pmatrix}
\end{eqnarray}
\begin{eqnarray}
\ket{V_3(t)}=
\begin{pmatrix}
\frac{-\sqrt{4 (S+\text{S12}) (S+\text{S21})+(\text{S11}-\text{S22})^2}+\text{S11}-\text{S22}}{2 (S+\text{S21})}, \\
1, \\
\frac{-\sqrt{4
   (S+\text{S12}) (S+\text{S21})+(\text{S11}-\text{S22})^2}+\text{S11}-\text{S22}}{2 (S+\text{S21})}, \\
   1
\end{pmatrix}
\end{eqnarray}

\begin{eqnarray}
\ket{V_4(t)}=
\begin{pmatrix}
\frac{\sqrt{4 (S+\text{S12}) (S+\text{S21})+(\text{S11}-\text{S22})^2}+\text{S11}-\text{S22}}{2 (S+\text{S21})}, \\
1, \\
\frac{\sqrt{4
   (S+\text{S12}) (S+\text{S21})+(\text{S11}-\text{S22})^2}+\text{S11}-\text{S22}}{2 (S+\text{S21})},\\
1
\end{pmatrix}
\end{eqnarray}
and $\ket{V_3(t)}$ and $\ket{V_4(t)}$ are physically justifiable in framework of epidemic model that can be used for classical entanglement.
We have 4 projectors corresponding to the measurement of $p_{1A}$, $p_{2A}$, $p_{1B}$ and $p_{2B}$ represented as matrices
\begin{eqnarray}
\hat{P}_{p1A}=
\begin{pmatrix}
1 & 0 & 0 & 0 \\
0 & 0 & 0 & 0 \\
0 & 0 & 1 & 0 \\
0 & 0 & 0 & 1 \\
\end{pmatrix},
\hat{P}_{p2A}=
\begin{pmatrix}
0 & 0 & 0 & 0 \\
0 & 1 & 0 & 0 \\
0 & 0 & 1 & 0 \\
0 & 0 & 0 & 1 \\
\end{pmatrix},
\hat{P}_{p1B}=
\begin{pmatrix}
1 & 0 & 0 & 0 \\
0 & 1 & 0 & 0 \\
0 & 0 & 1 & 0 \\
0 & 0 & 0 & 0 \\
\end{pmatrix},
\hat{P}_{p2B}=
\begin{pmatrix}
1 & 0 & 0 & 0 \\
0 & 1 & 0 & 0 \\
0 & 0 & 0 & 0 \\
0 & 0 & 0 & 1 \\
\end{pmatrix}
\end{eqnarray}
Measurement conducted on system A also brings the change of system B state what is due to presence of non-diagonal matrix elements in system evolution equations (generalized
epidemic model). It is therefore analogical to measurement of quantum entangled state. After the measurements we obtain the following classical states

\begin{eqnarray}
\hat{P}_{p1A}\ket{\psi_{classical}}=
\begin{pmatrix}
1 & 0 & 0 & 0 \\
0 & 0 & 0 & 0 \\
0 & 0 & 1 & 0 \\
0 & 0 & 0 & 1 \\
\end{pmatrix}
\begin{pmatrix}
p_{A1}(t) \\
p_{A2}(t) \\
p_{B1}(t) \\
p_{B2}(t) \\
\end{pmatrix}
=
\begin{pmatrix}
1 \\
0 \\
p_{B1}(t) \\
p_{B2}(t) \\
\end{pmatrix}
, \nonumber \\
\hat{P}_{p2A}\ket{\psi_{classical}}=
\begin{pmatrix}
0 & 0 & 0 & 0 \\
0 & 1 & 0 & 0 \\
0 & 0 & 1 & 0 \\
0 & 0 & 0 & 1 \\
\end{pmatrix}
\begin{pmatrix}
p_{A1}(t) \\
p_{A2}(t) \\
p_{B1}(t) \\
p_{B2}(t) \\
\end{pmatrix}
=
\begin{pmatrix}
0 \\
1 \\
p_{B1}(t) \\
p_{B2}(t) \\
\end{pmatrix},
\nonumber \\
\hat{P}_{p1B}\ket{\psi_{classical}}=
\begin{pmatrix}
1 & 0 & 0 & 0 \\
0 & 1 & 0 & 0 \\
0 & 0 & 1 & 0 \\
0 & 0 & 0 & 0 \\
\end{pmatrix}
\begin{pmatrix}
p_{A1}(t) \\
p_{A2}(t) \\
p_{B1}(t) \\
p_{B2}(t) \\
\end{pmatrix}
=
\begin{pmatrix}
p_{A1}(t) \\
p_{A2}(t) \\
1 \\
0 \\
\end{pmatrix}
, \nonumber \\
\hat{P}_{p2B}\ket{\psi_{classical}}=
\begin{pmatrix}
1 & 0 & 0 & 0 \\
0 & 1 & 0 & 0 \\
0 & 0 & 0 & 0 \\
0 & 0 & 0 & 1 \\
\end{pmatrix}
\begin{pmatrix}
p_{A1}(t) \\
p_{A2}(t) \\
p_{B1}(t) \\
p_{B2}(t) \\
\end{pmatrix}
=
\begin{pmatrix}
p_{A1}(t) \\
p_{A2}(t) \\
0 \\
1 \\
\end{pmatrix}
\end{eqnarray}
In such way the measurement process can be represented in epidemic model.
\section{2 classical statistical systems interacting in quantum mechanical way}
We are going to reformulate the description of 2 classical noninteracting systems A and B.
We have
\begin{eqnarray}
\hat{H}_0(t)=
\begin{pmatrix}
s_{11A}(t) & s_{12A}(t) \\
s_{21A}(t) & s_{22A}(t) \\
\end{pmatrix} \times
\begin{pmatrix}
1 & 0 \\
0 & 1 \\
\end{pmatrix} +
\begin{pmatrix}
1 & 0 \\
0 & 1 \\
\end{pmatrix} \times
\begin{pmatrix}
s_{11B}(t) & s_{12B}(t) \\
s_{21B}(t) & s_{22B}(t) \\
\end{pmatrix}= \nonumber \\
=
\begin{pmatrix}
s_{11A}(t)              & 0                   & s_{12A}(t) & 0          \\
0                       & s_{11A}(t)          & 0          & s_{12A}(t) \\
s_{21A}(t)              & 0                   & s_{22A}(t) & 0          \\
0                       & s_{21A}(t)          & 0          & s_{22A}(t) \\
\end{pmatrix} +
\begin{pmatrix}
s_{11B}(t)              & s_{12B}(t)          & 0          & 0          \\
s_{21B}(t)              & s_{22B}(t)          & 0          & 0          \\
0                       & 0                   & s_{11B}(t) & s_{12B}(t) \\
0                       & 0                   & s_{21B}(t) & s_{22B}(t) \\
\end{pmatrix}= \nonumber \\
=
\begin{pmatrix}
s_{11A}(t) +s_{11B}(t)  & s_{12B}(t)                     & s_{12A}(t)              & 0                     \\
s_{21B}(t)              & s_{11A}(t)+s_{22B}(t)          & 0                       & s_{12A}(t)            \\
s_{21A}(t)              & 0                              &  s_{22A}(t)+ s_{11B}(t) & s_{12B}(t)            \\
0                       & s_{21A}(t)                     & s_{21B}(t)              & s_{22B}(t)+s_{22A}(t) \\
\end{pmatrix}= \nonumber \\
=(E_{1A}(t)\ket{\psi_{E_{1A}}(t)}\bra{\psi_{E_{1A}}(t)}+E_{2A}(t)\ket{\psi_{E_{2A}}(t)}\bra{\psi_{E_{2A}}(t)})(\ket{\psi_{E_{1B}}(t)}\bra{\psi_{E_{1B}}(t)}+\ket{\psi_{E_{2B}}(t)}\bra{\psi_{E_{2B}}(t)})+\nonumber \\
+(\ket{\psi_{E_{1A}}(t)}\bra{\psi_{E_{1A}}(t)}+\ket{\psi_{E_{2A}}(t)}\bra{\psi_{E_{2A}}(t)})(E_{1B}(t)\ket{\psi_{E_{1B}}(t)}\bra{\psi_{E_{1B}}(t)}+E_{2B}(t)\ket{\psi_{E_{2B}}(t)}\bra{\psi_{E_{2B}}(t)}).
\end{eqnarray}
and we have the
\begin{eqnarray}
\ket{\psi(t)}=\gamma_1(t)\ket{\psi_{E_{1A}}(t)}\ket{\psi_{E_{1B}}(t)}+ \gamma_2(t)\ket{\psi_{E_{1A}}(t)}\ket{\psi_{E_{2B}}(t)}+\gamma_3(t)\ket{\psi_{E_{2A}}(t)}\ket{\psi_{E_{1B}}(t)}+\gamma_4(t)\ket{\psi_{E_{2A}}(t)}\ket{\psi_{E_{2B}}(t)}= \nonumber \\
=(p_{1A}(t)\ket{x_{1A}}+p_{2A}(t)\ket{x_{2A}})(p_{1B}(t)\ket{x_{1B}}+p_{2B}(t)\ket{x_{2B}})= \nonumber \\
=p_{1A}(t)p_{1B}(t)\ket{x_{1A}}\ket{x_{1B}}+p_{1A}(t)p_{2B}(t)\ket{x_{1A}}\ket{x_{2B}}+p_{2A}(t)p_{1B}(t)\ket{x_{2A}}\ket{x_{1B}}+p_{2A}(t)p_{2B}(t)\ket{x_{2A}}\ket{x_{2B}},
\end{eqnarray}
where $\ket{x_{kA}}\ket{x_{lB}}$ are time-independent and k and l are 1 or 2 and $\gamma_1(t)+\gamma_2(t)+\gamma_3(t)+\gamma_4(t)=1$. We observe that $\hat{H}_0(t)\ket{\psi}_t=\frac{d}{dt}\ket{\psi}_t$ explicitly given as
\begin{eqnarray}
\frac{d}{dt}
\begin{pmatrix}
p_{1A}(t)p_{1B}(t) \\
p_{1A}(t)p_{2B}(t) \\
p_{2A}(t)p_{1B}(t) \\
p_{2A}(t)p_{2B}(t) \\
\end{pmatrix}=
\begin{pmatrix}
s_{11A}(t) +s_{11B}(t)  & s_{12B}(t)                     & s_{12A}(t)              & 0                     \\
s_{21B}(t)              & s_{11A}(t)+s_{22B}(t)          & 0                       & s_{12A}(t)            \\
s_{21A}(t)              & 0                              &  s_{22A}(t)+ s_{11B}(t) & s_{12B}(t)            \\
0                       & s_{21A}(t)                     & s_{21B}(t)              & s_{22B}(t)+s_{22A}(t) \\
\end{pmatrix} 
\begin{pmatrix}
p_{1A}(t)p_{1B}(t) \\
p_{1A}(t)p_{2B}(t) \\
p_{2A}(t)p_{1B}(t) \\
p_{2A}(t)p_{2B}(t) \\
\end{pmatrix}= \nonumber \\
=\begin{pmatrix}
s_{11A}(t) +s_{11B}(t)  & s_{12B}(t)                     & s_{12A}(t)              & 0                     \\
s_{21B}(t)              & s_{11A}(t)+s_{22B}(t)          & 0                       & s_{12A}(t)            \\
s_{21A}(t)              & 0                              &  s_{22A}(t)+ s_{11B}(t) & s_{12B}(t)            \\
0                       & s_{21A}(t)                     & s_{21B}(t)              & s_{22B}(t)+s_{22A}(t) \\
\end{pmatrix} 
\begin{pmatrix}
p_{IQ}(t)   \\
p_{IIQ}(t)  \\
p_{IIIQ}(t) \\
p_{IVQ}(t)  \\
\end{pmatrix}
=\frac{d}{dt}
\begin{pmatrix}
p_{IQ}(t)   \\
p_{IIQ}(t)  \\
p_{IIIQ}(t) \\
p_{IVQ}(t)  \\
\end{pmatrix}
,
\end{eqnarray}
where $p_{IQ}(t)$, $p_{IIQ}(t)$, $p_{IIIQ}(t)$ and  $p_{IVQ}(t)$ (with $p_{IQ}(t)=p_{1A}(t)p_{1B}(t)$, $p_{IIQ}(t)=p_{1A}(t)p_{2B}(t)$, $p_{IIIQ}(t)=p_{2A}(t)p_{1B}(t)$, $p_{IVQ}(t)=p_{2A}(t)p_{2B}(t)$) are describing probabilities for 4 different states of statistic finite machine.
Similar situation occurs in the case of Schroedinger equation written for 2 non-interacting systems, but instead of probabilities we have square root of probabilities times phase factor.
In general case of interaction between A and B systems in classical epidemic model we have
\begin{eqnarray}
\hat{H}_0(t)+\hat{H}_{A-B}(t)
=E_{I}(t)\ket{\psi_{E_{1A}}(t)}\ket{\psi_{E_{1B}}(t)}\bra{\psi_{E_{1A}}(t)}\bra{\psi_{E_{1B}}(t)} 
+E_{II}(t)\ket{\psi_{E_{1A}}(t)}\ket{\psi_{E_{2B}}(t)}\bra{\psi_{E_{1A}}(t)}\bra{\psi_{E_{2B}}(t)}+ \nonumber \\
+E_{III}(t)\ket{\psi_{E_{2A}}(t)}\ket{\psi_{E_{1B}}(t)}\bra{\psi_{E_{2A}}(t)}\bra{\psi_{E_{1B}}(t)} 
+E_{IV}(t)\ket{\psi_{E_{2A}}(t)}\ket{\psi_{E_{2B}}(t)}\bra{\psi_{E_{2A}}(t)}\bra{\psi_{E_{2B}}(t)}+ \nonumber \\
+e_{(1A,1B) \rightarrow (1A,2B)}(t)\ket{\psi_{E_{1A}}(t)}\ket{\psi_{E_{1B}}(t)}\bra{\psi_{E_{1A}}(t)}\bra{\psi_{E_{2B}}(t)}
+e_{(1A,2B) \rightarrow (1A,1B)}(t)\ket{\psi_{E_{1A}}(t)}\ket{\psi_{E_{2B}}(t)}\bra{\psi_{E_{1A}}(t)}\bra{\psi_{E_{1B}}(t)}+ \nonumber \\
+e_{(2A,1B) \rightarrow (1A,1B)}(t)\ket{\psi_{E_{2A}}(t)}\ket{\psi_{E_{1B}}(t)}\bra{\psi_{E_{1A}}(t)}\bra{\psi_{E_{1B}}(t)}
+e_{(1A,1B) \rightarrow (2A,1B)}(t)\ket{\psi_{E_{1A}}(t)}\ket{\psi_{E_{1B}}(t)}\bra{\psi_{E_{2A}}(t)}\bra{\psi_{E_{1B}}(t)}+
\nonumber \\
+e_{(1A,2B) \rightarrow (2A,2B)}(t)\ket{\psi_{E_{1A}}(t)}\ket{\psi_{E_{2B}}(t)}\bra{\psi_{E_{2A}}(t)}\bra{\psi_{E_{2B}}(t)}+
e_{(2A,2B) \rightarrow (1A,2B)}(t)\ket{\psi_{E_{2A}}(t)}\ket{\psi_{E_{2B}}(t)}\bra{\psi_{E_{1A}}(t)}\bra{\psi_{E_{2B}}(t)}+ \nonumber \\
+e_{(2A,1B) \rightarrow (2A,2B)}(t)\ket{\psi_{E_{2A}}(t)}\ket{\psi_{E_{1B}}(t)}\bra{\psi_{E_{2A}}(t)}\bra{\psi_{E_{2B}}(t)}+
e_{(2A,2B) \rightarrow (2A,1B)}(t)\ket{\psi_{E_{2A}}(t)}\ket{\psi_{E_{2B}}(t)}\bra{\psi_{E_{2A}}(t)}\bra{\psi_{E_{1B}}(t)}+
\nonumber \\
+e_{(1A,1B) \rightarrow (2A,2B)}(t)\ket{\psi_{E_{1A}}(t)}\ket{\psi_{E_{1B}}(t)}\bra{\psi_{E_{2A}}(t)}\bra{\psi_{E_{2B}}(t)}+
e_{(2A,2B) \rightarrow (1A,1B)}(t)\ket{\psi_{E_{2A}}(t)}\ket{\psi_{E_{2B}}(t)}\bra{\psi_{E_{1A}}(t)}\bra{\psi_{E_{1B}}(t)}.
\end{eqnarray}
The presented analytical approach can be applied for system A with 4 distinct states as well as for system B with 4 distinct states since isolated system A(B) with 4 states can be described by evolution matrix 4 by 4 that has 4 analytical eigenvalues and eigenstates and becomes non-analytical for 5 and more distinct states due to fact that roots of polynomial of higher order than 4 becomes non-analytical and becomes numerical with some limited exceptions.
We can write the Hamiltonian matrix in eigenenergy representation
\begin{eqnarray}
\hat{H}_{E_{IQ},..,E_{IVQ}}=
\begin{pmatrix}
E_{IQ}(t) & e_{(1A,2B) \rightarrow (1A,1B)} & e_{(2A,1B) \rightarrow (1A,1B)} & e_{(2A,2B) \rightarrow (1A,1B)} \\
e_{(1A,1B) \rightarrow (1A,2B)} & E_{IIQ}(t) & e_{(2A,1B) \rightarrow (1A,2B)} & e_{(2A,2B) \rightarrow (1A,2B)} \\
e_{(1A,1B) \rightarrow (2A,1B)} & e_{(1A,1B) \rightarrow (2A,1B)} & E_{IIIQ}(t) & e_{(2A,2B) \rightarrow (1A,2B)} \\
e_{(1A,1B) \rightarrow (2A,2B)} & e_{(1A,1B) \rightarrow 2A,2B)} & e_{(2A,1B) \rightarrow (2A,2B)} & E_{IVQ}(t) \\
\end{pmatrix}.
\end{eqnarray}
where $e_{(1A,2B) \rightarrow (1A,1B)}$ is operator describing the transition from eigenenergy levels $(1A,2B)$ to eigenenergy $(2A,1B)$.
\section{From epidemic model to tight-binding equations}
\subsection{Case of 2 level classical stochastic finite state machine}
Let us be motivated by work on single electron devices by \cite{Fujisawa}, \cite{Likharev}, \cite{Petta}, \cite{Dirk} and \cite{Bashir19}.
Instead of probabilities it will be useful to operate with square root of probabilities as they are present in quantum mechanics and in Schroedinger or Dirac equation. Since
\begin{eqnarray}
\frac{d}{dt}(\sqrt{p_1}\sqrt{p_1})=2 \sqrt{p_1(t)}\frac{d}{dt}\sqrt{p_1(t)}, \nonumber \\
\frac{d}{dt}(\sqrt{p_2}\sqrt{p_2})=2 \sqrt{p_2(t)}\frac{d}{dt}\sqrt{p_2(t)}.
\end{eqnarray}
we can rewrite the epidemic equation as
\begin{eqnarray}
\begin{pmatrix}
s_{11}(t) & s_{12}(t) \\
s_{21}(t) & s_{22}(t) \\
\end{pmatrix}
\begin{pmatrix}
p_{1}(t) \\
p_{2}(t) \\
\end{pmatrix}=
\begin{pmatrix}
2 \sqrt{p_{1}(t)}\frac{d}{dt}\sqrt{p_1(t)} \\
2 \sqrt{p_{2}(t)}\frac{d}{dt}\sqrt{p_2(t)} \\
\end{pmatrix}
\end{eqnarray}
This equation is equivalent to
\begin{eqnarray}
\begin{pmatrix}
\frac{1}{2}s_{11}(t) & \frac{1}{2}\sqrt{\frac{p2(t)}{p1(t)}}s_{12}(t)\\ 
\frac{1}{2}\sqrt{\frac{p1(t)}{p2(t)}}s_{21}(t) & \frac{1}{2}s_{22}(t) \\ 
\end{pmatrix}
\begin{pmatrix}
\sqrt{p_1(t)} \\
\sqrt{p_2(t)} \\
\end{pmatrix}
=
\begin{pmatrix}
\frac{d}{dt}\sqrt{p_1(t)} \\
\frac{d}{dt}\sqrt{p_2(t)} \\
\end{pmatrix}=
\frac{i\hbar}{i\hbar}
\frac{d}{dt}
\begin{pmatrix}
\sqrt{p_1(t)} \\
\sqrt{p_2(t)} \\
\end{pmatrix}
\end{eqnarray}
and we set $t_1=i \hbar t$ so one has $dt_1=i \hbar dt$ and $t=\frac{t_1}{i\hbar}$ so we have \small
\begin{eqnarray}
\begin{pmatrix}
\frac{1}{2}s_{11}(\frac{t_1}{i\hbar}) & \frac{1}{2}\sqrt{\frac{p_1(t)}{p_2(t)}}s_{12}(\frac{t_1}{i\hbar})\\ 
\frac{1}{2}\sqrt{\frac{p_2(t)}{p_1(t)}}s_{21}(\frac{t_1}{i\hbar}) & \frac{1}{2}s_{22}(\frac{t_1}{i\hbar}) \\ 
\end{pmatrix} 
\begin{pmatrix}
\sqrt{p_1(\frac{t_1}{i\hbar})} \\
\sqrt{p_2(\frac{t_1}{i\hbar})} \\
\end{pmatrix}
=
i\hbar
\frac{d}{dt_1}
\begin{pmatrix}
\sqrt{p_1(\frac{t_1}{i\hbar})} \\
\sqrt{p_2(\frac{t_1}{i\hbar})} \\
\end{pmatrix}
\end{eqnarray}
\normalsize
The following notation is introduced:
\begin{eqnarray}
S_{12}[\frac{t_0}{i\hbar},t_1]=\frac{1}{2}\sqrt{\frac{p_1(t)}{p_2(t)}}s_{12}(\frac{t_1}{i\hbar})=S_{12R}[\frac{t_0}{i\hbar},t_1]+iS_{12I}[\frac{t_0}{i\hbar},t_1] \end{eqnarray}, 
\begin{eqnarray}
S_{21}[\frac{t_0}{i\hbar},t_1]=\frac{1}{2}\sqrt{\frac{p_2(t)}{p_1(t)}}s_{21}(\frac{t_1}{i\hbar})=S_{21R}[\frac{t_0}{i\hbar},t_1]+iS_{21I}[\frac{t_0}{i\hbar},t_1]
\end{eqnarray}
 and one obtains
\begin{eqnarray*}
\begin{pmatrix}
\frac{1}{2}s_{11}(\frac{t_1}{i\hbar})-\hbar \frac{d}{dt_1}\Theta_1(t) & S_{12R}[\frac{t_0}{i\hbar},t_1]+iS_{12R}[\frac{t_0}{i\hbar},t_1]e^{i(\Theta_1(t)-\Theta_2(t))}\\ 
S_{21R}[\frac{t_0}{i\hbar},t_1]+iS_{21I}[\frac{t_0}{i\hbar},t_1]e^{i((\Theta_2(t)-\Theta_1(t)))} & \frac{1}{2}s_{22}(\frac{t_1}{i\hbar})-\hbar \frac{d}{dt_1}\Theta_2(t) \\ 
\end{pmatrix}
\begin{pmatrix}
\sqrt{p_{1}(\frac{t_1}{i\hbar})}e^{i\Theta_1(t)} \\
\sqrt{p_{2}(\frac{t_1}{i\hbar})}e^{i\Theta_2(t)} \\
\end{pmatrix}
= \nonumber \\ =i \hbar \frac{d}{dt}
\begin{pmatrix}
\sqrt{p_{1}(\frac{t_1}{i\hbar})}e^{i\Theta_1(t)}\\
\sqrt{p_{2}(\frac{t_1}{i\hbar})}e^{i\Theta_2(t)} \\
\end{pmatrix}.
\end{eqnarray*}
Let us start from quantum mechanical perspective
\begin{eqnarray}
\begin{pmatrix}
E_{p1} & t_{sR}+it_{sI}\\
t_{sR}-it_{sI} & E_{p2}
\end{pmatrix}
\begin{pmatrix}
\sqrt{p_{1}}cos(\Theta_1)+i\sqrt{p_{1}}sin(\Theta_1)    \\
\sqrt{p_{2}}cos(\Theta_2)+i\sqrt{p_{1}}sin(\Theta_2)  \\
\end{pmatrix}=i\hbar \frac{d}{dt_1}
\begin{pmatrix}
\sqrt{p_{1}}cos(\Theta_1)+i\sqrt{p_{1}}sin(\Theta_1)  \\
\sqrt{p_{2}}cos(\Theta_2)+i\sqrt{p_{2}}sin(\Theta_2)  \\
\end{pmatrix}
\end{eqnarray}
and we obtain the set of 4 equations
\begin{eqnarray}
\begin{pmatrix}
E_{p1} & t_{sR}+it_{sI}\\
t_{sR}-it_{sI} & E_{p2}
\end{pmatrix}
\begin{pmatrix}
\sqrt{p_{1}}cos(\Theta_1)+i\sqrt{p_{1}}sin(\Theta_1)    \\
\sqrt{p_{2}}cos(\Theta_2)+i\sqrt{p_{2}}sin(\Theta_2)  \\
\end{pmatrix}=i\hbar \frac{d}{dt_1}
\begin{pmatrix}
\sqrt{p_{1}}cos(\Theta_1)+i\sqrt{p_{1}}sin(\Theta_1)  \\
\sqrt{p_{2}}cos(\Theta_2)+i\sqrt{p_{2}}sin(\Theta_2)  \\
\end{pmatrix}
\end{eqnarray}
that can be rewritten
\begin{eqnarray}
E_{p1}(\sqrt{p_{1}}cos(\Theta_1)+i\sqrt{p_{1}}sin(\Theta_1))+(t_{sR}+it_{sI})(\sqrt{p_{2}}cos(\Theta_1)+i\sqrt{p_{2}}sin\Theta_1))=\nonumber \\ =i\hbar \frac{d}{dt_1}(\sqrt{p_{1}}cos(\Theta_1)+i\sqrt{p_{1}}sin(\Theta_1)) \\
(t_{sR}-it_{sI})(\sqrt{p_{1}}cos(\Theta_1)+i\sqrt{p_{1}}sin\Theta_1)+E_{p2}(\sqrt{p_{2}}cos(\Theta_2)+i\sqrt{p_{2}}sin(\Theta_2))=\nonumber \\ =i\hbar \frac{d}{dt_1}(\sqrt{p_{2}}cos(\Theta_2)+i\sqrt{p_{2}}sin(\Theta_2)) \\
\end{eqnarray}
and that can be translated into
\begin{eqnarray}
E_{p1}\sqrt{p_{1}}cos(\Theta_1)+t_{sR}\sqrt{p_{2}}cos(\Theta_2)-t_{sI}\sqrt{p_{2}}sin(\Theta_2)=-\hbar \frac{d}{dt_1}\sqrt{p_{1}}sin(\Theta_1) \\
E_{p1}\sqrt{p_{1}}sin(\Theta_1)+t_{sR}\sqrt{p_{2}}sin(\Theta_2)+t_{sI}\sqrt{p_{2}}cos(\Theta_2)=+\hbar \frac{d}{dt_1}\sqrt{p_{1}}cos(\Theta_1) \\
E_{p2}\sqrt{p_{2}}cos(\Theta_2)+t_{sR}\sqrt{p_{1}}cos(\Theta_1)+t_{sI}\sqrt{p_{1}}sin(\Theta_1)=-\hbar \frac{d}{dt_1}\sqrt{p_{2}}sin(\Theta_2) \\
E_{p2}\sqrt{p_{2}}sin(\Theta_2)+t_{sR}\sqrt{p_{1}}sin(\Theta_1)-t_{sI}\sqrt{p_{1}}cos(\Theta_1)=+\hbar \frac{d}{dt_1}\sqrt{p_{2}}cos(\Theta_2) \\
\end{eqnarray}
so one can write

\begin{eqnarray}
\frac{1}{\hbar}
\begin{pmatrix}
0 & E_{p1} & +t_{sI} & t_{sR} \\
-E_{p1} & 0 & -t_{sR} & t_{sI}\\
-t_{sI} & t_{sR} & 0 & E_{p2}\\
-t_{sR} &
-it_{sI} &  -E_{p2} & 0\\
\end{pmatrix}
\begin{pmatrix}
\sqrt{p_{1}}cos(\Theta_1) \\
\sqrt{p_{1}}sin(\Theta_1)    \\
\sqrt{p_{2}}cos(\Theta_1)\\
\sqrt{p_{1}}sin(\Theta_1)  \\
\end{pmatrix}= \frac{d}{dt_1}
\begin{pmatrix}
\sqrt{p_{1}}cos(\Theta_1)\\
\sqrt{p_{1}}sin(\Theta_1)  \\
\sqrt{p_{2}}cos(\Theta_2) \\
\sqrt{p_{2}}sin(\Theta_2)  \\
\end{pmatrix}
\end{eqnarray}
what can be written as
\begin{eqnarray}
\hat{A}(t)
\begin{pmatrix}
\sqrt{p_{1}(t)}cos(\Theta_1(t)) \\
\sqrt{p_{1}(t)}sin(\Theta_1(t))    \\
\sqrt{p_{2}(t)}cos(\Theta_1(t))\\
\sqrt{p_{1}(t)}sin(\Theta_1(t))  \\
\end{pmatrix}= \frac{d}{dt}
\begin{pmatrix}
\sqrt{p_{1}(t)}cos(\Theta_1(t))\\
\sqrt{p_{1}(t)}sin(\Theta_1(t))  \\
\sqrt{p_{2}(t)}cos(\Theta_2(t)) \\
\sqrt{p_{2}(t)}sin(\Theta_2(t))  \\
\end{pmatrix},
\end{eqnarray}
what implies
\begin{eqnarray}
e^{\int_{t0}^{t}\hat{A}(t')dt'}
\begin{pmatrix}
\sqrt{p_{1}(t_0)}cos(\Theta_1(t_0))  \\
\sqrt{p_{1}(t_0)}sin(\Theta_1(t_0))  \\
\sqrt{p_{2}(t_0)}cos(\Theta_2(t_0))  \\
\sqrt{p_{2}(t_0)}sin(\Theta_2(t_0))  \\
\end{pmatrix}=
\begin{pmatrix}
\sqrt{p_{1}(t)}cos(\Theta_1(t))\\
\sqrt{p_{1}(t)}sin(\Theta_1(t))  \\
\sqrt{p_{2}(t)}cos(\Theta_2(t)) \\
\sqrt{p_{2}(t)}sin(\Theta_2(t))  \\
\end{pmatrix} 
\end{eqnarray}
Furthermore we can obtain
\small
\begin{eqnarray*}
2
\begin{pmatrix}
\sqrt{p_{1}(t)cos(\Theta_1(t))} & 0 & 0 & 0 \\
0 & \sqrt{p_{1}(t)sin(\Theta_1(t))}& 0 & 0   \\
0 & 0 & \sqrt{p_{2}(t)cos(\Theta_2(t))} & 0 \\
0 & 0 & 0 & \sqrt{p_{2}(t)sin(\Theta_2(t))}  \\
\end{pmatrix}
e^{\int_{t0}^{t}\hat{A}(t')dt'} \times \\ \times
\begin{pmatrix}
\frac{1}{\sqrt{p_{1}(t)cos(\Theta_1(t))^2}} & 0 & 0 & 0 \\
0 & \frac{1}{\sqrt{p_{1}(t)sin(\Theta_1(t))^2}} & 0 & 0   \\
0 & 0 & \frac{1}{\sqrt{p_{2}(t)cos(\Theta_2(t))^2}} & 0 \\
0 & 0 & 0 & \frac{1}{\sqrt{p_{2}(t)sin(\Theta_2(t))^2}}  \\
\end{pmatrix} \times \nonumber \\
\begin{pmatrix}
\sqrt{p_{1}(t)cos(\Theta_1(t))^2} & 0 & 0 & 0 \\
0 & \sqrt{p_{1}(t)sin(\Theta_1(t))^2}& 0 & 0   \\
0 & 0 & \sqrt{p_{2}(t)cos(\Theta_2(t))^2} & 0 \\
0 & 0 & 0 & \sqrt{p_{2}(t)sin(\Theta_2(t))^2}  \\
\end{pmatrix}
\begin{pmatrix}
\sqrt{p_{1}(t_0)cos(\Theta_1(t_0))}  \\
\sqrt{p_{1}(t_0)sin(\Theta_1(t_0))}  \\
\sqrt{p_{2}(t_0)cos(\Theta_2(t_0))}  \\
\sqrt{p_{2}(t_0)sin(\Theta_2(t_0))}  \\
\end{pmatrix}=\nonumber \\
2
\begin{pmatrix}
\sqrt{p_{1}(t)cos(\Theta_1(t))^2} & 0 & 0 & 0 \\
0 & \sqrt{p_{1}(t)sin(\Theta_1(t))^2}& 0 & 0   \\
0 & 0 & \sqrt{p_{2}(t)cos(\Theta_2(t)^2)} & 0 \\
0 & 0 & 0 & \sqrt{p_{2}(t)sin(\Theta_2(t)^2)}  \\
\end{pmatrix}
\begin{pmatrix}
\frac{d}{dt}\sqrt{p_{1}(t)cos(\Theta_1(t))^2}\\
\frac{d}{dt}\sqrt{p_{1}(t)sin(\Theta_1(t))^2}  \\
\frac{d}{dt}\sqrt{p_{2}(t)cos(\Theta_2(t))^2} \\
\frac{d}{dt}\sqrt{p_{2}(t)sin(\Theta_2(t))^2}  \\
\end{pmatrix}.
\end{eqnarray*}
\normalsize
The last set of equations is equivalent to
\small
\begin{eqnarray*}
\Bigg[
2
\begin{pmatrix}
\sqrt{p_{1}(t)(cos(\Theta_1(t)))^2} & 0 & 0 & 0 \\
0 & \sqrt{p_{1}(t)(sin(\Theta_1(t)))^2}& 0 & 0   \\
0 & 0 & \sqrt{p_{2}(t)(cos(\Theta_2(t)))^2} & 0 \\
0 & 0 & 0 & \sqrt{p_{2}(t)(sin(\Theta_2(t)))^2}  \\
\end{pmatrix}
\Bigg[
e^{\int_{t0}^{t}\hat{A}(t')dt'} \Bigg] \times \\ \times
\begin{pmatrix}
\frac{1}{\sqrt{p_{1}(t)(cos(\Theta_1(t)))^2}} & 0 & 0 & 0 \\
0 & \frac{1}{\sqrt{p_{1}(t)(sin(\Theta_1(t)))^2}} & 0 & 0   \\
0 & 0 & \frac{1}{\sqrt{p_{2}(t)(cos(\Theta_2(t)))^2}} & 0 \\
0 & 0 & 0 & \frac{1}{\sqrt{p_{2}(t)(sin(\Theta_2(t)))^2}}  \\
\end{pmatrix}
\Bigg]
\begin{pmatrix}
p_{1}(t_0)(cos(\Theta_1(t_0)))^2=p_{1R}(t)  \\
p_{1}(t_0)(sin(\Theta_1(t_0)))^2=p_{1I}(t)  \\
p_{2}(t_0)(cos(\Theta_2(t_0)))^2=p_{2R}(t)  \\
p_{2}(t_0)(sin(\Theta_2(t_0)))^2=p_{2I}(t)  \\
\end{pmatrix}=\nonumber \\
=\frac{d}{dt}
\begin{pmatrix}
p_{1}(t)(cos(\Theta_1(t))^2)\\
p_{1}(t)(sin(\Theta_1(t))^2)  \\
p_{2}(t)(cos(\Theta_2(t))^2) \\
p_{2}(t)(sin(\Theta_2(t))^2)  \\
\end{pmatrix} =
\frac{d}{dt}
\begin{pmatrix}
p_{1R}(t)  \\
p_{1I}(t)  \\
p_{2R}(t)  \\
p_{2I}(t)  \\
\end{pmatrix}
\end{eqnarray*}
\normalsize




\subsection{Case of 2 coupled 2 level classical stochastic finite state machines}
Instead of probabilities it will be useful to operate with square root of probabilities as they are present in quantum mechanics and in Schroedinger or Dirac equation. Since
\begin{eqnarray}
\frac{d}{dt}(\sqrt{p_1}\sqrt{p_1})=2 \sqrt{p_1(t)}\frac{d}{dt}\sqrt{p_1(t)}, \nonumber \\
\frac{d}{dt}(\sqrt{p_2}\sqrt{p_2})=2 \sqrt{p_2(t)}\frac{d}{dt}\sqrt{p_2(t)}, \nonumber \\
\frac{d}{dt}(\sqrt{p_3}\sqrt{p_3})=2 \sqrt{p_3(t)}\frac{d}{dt}\sqrt{p_3(t)}, \nonumber \\
\frac{d}{dt}(\sqrt{p_4}\sqrt{p_4})=2 \sqrt{p_4(t)}\frac{d}{dt}\sqrt{p_4(t)}, \nonumber \\
\end{eqnarray}
we can rewrite the epidemic equation as
\begin{eqnarray}
\begin{pmatrix}
s_{11}(t) & s_{12}(t) & s_{13}(t) & s_{14}(t)\\
s_{21}(t) & s_{22}(t) & s_{23}(t) & s_{24}(t)\\
s_{31}(t) & s_{32}(t) & s_{33}(t) & s_{34}(t)\\
s_{41}(t) & s_{42}(t) & s_{43}(t) & s_{44}(t)\\
\end{pmatrix}
\begin{pmatrix}
p_{1}(t) \\
p_{2}(t) \\
p_{3}(t) \\
p_{4}(t) \\
\end{pmatrix}=
\begin{pmatrix}
2 \sqrt{p_{1}(t)}\frac{d}{dt}\sqrt{p_1(t)} \\
2 \sqrt{p_{2}(t)}\frac{d}{dt}\sqrt{p_2(t)} \\
2 \sqrt{p_{3}(t)}\frac{d}{dt}\sqrt{p_3(t)} \\
2 \sqrt{p_{4}(t)}\frac{d}{dt}\sqrt{p_4(t)} \\
\end{pmatrix}
\end{eqnarray}
Equivalently we have
\begin{eqnarray}
\begin{pmatrix}
s_{11}(t) & s_{12}(t) & s_{13}(t) & s_{14}(t)\\
s_{21}(t) & s_{22}(t) & s_{23}(t) & s_{24}(t)\\
s_{31}(t) & s_{32}(t) & s_{33}(t) & s_{34}(t)\\
s_{41}(t) & s_{42}(t) & s_{43}(t) & s_{44}(t)\\
\end{pmatrix}
\begin{pmatrix}
\sqrt{p_1} & 0 & 0 & 0\\
0 & \sqrt{p_{2}(t)} & 0 & 0\\
0 & 0 & \sqrt{p_{3}(t)} & 0\\
0 & 0 & 0 & \sqrt{p_{4}(t)}\\
\end{pmatrix}
\begin{pmatrix}
\sqrt{p_{1}(t)} \\
\sqrt{p_{2}(t)} \\
\sqrt{p_{3}(t)} \\
\sqrt{p_{4}(t)} \\
\end{pmatrix}=
\begin{pmatrix}
2 \sqrt{p_{1}(t)}\frac{d}{dt}\sqrt{p_1(t)} \\
2 \sqrt{p_{2}(t)}\frac{d}{dt}\sqrt{p_2(t)} \\
2 \sqrt{p_{3}(t)}\frac{d}{dt}\sqrt{p_3(t)} \\
2 \sqrt{p_{4}(t)}\frac{d}{dt}\sqrt{p_4(t)} \\
\end{pmatrix}
\end{eqnarray}
and we have
\begin{eqnarray}
\frac{1}{2}
\begin{pmatrix}
\frac{1}{\sqrt{p_1}} & 0 & 0 & 0\\
0 & \frac{1}{\sqrt{p_{2}(t)}} & 0 & 0\\
0 & 0 & \frac{1}{\sqrt{p_{3}(t)}} & 0\\
0 & 0 & 0 & \frac{1}{\sqrt{p_{4}(t)}}\\
\end{pmatrix}
\begin{pmatrix}
s_{11}(t) & s_{12}(t) & s_{13}(t) & s_{14}(t)\\
s_{21}(t) & s_{22}(t) & s_{23}(t) & s_{24}(t)\\
s_{31}(t) & s_{32}(t) & s_{33}(t) & s_{34}(t)\\
s_{41}(t) & s_{42}(t) & s_{43}(t) & s_{44}(t)\\
\end{pmatrix}
\begin{pmatrix}
\sqrt{p_1} & 0 & 0 & 0\\
0 & \sqrt{p_{2}(t)} & 0 & 0\\
0 & 0 & \sqrt{p_{3}(t)} & 0\\
0 & 0 & 0 & \sqrt{p_{4}(t)}\\
\end{pmatrix}
\begin{pmatrix}
\sqrt{p_{1}(t)} \\
\sqrt{p_{2}(t)} \\
\sqrt{p_{3}(t)} \\
\sqrt{p_{4}(t)} \\
\end{pmatrix}= \nonumber \\ = \frac{1}{2}
\begin{pmatrix}
\frac{1}{\sqrt{p_1}} & 0 & 0 & 0\\
0 & \frac{1}{\sqrt{p_{2}(t)}} & 0 & 0\\
0 & 0 & \frac{1}{\sqrt{p_{3}(t)}} & 0\\
0 & 0 & 0 & \frac{1}{\sqrt{p_{4}(t)}}\\
\end{pmatrix}
\begin{pmatrix}
2 \sqrt{p_{1}(t)}\frac{d}{dt}\sqrt{p_1(t)} \\
2 \sqrt{p_{2}(t)}\frac{d}{dt}\sqrt{p_2(t)} \\
2 \sqrt{p_{3}(t)}\frac{d}{dt}\sqrt{p_3(t)} \\
2 \sqrt{p_{4}(t)}\frac{d}{dt}\sqrt{p_4(t)} \\
\end{pmatrix}
\end{eqnarray}

In very real way we obtain the following classical physics Hamiltonian and we obtain

\begin{eqnarray}
\hat{H}_{classical}(t)=
\frac{1}{2}
\begin{pmatrix}
\frac{1}{\sqrt{p_1(t)}} & 0 & 0 & 0\\
0 & \frac{1}{\sqrt{p_{2}(t)}} & 0 & 0\\
0 & 0 & \frac{1}{\sqrt{p_{3}(t)}} & 0\\
0 & 0 & 0 & \frac{1}{\sqrt{p_{4}(t)}}\\
\end{pmatrix}
\begin{pmatrix}
s_{11}(t) & s_{12}(t) & s_{13}(t) & s_{14}(t)\\
s_{21}(t) & s_{22}(t) & s_{23}(t) & s_{24}(t)\\
s_{31}(t) & s_{32}(t) & s_{33}(t) & s_{34}(t)\\
s_{41}(t) & s_{42}(t) & s_{43}(t) & s_{44}(t)\\
\end{pmatrix}
\begin{pmatrix}
\sqrt{p_1(t)} & 0 & 0 & 0\\
0 & \sqrt{p_{2}(t)} & 0 & 0\\
0 & 0 & \sqrt{p_{3}(t)} & 0\\
0 & 0 & 0 & \sqrt{p_{4}(t)}\\
\end{pmatrix}= \nonumber \\
=
\frac{1}{2}
\begin{pmatrix}
\frac{1}{\sqrt{p_1}(t)} & 0 & 0 & 0\\
0 & \frac{1}{\sqrt{p_{2}(t)}} & 0 & 0\\
0 & 0 & \frac{1}{\sqrt{p_{3}(t)}} & 0\\
0 & 0 & 0 & \frac{1}{\sqrt{p_{4}(t)}}\\
\end{pmatrix}
\begin{pmatrix}
\sqrt{p_1(t)}s_{11}(t) & \sqrt{p_2}(t)s_{12}(t) & \sqrt{p_3(t)}s_{13}(t) & \sqrt{p_4(t)}s_{14}(t)\\
\sqrt{p_1(t)}s_{21}(t) & \sqrt{p_2(t)}s_{22}(t) & \sqrt{p_3(t)}s_{23}(t) & \sqrt{p_4}(t)s_{24}(t)\\
\sqrt{p_1(t)}s_{31}(t) & \sqrt{p_2(t)}s_{32}(t) & \sqrt{p_3(t)}s_{33}(t) & \sqrt{p_4}(t)s_{34}(t)\\
\sqrt{p_1(t)}s_{41}(t) & \sqrt{p_2(t)}s_{42}(t) & \sqrt{p_3(t)}s_{43}(t) & \sqrt{p_4}(t)s_{44}(t)\\
\end{pmatrix}= \nonumber \\
=\frac{1}{2}
\begin{pmatrix}
s_{11}(t) & \frac{\sqrt{p_2}(t)}{\sqrt{p_1}(t)}s_{12}(t) & \frac{\sqrt{p_3}(t)}{\sqrt{p_1}(t)}s_{13}(t) & \frac{\sqrt{p_4}(t)}{\sqrt{p_1}(t)}s_{14}(t)\\
\frac{\sqrt{p_1(t)}s_{21}(t)}{\sqrt{p_2}} & s_{22}(t) & \frac{\sqrt{p_3}(t)}{\sqrt{p_2(t)}}s_{23}(t) & \frac{\sqrt{p_4(t)}}{\sqrt{p_2(t)}}s_{24}(t)\\
\frac{\sqrt{p_1(t)}s_{31}(t)}{\sqrt{p_3(t)}} & \frac{\sqrt{p_2(t)}s_{32}(t)}{\sqrt{p_3(t)}} & s_{33}(t) & \frac{\sqrt{p_4(t)}s_{34}(t)}{\sqrt{p_3(t)}}\\
\frac{\sqrt{p_1(t)}s_{41}(t)}{\sqrt{p_4(t)}} & \frac{\sqrt{p_2(t)}s_{42}(t)}{\sqrt{p_4(t)}} & \frac{\sqrt{p_3(t)}s_{43}(t)}{\sqrt{p_4(t)}} & s_{44}(t)\\
\end{pmatrix}
\end{eqnarray}

This equation is equivalent to
\begin{eqnarray}
\begin{pmatrix}
\frac{1}{2}s_{11}(t) & \frac{1}{2}\sqrt{\frac{p_2(t)}{p_1(t)}}s_{12}(t) & \frac{1}{2}\sqrt{\frac{p_3(t)}{p_1(t)}}s_{13}(t) & \frac{1}{2}\sqrt{\frac{p_4(t)}{p_1(t)}}s_{14}(t) \\ 
\frac{1}{2}\sqrt{\frac{p_1(t)}{p_2(t)}}s_{21}(t) & \frac{1}{2}s_{22}(t) & \frac{1}{2}\sqrt{\frac{p_3(t)}{p_2(t)}}s_{23}(t) & \frac{1}{2}\sqrt{\frac{p_4(t)}{p_2(t)}}s_{24}(t) \\ 
\frac{1}{2}\sqrt{\frac{p_1(t)}{p_3(t)}}s_{31}(t) & \frac{1}{2}\sqrt{\frac{p_2(t)}{p_3(t)}}s_{32}(t) & \frac{1}{2}s_{33}(t) & \frac{1}{2}\sqrt{\frac{p_4(t)}{p_3(t)}}s_{34}(t) \\ 
\frac{1}{2}\sqrt{\frac{p_1(t)}{p_4(t)}}s_{41}(t) & \frac{1}{2}\sqrt{\frac{p_2(t)}{p_4(t)}}s_{42}(t) & \frac{1}{2}\sqrt{\frac{p_3(t)}{p_4(t)}}s_{43}(t) & s_{44}(t) \\ 
\end{pmatrix}
\begin{pmatrix}
\sqrt{p_1(t)} \\
\sqrt{p_2(t)} \\
\sqrt{p_3(t)} \\
\sqrt{p_4(t)} \\
\end{pmatrix}
=\frac{d}{dt}
\begin{pmatrix}
\sqrt{p_1(t)} \\
\sqrt{p_2(t)} \\
\sqrt{p_3(t)} \\
\sqrt{p_4(t)} \\
\end{pmatrix}=
\frac{i\hbar}{i\hbar}
\frac{d}{dt}
\begin{pmatrix}
\sqrt{p_1(t)} \\
\sqrt{p_2(t)} \\
\sqrt{p_3(t)} \\
\sqrt{p_4(t)} \\
\end{pmatrix} = \nonumber \\
=\frac{d}{dt}\ket{\psi}_{classical}(t)=\hat{H}_{classical}(t)\ket{\psi}_{classical}(t).
\end{eqnarray}
At this stage we can identify analytical solutions given as

\begin{eqnarray}
exp \Bigg[
\begin{pmatrix}
\int_{t_0}^{t}dt'\frac{1}{2}s_{11}(t') & \int_{t_0}^{t}dt'\frac{1}{2}\sqrt{\frac{p_2(t')}{p_1(t')}}s_{12}(t') & \int_{t_0}^{t}dt'\frac{1}{2}\sqrt{\frac{p_3(t')}{p_1(t')}}s_{13}(t') & \int_{t_0}^{t}dt'\frac{1}{2}\sqrt{\frac{p_4(t')}{p_1(t')}}s_{14}(t') \\ 
\int_{t_0}^{t}dt'\frac{1}{2}\sqrt{\frac{p_1(t')}{p_2(t')}}s_{21}(t') & \int_{t_0}^{t}dt'\frac{1}{2}s_{22}(t') &\int_{t_0}^{t}dt' \frac{1}{2}\sqrt{\frac{p_3(t')}{p_2(t')}}s_{23}(t') & \int_{t_0}^{t}dt'\frac{1}{2}\sqrt{\frac{p_4(t')}{p_2(t')}}s_{24}(t') \\ 
\int_{t_0}^{t}dt'\frac{1}{2}\sqrt{\frac{p_1(t')}{p_3(t')}}s_{31}(t') & \int_{t_0}^{t}dt'\frac{1}{2}\sqrt{\frac{p_2(t')}{p_3(t')}}s_{32}(t') & \int_{t_0}^{t}dt'\frac{1}{2}s_{33}(t') & \int_{t_0}^{t}dt'\frac{1}{2}\sqrt{\frac{p_4(t')}{p_3(t')}}s_{34}(t') \\ 
\int_{t_0}^{t}dt'\frac{1}{2}\sqrt{\frac{p_1(t')}{p_4(t')}}s_{41}(t') & \int_{t_0}^{t}dt'\frac{1}{2}\sqrt{\frac{p_2(t')}{p_4(t')}}s_{42}(t') & \int_{t_0}^{t}dt'\frac{1}{2}\sqrt{\frac{p_3(t')}{p_4(t')}}s_{43}(t') & \int_{t_0}^{t}dt's_{44}(t') \\ 
\end{pmatrix} \Bigg]
\begin{pmatrix}
\sqrt{p_1(t_0)} \\
\sqrt{p_2(t_0)} \\
\sqrt{p_3(t_0)} \\
\sqrt{p_4(t_0)} \\
\end{pmatrix}  
=
\begin{pmatrix}
\sqrt{p_1(t)} \\
\sqrt{p_2(t)} \\
\sqrt{p_3(t)} \\
\sqrt{p_4(t)} \\
\end{pmatrix}
\end{eqnarray}

We have
\begin{eqnarray}
\frac{d}{dt}\bra{\psi(t)}_{classical}=\bra{\psi(t)}_{classical}[\hat{H(t)}_{classical}]^{T}.
\end{eqnarray}
and it implies that
\begin{eqnarray}
\frac{d}{dt}(\ket{\psi(t)}(t)\bra{\psi}_{classical})=(\frac{d}{dt}(\ket{\psi(t)})\bra{\psi}_{classical}+(\ket{\psi(t)})\frac{d}{dt}(\bra{\psi}_{classical})) = \nonumber \\=\hat{H(t)}_{classical}\ket{\psi(t)}\bra{\psi(t)}_{classical}+\ket{\psi(t)}\bra{\psi(t)}_{classical}[\hat{H(t)}_{classical}]^{T}=\Bigg[\hat{H(t)}_{classical}, \ket{\psi(t)}\bra{\psi(t)}_{classical} \Bigg]_{+},
\end{eqnarray}
where $\Bigg[\hat{A},\hat{B}\Bigg]_{+}=\hat{A}\hat{B}+\hat{B}\hat{A}$ is anticommutator. We notice that in such situation we have commutator in Quantum Mechanics defined as
$\Bigg[\hat{A},\hat{B}\Bigg]_{-}=\hat{A}\hat{B}-\hat{B}\hat{A}$ and thus
\begin{eqnarray}
i \hbar \frac{d}{dt}(\ket{\psi(t)}(t)\bra{\psi}_{quantum})=i \hbar(\frac{d}{dt}(\ket{\psi(t)})\bra{\psi}_{quantum}+i \hbar(\ket{\psi(t)})\frac{d}{dt}(\bra{\psi}_{classical})) = \nonumber \\=\hat{H}(t)_{quantum}\ket{\psi(t)}\bra{\psi(t)}_{quantum}-\ket{\psi(t)}\bra{\psi(t)}_{quantum}[\hat{H}(t)_{quantum}]^{\dag}=\Bigg[\hat{H}(t)_{quantum}, \ket{\psi(t)}\bra{\psi(t)}_{quantum} \Bigg]_{-}= \nonumber \\
=\Bigg[\hat{H}(t)_{quantum}, \rho_{quantum} \Bigg]_{-}.
\end{eqnarray}

.
We recognize analogies with Schroedinger equation that are occurring only in case of real valued wave functions.
We can introduce the object similar to quantum density matrix that is classical density matrix. We have
\begin{eqnarray}
\hat{\rho}_{classical}(t)=
\begin{pmatrix}
\sqrt{p_1(t)} \\
\sqrt{p_2(t)} \\
\sqrt{p_3(t)} \\
\sqrt{p_4(t)} \\
\end{pmatrix}
\begin{pmatrix}
\sqrt{p_1(t)} & \sqrt{p_2(t)} & \sqrt{p_3(t)} & \sqrt{p_4(t)} \\
\end{pmatrix}=\nonumber \\ =
\begin{pmatrix}
p_1(t) & \sqrt{p_1(t)}\sqrt{p_2(t)} & \sqrt{p_1(t)}\sqrt{p_3(t)} & \sqrt{p_1(t)}\sqrt{p_4(t)}\\
\sqrt{p_2(t)}\sqrt{p_1(t)} & p_2(t)& \sqrt{p_2(t)}\sqrt{p_3(t)} & \sqrt{p_2(t)}\sqrt{p_4(t)} \\
\sqrt{p_3(t)}\sqrt{p_1(t)} & \sqrt{p_3(t)}\sqrt{p_2(t)}& p_3(t) & \sqrt{p_3(t)}\sqrt{p_4(t)} \\
\sqrt{p_4(t)}\sqrt{p_1(t)}& \sqrt{p_4(t)}\sqrt{p_2(t)} & \sqrt{p_4(t)}\sqrt{p_3(t)} & p_4(t)\\
\end{pmatrix}=\hat{\rho}_{classical}(t)^{T}
\end{eqnarray}
We can obtain the analytical solution for such system in quite straighforward way
\begin{eqnarray}
\hat{\rho}_{classical}(t)=
exp \Bigg[
\begin{pmatrix}
\int_{t_0}^{t}dt'\frac{1}{2}s_{11}(t') & \int_{t_0}^{t}dt'\frac{1}{2}\sqrt{\frac{p_2(t')}{p_1(t')}}s_{12}(t') & \int_{t_0}^{t}dt'\frac{1}{2}\sqrt{\frac{p_3(t')}{p_1(t')}}s_{13}(t') & \int_{t_0}^{t}dt'\frac{1}{2}\sqrt{\frac{p_4(t')}{p_1(t')}}s_{14}(t') \\ 
\int_{t_0}^{t}dt'\frac{1}{2}\sqrt{\frac{p_1(t')}{p_2(t')}}s_{21}(t') & \int_{t_0}^{t}dt'\frac{1}{2}s_{22}(t') &\int_{t_0}^{t}dt' \frac{1}{2}\sqrt{\frac{p_3(t')}{p_2(t')}}s_{23}(t') & \int_{t_0}^{t}dt'\frac{1}{2}\sqrt{\frac{p_4(t')}{p_2(t')}}s_{24}(t') \\ 
\int_{t_0}^{t}dt'\frac{1}{2}\sqrt{\frac{p_1(t')}{p_3(t')}}s_{31}(t') & \int_{t_0}^{t}dt'\frac{1}{2}\sqrt{\frac{p_2(t')}{p_3(t')}}s_{32}(t') & \int_{t_0}^{t}dt'\frac{1}{2}s_{33}(t') & \int_{t_0}^{t}dt'\frac{1}{2}\sqrt{\frac{p_4(t')}{p_3(t')}}s_{34}(t') \\ 
\int_{t_0}^{t}dt'\frac{1}{2}\sqrt{\frac{p_1(t')}{p_4(t')}}s_{41}(t') & \int_{t_0}^{t}dt'\frac{1}{2}\sqrt{\frac{p_2(t')}{p_4(t')}}s_{42}(t') & \int_{t_0}^{t}dt'\frac{1}{2}\sqrt{\frac{p_3(t')}{p_4(t')}}s_{43}(t') & \int_{t_0}^{t}dt's_{44}(t') \\ 
\end{pmatrix} \Bigg] \times
\hat{\rho}_{classical}(t_0) \times \nonumber \\
\times \exp \Bigg[ \Bigg[
\begin{pmatrix}
\int_{t_0}^{t}dt'\frac{1}{2}s_{11}(t') & \int_{t_0}^{t}dt'\frac{1}{2}\sqrt{\frac{p_2(t')}{p_1(t')}}s_{12}(t') & \int_{t_0}^{t}dt'\frac{1}{2}\sqrt{\frac{p_3(t')}{p_1(t')}}s_{13}(t') & \int_{t_0}^{t}dt'\frac{1}{2}\sqrt{\frac{p_4(t')}{p_1(t')}}s_{14}(t') \\ 
\int_{t_0}^{t}dt'\frac{1}{2}\sqrt{\frac{p_1(t')}{p_2(t')}}s_{21}(t') & \int_{t_0}^{t}dt'\frac{1}{2}s_{22}(t') &\int_{t_0}^{t}dt' \frac{1}{2}\sqrt{\frac{p_3(t')}{p_2(t')}}s_{23}(t') & \int_{t_0}^{t}dt'\frac{1}{2}\sqrt{\frac{p_4(t')}{p_2(t')}}s_{24}(t') \\ 
\int_{t_0}^{t}dt'\frac{1}{2}\sqrt{\frac{p_1(t')}{p_3(t')}}s_{31}(t') & \int_{t_0}^{t}dt'\frac{1}{2}\sqrt{\frac{p_2(t')}{p_3(t')}}s_{32}(t') & \int_{t_0}^{t}dt'\frac{1}{2}s_{33}(t') & \int_{t_0}^{t}dt'\frac{1}{2}\sqrt{\frac{p_4(t')}{p_3(t')}}s_{34}(t') \\ 
\int_{t_0}^{t}dt'\frac{1}{2}\sqrt{\frac{p_1(t')}{p_4(t')}}s_{41}(t') & \int_{t_0}^{t}dt'\frac{1}{2}\sqrt{\frac{p_2(t')}{p_4(t')}}s_{42}(t') & \int_{t_0}^{t}dt'\frac{1}{2}\sqrt{\frac{p_3(t')}{p_4(t')}}s_{43}(t') & \int_{t_0}^{t}dt's_{44}(t') \\ 
\end{pmatrix} \Bigg]^T \Bigg]
\end{eqnarray}
We can compute the reduced density matrix of subsystem A and B in standard way and we can establish von-Neumann entropy relation.

By introducing $t_1=i\hbar t$ we have $\frac{t_1}{i\hbar}=t$ and it leads to
\begin{eqnarray}
\begin{pmatrix}
\frac{1}{2}s_{11}(\frac{t_1}{i\hbar}) & \frac{1}{2}\sqrt{\frac{p_2(\frac{t_1}{i\hbar})}{p_1(\frac{t_1}{i\hbar})}}s_{12}(\frac{t_1}{i\hbar}) & \frac{1}{2}\sqrt{\frac{p_3(\frac{t_1}{i\hbar})}{p_1(\frac{t_1}{i\hbar})}}s_{13}(\frac{t_1}{i\hbar}) & \frac{1}{2}\sqrt{\frac{p_4(\frac{t_1}{i\hbar})}{p_1(\frac{t_1}{i\hbar})}}s_{14}(\frac{t_1}{i\hbar}) \\ 
\frac{1}{2}\sqrt{\frac{p_1(\frac{t_1}{i\hbar})}{p_2(\frac{t_1}{i\hbar})}}s_{21}(\frac{t_1}{i\hbar}) & \frac{1}{2}s_{22}(\frac{t_1}{i\hbar}) & \frac{1}{2}\sqrt{\frac{p_3(\frac{t_1}{i\hbar})}{p_2(t)}}s_{23}(\frac{t_1}{i\hbar}) & \frac{1}{2}\sqrt{\frac{p_4(\frac{t_1}{i\hbar})}{p_2(\frac{t_1}{i\hbar})}}s_{24}(\frac{t_1}{i\hbar}) \\ 
\frac{1}{2}\sqrt{\frac{p_1(\frac{t_1}{i\hbar})}{p_3(\frac{t_1}{i\hbar})}}s_{31}(\frac{t_1}{i\hbar}) & \frac{1}{2}\sqrt{\frac{p_2(\frac{t_1}{i\hbar})}{p_3(\frac{t_1}{i\hbar})}}s_{32}(\frac{t_1}{i\hbar}) & \frac{1}{2}s_{33}(\frac{t_1}{i\hbar}) & \frac{1}{2}\sqrt{\frac{p_4(\frac{t_1}{i\hbar})}{p_3(\frac{t_1}{i\hbar})}}s_{34}(\frac{t_1}{i\hbar}) \\ 
\frac{1}{2}\sqrt{\frac{p_1(\frac{t_1}{i\hbar})}{p_4(\frac{t_1}{i\hbar})}}s_{41}(\frac{t_1}{i\hbar}) & \frac{1}{2}\sqrt{\frac{p_2(\frac{t_1}{i\hbar})}{p_4(\frac{t_1}{i\hbar})}}s_{42}(\frac{t_1}{i\hbar}) & \frac{1}{2}\sqrt{\frac{p_3(\frac{t_1}{i\hbar})}{p_4(\frac{t_1}{i\hbar})}}s_{43}(\frac{t_1}{i\hbar}) & s_{44}(\frac{t_1}{i\hbar}) \\ 
\end{pmatrix}
\begin{pmatrix}
\sqrt{p_1(\frac{t_1}{i\hbar})} \\
\sqrt{p_2(\frac{t_1}{i\hbar})} \\
\sqrt{p_3(\frac{t_1}{i\hbar})} \\
\sqrt{p_4(\frac{t_1}{i\hbar})} \\
\end{pmatrix}
=
i\hbar
\frac{d}{dt_1}
\begin{pmatrix}
\sqrt{p_1(\frac{t_1}{i\hbar})} \\
\sqrt{p_2(\frac{t_1}{i\hbar})} \\
\sqrt{p_3(\frac{t_1}{i\hbar})} \\
\sqrt{p_4(\frac{t_1}{i\hbar})} \\
\end{pmatrix}
\end{eqnarray}
It leads us towards analytical solution

\begin{eqnarray}
\begin{pmatrix}
\sqrt{p_1(\frac{t_1}{i\hbar})} \\
\sqrt{p_2(\frac{t_1}{i\hbar})} \\
\sqrt{p_3(\frac{t_1}{i\hbar})} \\
\sqrt{p_4(\frac{t_1}{i\hbar})} \\
\end{pmatrix}
= \nonumber \\ =
\exp [ \Bigg[
\begin{pmatrix}
\frac{1}{i\hbar}\int_{0}^{t_1}dt_1'\frac{1}{2}s_{11}(\frac{t_1}{i\hbar}) & \frac{1}{i\hbar}\int_{0}^{t_1}dt_1\frac{1}{2}\sqrt{\frac{p_2(\frac{t_1}{i\hbar})}{p_1(\frac{t_1}{i\hbar})}}s_{12}(\frac{t_1}{i\hbar}) & \frac{1}{i\hbar}\int_{0}^{t_1}dt_1'\frac{1}{2}\sqrt{\frac{p_3(\frac{t_1}{i\hbar})}{p_1(\frac{t_1}{i\hbar})}}s_{13}(\frac{t_1}{i\hbar}) & \frac{1}{i\hbar}\int_{0}^{t_1}dt_1'\frac{1}{2}\sqrt{\frac{p_4(\frac{t_1}{i\hbar})}{p_1(\frac{t_1}{i\hbar})}}s_{14}(\frac{t_1}{i\hbar}) \\ 
\frac{1}{i\hbar}\int_{0}^{t_1}dt_1'\frac{1}{2}\sqrt{\frac{p_1(\frac{t_1}{i\hbar})}{p_2(\frac{t_1}{i\hbar})}}s_{21}(\frac{t_1}{i\hbar}) & \frac{1}{2}s_{22}(\frac{t_1}{i\hbar}) & \frac{1}{i\hbar}\int_{0}^{t_1}dt_1'\frac{1}{2}\sqrt{\frac{p_3(\frac{t_1}{i\hbar})}{p_2(t)}}s_{23}(\frac{t_1}{i\hbar}) & \frac{1}{i\hbar}\int_{0}^{t_1}dt_1'\frac{1}{2}\sqrt{\frac{p_4(\frac{t_1}{i\hbar})}{p_2(\frac{t_1}{i\hbar})}}s_{24}(\frac{t_1}{i\hbar}) \\ 
\frac{1}{i\hbar}\int_{0}^{t_1}dt_1'\frac{1}{2}\sqrt{\frac{p_1(\frac{t_1}{i\hbar})}{p_3(\frac{t_1}{i\hbar})}}s_{31}(\frac{t_1}{i\hbar}) & \frac{1}{i\hbar}\int_{0}^{t_1}dt_1'\frac{1}{2}\sqrt{\frac{p_2(\frac{t_1}{i\hbar})}{p_3(\frac{t_1}{i\hbar})}}s_{32}(\frac{t_1}{i\hbar}) & \frac{1}{2}s_{33}(\frac{t_1}{i\hbar}) & \frac{1}{i\hbar}\int_{0}^{t_1}dt_1'\frac{1}{2}\sqrt{\frac{p_4(\frac{t_1}{i\hbar})}{p_3(\frac{t_1}{i\hbar})}}s_{34}(\frac{t_1}{i\hbar}) \\ 
\frac{1}{i\hbar}\int_{0}^{t_1}dt_1'\frac{1}{2}\sqrt{\frac{p_1(\frac{t_1}{i\hbar})}{p_4(\frac{t_1}{i\hbar})}}s_{41}(\frac{t_1}{i\hbar}) & \frac{1}{i\hbar}\int_{0}^{t_1}dt_1'\frac{1}{2}\sqrt{\frac{p_2(\frac{t_1}{i\hbar})}{p_4(\frac{t_1}{i\hbar})}}s_{42}(\frac{t_1}{i\hbar}) & \frac{1}{i\hbar}\int_{0}^{t_1}dt_1'\frac{1}{2}\sqrt{\frac{p_3(\frac{t_1}{i\hbar})}{p_4(\frac{t_1}{i\hbar})}}s_{43}(\frac{t_1}{i\hbar}) & \frac{1}{i\hbar}\int_{0}^{t_1}dt_1s_{44}(\frac{t_1}{i\hbar}) \\ 
\end{pmatrix}
\Bigg]
\begin{pmatrix}
\sqrt{p_1(\frac{0}{i\hbar})} \\
\sqrt{p_2(\frac{0}{i\hbar})} \\
\sqrt{p_3(\frac{0}{i\hbar})} \\
\sqrt{p_4(\frac{0}{i\hbar})} \\
\end{pmatrix}
\end{eqnarray}
Dealing in similar fashion we can obtain quantum density matrix given as
\begin{eqnarray}
\begin{pmatrix}
\sqrt{p_1(\frac{t_1}{i\hbar})} \\
\sqrt{p_2(\frac{t_1}{i\hbar})} \\
\sqrt{p_3(\frac{t_1}{i\hbar})} \\
\sqrt{p_4(\frac{t_1}{i\hbar})} \\
\end{pmatrix}
\begin{pmatrix}
\sqrt{p_1(\frac{t_1}{i\hbar})}^{*} & \sqrt{p_2(\frac{t_1}{i\hbar})}^{*} & \sqrt{p_3(\frac{t_1}{i\hbar})}^{*} & \sqrt{p_4(\frac{t_1}{i\hbar})}^{*} \\
\end{pmatrix}
= \nonumber \\ =
\exp [ \Bigg[
\begin{pmatrix}
\frac{1}{i\hbar}\int_{0}^{t_1}dt_1'\frac{1}{2}s_{11}(\frac{t_1}{i\hbar}) & \frac{1}{i\hbar}\int_{0}^{t_1}dt_1\frac{1}{2}\sqrt{\frac{p_2(\frac{t_1}{i\hbar})}{p_1(\frac{t_1}{i\hbar})}}s_{12}(\frac{t_1}{i\hbar}) & \frac{1}{i\hbar}\int_{0}^{t_1}dt_1'\frac{1}{2}\sqrt{\frac{p_3(\frac{t_1}{i\hbar})}{p_1(\frac{t_1}{i\hbar})}}s_{13}(\frac{t_1}{i\hbar}) & \frac{1}{i\hbar}\int_{0}^{t_1}dt_1'\frac{1}{2}\sqrt{\frac{p_4(\frac{t_1}{i\hbar})}{p_1(\frac{t_1}{i\hbar})}}s_{14}(\frac{t_1}{i\hbar}) \\ 
\frac{1}{i\hbar}\int_{0}^{t_1}dt_1'\frac{1}{2}\sqrt{\frac{p_1(\frac{t_1}{i\hbar})}{p_2(\frac{t_1}{i\hbar})}}s_{21}(\frac{t_1}{i\hbar}) & \frac{1}{2}s_{22}(\frac{t_1}{i\hbar}) & \frac{1}{i\hbar}\int_{0}^{t_1}dt_1'\frac{1}{2}\sqrt{\frac{p_3(\frac{t_1}{i\hbar})}{p_2(t)}}s_{23}(\frac{t_1}{i\hbar}) & \frac{1}{i\hbar}\int_{0}^{t_1}dt_1'\frac{1}{2}\sqrt{\frac{p_4(\frac{t_1}{i\hbar})}{p_2(\frac{t_1}{i\hbar})}}s_{24}(\frac{t_1}{i\hbar}) \\ 
\frac{1}{i\hbar}\int_{0}^{t_1}dt_1'\frac{1}{2}\sqrt{\frac{p_1(\frac{t_1}{i\hbar})}{p_3(\frac{t_1}{i\hbar})}}s_{31}(\frac{t_1}{i\hbar}) & \frac{1}{i\hbar}\int_{0}^{t_1}dt_1'\frac{1}{2}\sqrt{\frac{p_2(\frac{t_1}{i\hbar})}{p_3(\frac{t_1}{i\hbar})}}s_{32}(\frac{t_1}{i\hbar}) & \frac{1}{2}s_{33}(\frac{t_1}{i\hbar}) & \frac{1}{i\hbar}\int_{0}^{t_1}dt_1'\frac{1}{2}\sqrt{\frac{p_4(\frac{t_1}{i\hbar})}{p_3(\frac{t_1}{i\hbar})}}s_{34}(\frac{t_1}{i\hbar}) \\ 
\frac{1}{i\hbar}\int_{0}^{t_1}dt_1'\frac{1}{2}\sqrt{\frac{p_1(\frac{t_1}{i\hbar})}{p_4(\frac{t_1}{i\hbar})}}s_{41}(\frac{t_1}{i\hbar}) & \frac{1}{i\hbar}\int_{0}^{t_1}dt_1'\frac{1}{2}\sqrt{\frac{p_2(\frac{t_1}{i\hbar})}{p_4(\frac{t_1}{i\hbar})}}s_{42}(\frac{t_1}{i\hbar}) & \frac{1}{i\hbar}\int_{0}^{t_1}dt_1'\frac{1}{2}\sqrt{\frac{p_3(\frac{t_1}{i\hbar})}{p_4(\frac{t_1}{i\hbar})}}s_{43}(\frac{t_1}{i\hbar}) & \frac{1}{i\hbar}\int_{0}^{t_1}dt_1s_{44}(\frac{t_1}{i\hbar}) \\ 
\end{pmatrix}
\Bigg] \times \nonumber \\ \times
\begin{pmatrix}
\sqrt{p_1(\frac{0}{i\hbar})} \\
\sqrt{p_2(\frac{0}{i\hbar})} \\
\sqrt{p_3(\frac{0}{i\hbar})} \\
\sqrt{p_4(\frac{0}{i\hbar})} \\
\end{pmatrix}
\begin{pmatrix}
\sqrt{p_1(\frac{t_1}{i\hbar})}^{*} & \sqrt{p_2(\frac{t_1}{i\hbar})}^{*} & \sqrt{p_3(\frac{t_1}{i\hbar})}^{*} & \sqrt{p_4(\frac{t_1}{i\hbar})}^{*} \\
\end{pmatrix}
\times \nonumber \\
\times \exp [ -\Bigg[
\begin{pmatrix}
\frac{1}{i\hbar}\int_{0}^{t_1}dt_1'\frac{1}{2}s_{11}(\frac{t_1}{i\hbar}) & \frac{1}{i\hbar}\int_{0}^{t_1}dt_1\frac{1}{2}\sqrt{\frac{p_2(\frac{t_1}{i\hbar})}{p_1(\frac{t_1}{i\hbar})}}s_{12}(\frac{t_1}{i\hbar}) & \frac{1}{i\hbar}\int_{0}^{t_1}dt_1'\frac{1}{2}\sqrt{\frac{p_3(\frac{t_1}{i\hbar})}{p_1(\frac{t_1}{i\hbar})}}s_{13}(\frac{t_1}{i\hbar}) & \frac{1}{i\hbar}\int_{0}^{t_1}dt_1'\frac{1}{2}\sqrt{\frac{p_4(\frac{t_1}{i\hbar})}{p_1(\frac{t_1}{i\hbar})}}s_{14}(\frac{t_1}{i\hbar}) \\ 
\frac{1}{i\hbar}\int_{0}^{t_1}dt_1'\frac{1}{2}\sqrt{\frac{p_1(\frac{t_1}{i\hbar})}{p_2(\frac{t_1}{i\hbar})}}s_{21}(\frac{t_1}{i\hbar}) & \frac{1}{2}s_{22}(\frac{t_1}{i\hbar}) & \frac{1}{i\hbar}\int_{0}^{t_1}dt_1'\frac{1}{2}\sqrt{\frac{p_3(\frac{t_1}{i\hbar})}{p_2(t)}}s_{23}(\frac{t_1}{i\hbar}) & \frac{1}{i\hbar}\int_{0}^{t_1}dt_1'\frac{1}{2}\sqrt{\frac{p_4(\frac{t_1}{i\hbar})}{p_2(\frac{t_1}{i\hbar})}}s_{24}(\frac{t_1}{i\hbar}) \\ 
\frac{1}{i\hbar}\int_{0}^{t_1}dt_1'\frac{1}{2}\sqrt{\frac{p_1(\frac{t_1}{i\hbar})}{p_3(\frac{t_1}{i\hbar})}}s_{31}(\frac{t_1}{i\hbar}) & \frac{1}{i\hbar}\int_{0}^{t_1}dt_1'\frac{1}{2}\sqrt{\frac{p_2(\frac{t_1}{i\hbar})}{p_3(\frac{t_1}{i\hbar})}}s_{32}(\frac{t_1}{i\hbar}) & \frac{1}{2}s_{33}(\frac{t_1}{i\hbar}) & \frac{1}{i\hbar}\int_{0}^{t_1}dt_1'\frac{1}{2}\sqrt{\frac{p_4(\frac{t_1}{i\hbar})}{p_3(\frac{t_1}{i\hbar})}}s_{34}(\frac{t_1}{i\hbar}) \\ 
\frac{1}{i\hbar}\int_{0}^{t_1}dt_1'\frac{1}{2}\sqrt{\frac{p_1(\frac{t_1}{i\hbar})}{p_4(\frac{t_1}{i\hbar})}}s_{41}(\frac{t_1}{i\hbar}) & \frac{1}{i\hbar}\int_{0}^{t_1}dt_1'\frac{1}{2}\sqrt{\frac{p_2(\frac{t_1}{i\hbar})}{p_4(\frac{t_1}{i\hbar})}}s_{42}(\frac{t_1}{i\hbar}) & \frac{1}{i\hbar}\int_{0}^{t_1}dt_1'\frac{1}{2}\sqrt{\frac{p_3(\frac{t_1}{i\hbar})}{p_4(\frac{t_1}{i\hbar})}}s_{43}(\frac{t_1}{i\hbar}) & \frac{1}{i\hbar}\int_{0}^{t_1}dt_1s_{44}(\frac{t_1}{i\hbar}) \\ 
\end{pmatrix}
\Bigg]
\end{eqnarray}
It is quite straightforward to establish the reduced density matrix of system A and B and hence obtain von-Neumann entropy with time that characterizes the quantum like entropy in time in classical epidemic model in 2 coupled systems. We can extract density of matrix for system A given as
\begin{eqnarray}
\rho_A=
\begin{pmatrix}
\rho(1,1)+\rho(2,2) & \rho(1,3)+\rho(2,4) \\
\rho(3,1)+\rho(4,2) & \rho(3,3)+\rho(4,4) \\
\end{pmatrix}
\end{eqnarray}
and for system B given as
\begin{eqnarray}
\rho_B=
\begin{pmatrix}
\rho(1,1)+\rho(3,3) & \rho(1,2)+\rho(3,4) \\
\rho(2,1)+\rho(4,3) & \rho(2,2)+\rho(4,4) \\
\end{pmatrix}
\end{eqnarray}.
Those density matrices needs to be normalized so finally we obtain
\begin{eqnarray}
\rho_A=\frac{1}{\rho(1,1)+\rho(2,2)+\rho(3,3)+\rho(4,4)}
\begin{pmatrix}
\rho(1,1)+\rho(2,2) & \rho(1,3)+\rho(2,4) \\
\rho(3,1)+\rho(4,2) & \rho(3,3)+\rho(4,4) \\
\end{pmatrix}
\end{eqnarray}
and for system B given as
\begin{eqnarray}
\rho_B=\frac{1}{\rho(1,1)+\rho(2,2)+\rho(3,3)+\rho(4,4)}
\begin{pmatrix}
\rho(1,1)+\rho(3,3) & \rho(1,2)+\rho(3,4) \\
\rho(2,1)+\rho(4,3) & \rho(2,2)+\rho(4,4) \\
\end{pmatrix}
\end{eqnarray}.
In such case we can apply von-Neumman entropy measure to characterize the density matrix of classical state so we have $S_A=Tr(\rho_A log(\rho_A))$ and $S_B=Tr(\rho_B log(\rho_B))$. In our general case $S_A \neq S_B$, while in case of two quantum systems with mutual interaction, but isolated from external environment we have $S_A = S_B$.

 We conduct reverse reasoning and show that any quantum system describing by tight binding model can be modeled by stochastic finite state machine. This proves the fact that classical electronics can model quantum system made from single-electron devices as in CMOS quantum computer.
This reasoning can be conduced for N coupled quantum dots implementing CMOS quantum computer. It has therefore its meaning in quantum machine learning implemented in CMOS quantum
computer.
\section{From quantum tight-binding model to classical epidemic model}
We are considering two position-based qubits interacting by means of electrostatic Coulomb force as depicted in Fig.\ref{fig:FiniteStates}.
\begin{figure}
\centering
\includegraphics[scale=0.7]{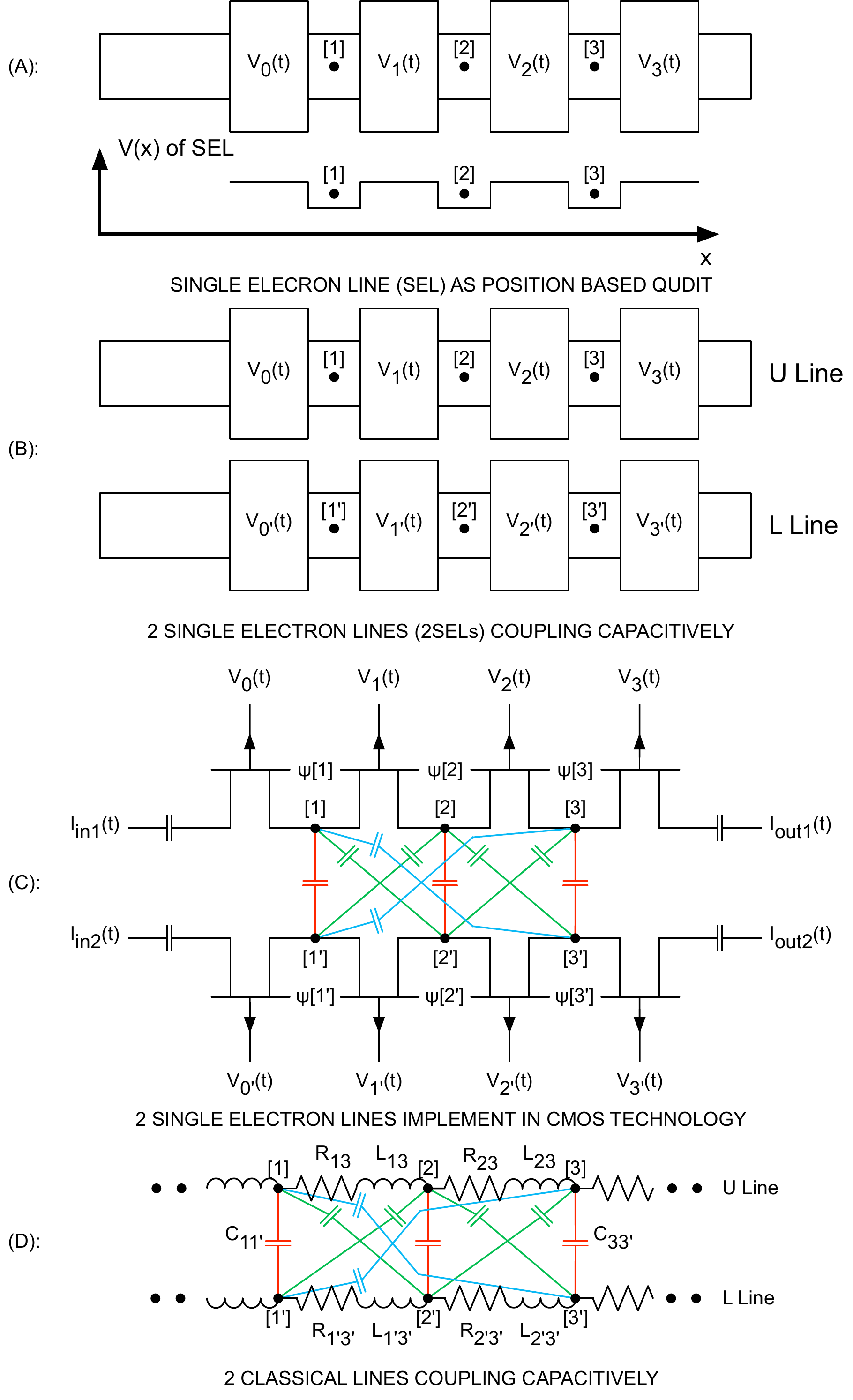} 
\caption{2 interacting position-based qubits implemented in CMOS semicondunctor technology \cite{2SEL} that can be parameterized by 6 or 12 state stochastic finite state machine.}
\end{figure}
We define the Hamiltonian of 2 interacting Wannier qubits in the simplistic form given by tight-binding model. We have
\begin{eqnarray}
\hat{H}=(\hat{H}_A)_0+(\hat{H}_B)_0+\hat{H}_{A-B}= \nonumber \\
=(E_{p1A}\ket{1_A}\bra{1_A}+E_{p2A}\ket{2_A}\bra{2_A}+t_{s(1_A \rightarrow 2_A)}\ket{1_A}\bra{2_A}+t_{s(2_A \rightarrow 1_A)}\ket{2_A}\bra{1_A})_0 \times \hat{I}_B+ \nonumber \\
+\hat{I}_A\times(E_{p1B}\ket{2_B}\bra{2_B}+E_{p2B}\ket{2_B}\bra{2_B}+t_{s(1_B \rightarrow 2_B)}\ket{1_B}\bra{2_B}+t_{s(2_B \rightarrow 1_B)}\ket{2_B}\bra{1_B})_0+\nonumber \\
+\frac{q^2}{d_{1A-1B}}(\ket{1_A,1_B}\bra{1_A,1_B})+\frac{q^2}{d_{1A-2B}}\ket{1_A,2_B}\bra{1_A,2_B}+\frac{q^2}{d_{2A-1B}}\ket{2_A,1_B}\bra{2_A,1_B}+\frac{q^2}{d_{2A-2B}}\ket{2_A,2_B}\bra{2_A,2_B}, \nonumber \\
\hat{I}_A=\ket{1_A}\bra{1_A}+\ket{2_A}\bra{2_A},\hat{I}_B=\ket{1_B}\bra{1_B}+\ket{2_B}\bra{2_B}
\end{eqnarray}
and for quantum state of the form
\begin{eqnarray}
\ket{\psi(t)}=\gamma_1(t)\ket{1_A}\ket{1_B}+\gamma_2(t)\ket{1_A}\ket{2_B}+\gamma_3(t)\ket{2_A}\ket{1_B}+\gamma_4(t)\ket{2_A}\ket{2_B}= 
\begin{pmatrix}
\gamma_1(t) \\
\gamma_2(t) \\
\gamma_3(t) \\
\gamma_4(t) \\
\end{pmatrix}, 
|\gamma_1(t)|^2+|\gamma_2(t)|^2+|\gamma_3(t)|^2+|\gamma_4(t)|^2=1, \nonumber \\
.
\end{eqnarray}
and with $\gamma_1(t)=\sqrt{p_{I}(t)}e^{i\Theta_1(t)},..,\gamma_{IV}(t)=\sqrt{p_{IV}(t)}e^{i\Theta_{IV}(t)} $. We have the following evolution of non-disspative quantum state of isolated system in tight-binding model given as
\begin{eqnarray}
\begin{pmatrix}
E_{p1A}+E_{p1B}+\frac{q^2}{d_{1A-1B}} & t_{s(2_B \rightarrow 1_B)} & t_{s(2_A \rightarrow 1_A)} & 0 \\
t_{s(1_B \rightarrow 2_B)}      & E_{p1A}+E_{p2B}+\frac{q^2}{d_{1A-2B}} & 0 & t_{s(2_A \rightarrow 1_A)} \\
t_{s(1_A \rightarrow 2_A)}      & 0 & E_{p2A}+E_{p1B}+\frac{q^2}{d_{2A-1B}} & t_{s(2_B \rightarrow 1_B)} \\
0      & t_{s(1_A \rightarrow 2_A)} & t_{s(1_B \rightarrow 2_B)} & E_{p2A}+E_{p2B}+\frac{q^2}{d_{2A-2B}} \\
\end{pmatrix}
\begin{pmatrix}
\sqrt{p_{I}(t)}e^{i\Theta_{I}(t)} \\
\sqrt{p_{II}(t)}e^{i\Theta_{II}(t)} \\
\sqrt{p_{III}(t)}e^{i\Theta_{III}(t)} \\
\sqrt{p_{IV}(t)}e^{i\Theta_{IV}(t)} \\
\end{pmatrix}=\nonumber \\
=i\hbar \frac{d}{dt}
\begin{pmatrix}
\sqrt{p_{I}(t)}e^{i\Theta_{I}(t)} \\
\sqrt{p_{II}(t)}e^{i\Theta_{II}(t)} \\
\sqrt{p_{III}(t)}e^{i\Theta_{III}(t)} \\
\sqrt{p_{IV}(t)}e^{i\Theta_{IV}(t)} \\
\end{pmatrix}=\hbar
\begin{pmatrix}
(-\frac{d}{dt}\Theta_{I}(t)\sqrt{p_{I}(t)}+i\frac{1}{2 \sqrt{p_{I}(t)}}[\frac{d}{dt}p_{I}(t)])e^{i\Theta_{I}(t)} \\
(-\frac{d}{dt}\Theta_{II}(t)\sqrt{p_{II}(t)}+i\frac{1}{2 \sqrt{p_{II}(t)}}[\frac{d}{dt}p_{II}(t)])e^{i\Theta_{II}(t)} \\
(-\frac{d}{dt}\Theta_{III}(t)\sqrt{p_{III}(t)}+i\frac{1}{2 \sqrt{p_{III}(t)}}[\frac{d}{dt}p_{III}(t)])e^{i\Theta_{III}(t)} \\
(-\frac{d}{dt}\Theta_{IV}(t)\sqrt{p_{IV}(t)}+i\frac{1}{2 \sqrt{p_{IV}(t)}}[\frac{d}{dt}p_{IV}(t)])e^{i\Theta_{IV}(t)} \\
\end{pmatrix}= \nonumber \\
=\hbar
\begin{pmatrix}
(-\frac{d}{dt}\Theta_{I}(t)+i\frac{1}{2 p_{I}(t)}[\frac{d}{dt}p_{I}(t)])(\sqrt{p_{I}(t)}e^{i\Theta_{I}(t)}) \\
(-\frac{d}{dt}\Theta_{II}(t)+i\frac{1}{2 p_{II}(t)}[\frac{d}{dt}p_{II}(t)])(\sqrt{p_{II}(t)}e^{i\Theta_{II}(t)}) \\
(-\frac{d}{dt}\Theta_{III}(t)+i\frac{1}{2 p_{III}(t)}[\frac{d}{dt}p_{III}(t)])(\sqrt{p_{III}(t)}e^{i\Theta_{III}(t)}) \\
(-\frac{d}{dt}\Theta_{IV}(t)+i\frac{1}{2  p_{IV}(t)}[\frac{d}{dt}p_{IV}(t)])(\sqrt{p_{IV}(t)}e^{i\Theta_{IV}(t)}) \\
\end{pmatrix}
.
\end{eqnarray}
After rearrangement of expressions we obtain
\small
\begin{eqnarray*}
\frac{1}{i\hbar}
\begin{pmatrix}
E_{p1A}+E_{p1B}+\frac{q^2}{d_{1A-1B}}+\hbar\frac{d}{dt}\Theta_{I}(t) & t_{s(2_B \rightarrow 1_B)} & t_{s(2_A \rightarrow 1_A)} & 0 \\
t_{s(1_B \rightarrow 2_B)}      & E_{p1A}+E_{p2B}+\frac{q^2}{d_{1A-2B}}+\hbar\frac{d}{dt}\Theta_{II}(t) & 0 & t_{s(2_A \rightarrow 1_A)} \\
t_{s(1_A \rightarrow 2_A)}      & 0 & E_{p2A}+E_{p1B}+\frac{q^2}{d_{2A-1B}}+\hbar\frac{d}{dt}\Theta_{III}(t) & t_{s(2_B \rightarrow 1_B)} \\
0      & t_{s(1_A \rightarrow 2_A)} & t_{s(1_B \rightarrow 2_B)} & E_{p2A}+E_{p2B}+\frac{q^2}{d_{2A-2B}}+\hbar\frac{d}{dt}\Theta_{IV}(t) \\
\end{pmatrix}\times \nonumber \\
\begin{pmatrix}
\sqrt{p_{I}(t)}e^{i\Theta_{I}(t)} \\
\sqrt{p_{II}(t)}e^{i\Theta_{II}(t)} \\
\sqrt{p_{III}(t)}e^{i\Theta_{III}(t)} \\
\sqrt{p_{IV}(t)}e^{i\Theta_{IV}(t)} \\
\end{pmatrix}= 
\begin{pmatrix}
\frac{e^{i\Theta_{I}(t)}}{2\sqrt{p_{I}(t)}}\frac{d}{dt}p_{I}(t) \\
\frac{e^{i\Theta_{II}(t)}}{2\sqrt{p_{II}(t)}}\frac{d}{dt}p_{II}(t) \\
\frac{e^{i\Theta_{III}(t)}}{2\sqrt{p_{III}(t)}}\frac{d}{dt}p_{III}(t) \\
\frac{e^{i\Theta_{IV}(t)}}{2\sqrt{p_{IV}(t)}}\frac{d}{dt}p_{IV}(t) \\
\end{pmatrix}.
\end{eqnarray*}
\normalsize
Equivalently we have
\small
\begin{eqnarray*}
\frac{1}{i\hbar}
\begin{pmatrix}
E_{p1A}+E_{p1B}+\frac{q^2}{d_{1A-1B}}+\hbar\frac{d}{dt}\Theta_{I}(t) & t_{s(2_B \rightarrow 1_B)} & t_{s(2_A \rightarrow 1_A)} & 0 \\
t_{s(1_B \rightarrow 2_B)}      & E_{p1A}+E_{p2B}+\frac{q^2}{d_{1A-2B}}+\hbar\frac{d}{dt}\Theta_{II}(t) & 0 & t_{s(2_A \rightarrow 1_A)} \\
t_{s(1_A \rightarrow 2_A)}      & 0 & E_{p2A}+E_{p1B}+\frac{q^2}{d_{2A-1B}}+\hbar\frac{d}{dt}\Theta_{III}(t) & t_{s(2_B \rightarrow 1_B)} \\
0      & t_{s(1_A \rightarrow 2_A)} & t_{s(1_B \rightarrow 2_B)} & E_{p2A}+E_{p2B}+\frac{q^2}{d_{2A-2B}}+\hbar\frac{d}{dt}\Theta_{IV}(t) \\
\end{pmatrix} \times \nonumber \\
\times
\begin{pmatrix}
\frac{e^{i\Theta_{I}(t)}}{\sqrt{p_{I}(t)}} & 0 & 0 & 0 \\
0 & \frac{e^{i\Theta_{II}(t)}}{\sqrt{p_{II}(t)}} & 0 & 0 \\
0 & 0 & \frac{e^{i\Theta_{III}(t)}}{\sqrt{p_{III}(t)}} & 0  \\
0 & 0 & 0 & \frac{e^{i\Theta_{IV}(t)}}{\sqrt{p_{IV}(t)}}\\
\end{pmatrix}
\begin{pmatrix}
\sqrt{p_{I}(t)}e^{-i\Theta_{I}(t)} & 0 & 0 & 0 \\
0 & \sqrt{p_{II}(t)}e^{-i\Theta_{II}(t)} & 0 & 0 \\
0 & 0 & \sqrt{p_{III}(t)}e^{-i\Theta_{III}(t)} & 0  \\
0 & 0 & 0 & \sqrt{p_{IV}(t)}e^{-i\Theta_{IV}(t)}\\
\end{pmatrix}
\begin{pmatrix}
\sqrt{p_{I}(t)}e^{i\Theta_{I}(t)} \\
\sqrt{p_{II}(t)}e^{i\Theta_{II}(t)} \\
\sqrt{p_{III}(t)}e^{i\Theta_{III}(t)} \\
\sqrt{p_{IV}(t)}e^{i\Theta_{IV}(t)} \\
\end{pmatrix}= \nonumber \\
\begin{pmatrix}
\frac{e^{i\Theta_{I}(t)}}{2\sqrt{p_{I}(t)}}\frac{d}{dt}p_{I}(t) \\
\frac{e^{i\Theta_{II}(t)}}{2\sqrt{p_{II}(t)}}\frac{d}{dt}p_{II}(t) \\
\frac{e^{i\Theta_{III}(t)}}{2\sqrt{p_{III}(t)}}\frac{d}{dt}p_{III}(t) \\
\frac{e^{i\Theta_{IV}(t)}}{2\sqrt{p_{IV}(t)}}\frac{d}{dt}p_{IV}(t) \\
\end{pmatrix}.
\end{eqnarray*}
\normalsize
and final form of classical epidemic model mimicking the quantum tight-binding model in the form as
\small
\begin{eqnarray*}
\begin{pmatrix}
e^{-i\Theta_{I}(t)}\sqrt{p_{I}(t)} & 0 & 0 & 0 \\
0 & e^{-i\Theta_{II}(t)}\sqrt{p_{II}(t)} & 0 & 0 \\
0 & 0 & e^{-i\Theta_{III}(t)}\sqrt{p_{III}(t)} & 0 \\
0 & 0 & 0 & e^{-i\Theta_{IV}(t)}\sqrt{p_{IV}(t)} \\
\end{pmatrix}.
\frac{2}{i\hbar} \times \nonumber \\
\times
\begin{pmatrix}
E_{p1A}+E_{p1B}+\frac{q^2}{d_{1A-1B}}+\hbar\frac{d}{dt}\Theta_{I}(t) & t_{s(2_B \rightarrow 1_B)} & t_{s(2_A \rightarrow 1_A)} & 0 \\
t_{s(1_B \rightarrow 2_B)}      & E_{p1A}+E_{p2B}+\frac{q^2}{d_{1A-2B}}+\hbar\frac{d}{dt}\Theta_{II}(t) & 0 & t_{s(2_A \rightarrow 1_A)} \\
t_{s(1_A \rightarrow 2_A)}      & 0 & E_{p2A}+E_{p1B}+\frac{q^2}{d_{2A-1B}}+\hbar\frac{d}{dt}\Theta_{III}(t) & t_{s(2_B \rightarrow 1_B)} \\
0      & t_{s(1_A \rightarrow 2_A)} & t_{s(1_B \rightarrow 2_B)} & E_{p2A}+E_{p2B}+\frac{q^2}{d_{2A-2B}}+\hbar\frac{d}{dt}\Theta_{IV}(t) \\
\end{pmatrix} \times \nonumber \\
\times
\begin{pmatrix}
\frac{e^{i\Theta_{I}(t)}}{\sqrt{p_{I}(t)}} & 0 & 0 & 0 \\
0 & \frac{e^{i\Theta_{II}(t)}}{\sqrt{p_{II}(t)}} & 0 & 0 \\
0 & 0 & \frac{e^{i\Theta_{III}(t)}}{\sqrt{p_{III}(t)}} & 0  \\
0 & 0 & 0 & \frac{e^{i\Theta_{IV}(t)}}{\sqrt{p_{IV}(t)}}\\
\end{pmatrix}
\begin{pmatrix}
p_{I}(t) \\
p_{II}(t) \\
p_{III}(t) \\
p_{IV}(t) \\
\end{pmatrix}= 
\frac{d}{dt}
\begin{pmatrix}
p_{I}(t) \\
p_{II}(t) \\
p_{III}(t) \\
p_{IV}(t) \\
\end{pmatrix}.
\end{eqnarray*}
\normalsize
We obtain
\tiny
\begin{eqnarray*}
\frac{d}{dt}
\begin{pmatrix}
p_{I}(t) \\
p_{II}(t) \\
p_{III}(t) \\
p_{IV}(t) \\
\end{pmatrix}=
\begin{pmatrix}
e^{-i\Theta_{I}(t)}\sqrt{p_{I}(t)} & 0 & 0 & 0 \\
0 & e^{-i\Theta_{II}(t)}\sqrt{p_{II}(t)} & 0 & 0 \\
0 & 0 & e^{-i\Theta_{III}(t)}\sqrt{p_{III}(t)} & 0 \\
0 & 0 & 0 & e^{-i\Theta_{IV}(t)}\sqrt{p_{IV}(t)} \\
\end{pmatrix}.
\frac{2}{i\hbar} \times \nonumber \\
\times
\begin{pmatrix}
[E_{p1A}+E_{p1B}+\frac{q^2}{d_{1A-1B}}+\hbar\frac{d}{dt}\Theta_{I}(t)]\frac{e^{i\Theta_{I}(t)}}{\sqrt{p_{I}(t)}} & [t_{s(2_B \rightarrow 1_B)}]\frac{e^{i\Theta_{II}(t)}}{\sqrt{p_{II}(t)}} & [t_{s(2_A \rightarrow 1_A)}]\frac{e^{i\Theta_{III}(t)}}{\sqrt{p_{III}(t)}} & 0 \\
t_{s(1_B \rightarrow 2_B)}\frac{e^{i\Theta_{I}(t)}}{\sqrt{p_{I}(t)}}      & [E_{p1A}+E_{p2B}+\frac{q^2}{d_{1A-2B}}+\hbar\frac{d}{dt}\Theta_{II}(t)]\frac{e^{i\Theta_{II}(t)}}{\sqrt{p_{II}(t)}}  & 0 & [t_{s(2_A \rightarrow 1_A)}]\frac{e^{i\Theta_{IV}(t)}}{\sqrt{p_{IV}(t)}} \\
t_{s(1_A \rightarrow 2_A)}\frac{e^{i\Theta_{I}(t)}}{\sqrt{p_{I}(t)}}      & 0 & [E_{p2A}+E_{p1B}+\frac{q^2}{d_{2A-1B}}+\hbar\frac{d}{dt}\Theta_{III}(t)]\frac{e^{i\Theta_{III}(t)}}{\sqrt{p_{III}(t)}} & t_{s(2_B \rightarrow 1_B)}\frac{e^{i\Theta_{IV}(t)}}{\sqrt{p_{IV}(t)}} \\
0      & t_{s(1_A \rightarrow 2_A)}\frac{e^{i\Theta_{II}(t)}}{\sqrt{p_{II}(t)}} & t_{s(1_B \rightarrow 2_B)}\frac{e^{i\Theta_{III}(t)}}{\sqrt{p_{III}(t)}} & [E_{p2A}+E_{p2B}+\frac{q^2}{d_{2A-2B}}+\hbar\frac{d}{dt}\Theta_{IV}(t)]\frac{e^{i\Theta_{IV}(t)}}{\sqrt{p_{IV}(t)}} \\
\end{pmatrix} 
\begin{pmatrix}
p_{I}(t) \\
p_{II}(t) \\
p_{III}(t) \\
p_{IV}(t) \\
\end{pmatrix} 
\end{eqnarray*}
\normalsize
and in compact form we have \small
\begin{eqnarray*}
\frac{d}{dt}
\begin{pmatrix}
p_{I}(t) \\
p_{II}(t) \\
p_{III}(t) \\
p_{IV}(t) \\
\end{pmatrix}= \nonumber \\
\frac{2}{i\hbar} 
\begin{pmatrix}
[E_{p1A}+E_{p1B}+\frac{q^2}{d_{1A-1B}}+\hbar\frac{d}{dt}\Theta_{I}(t)] & [t_{s(2_B \rightarrow 1_B)}]\frac{e^{i(\Theta_{II}(t)-\Theta_{I}(t))}\sqrt{p_{I}(t)}}{\sqrt{p_{II}(t)}} & [t_{s(2_A \rightarrow 1_A)}]\frac{e^{i(\Theta_{III}(t)-\Theta_{I}(t))}\sqrt{p_{I}(t)}}{\sqrt{p_{III}(t)}} & 0 \\
t_{s(1_B \rightarrow 2_B)}\frac{e^{i(\Theta_{I}(t)-\Theta_{II}(t))}\sqrt{p_{II}(t)}}{\sqrt{p_{I}(t)}}      & [E_{p1A}+E_{p2B}+\frac{q^2}{d_{1A-2B}}+\hbar\frac{d}{dt}\Theta_{II}(t)]  & 0 & [t_{s(2_A \rightarrow 1_A)}]\frac{e^{i(\Theta_{IV}(t)-\Theta_{II}(t))}}{(\sqrt{p_{IV}(t)}-\sqrt{p_{II}(t)})} \\
t_{s(1_A \rightarrow 2_A)}\frac{e^{i(\Theta_{I}(t)-\Theta_{III}(t))}\sqrt{p_{III}(t)}}{\sqrt{p_{I}(t)}}      & 0 & [E_{p2A}+E_{p1B}+\frac{q^2}{d_{2A-1B}}+\hbar\frac{d}{dt}\Theta_{III}(t)]& t_{s(2_B \rightarrow 1_B)}\frac{e^{i(\Theta_{IV}(t)-\Theta_{III}(t))}\sqrt{p_{III}(t)}}{\sqrt{p_{IV}(t)}} \\
0      & t_{s(1_A \rightarrow 2_A)}\frac{e^{i(\Theta_{II}(t)-\Theta_{IV}(t))}\sqrt{p_{IV}(t)}}{\sqrt{p_{II}(t)}} & t_{s(1_B \rightarrow 2_B)}\frac{e^{i(\Theta_{III}(t)-\Theta_{IV}(t))}\sqrt{p_{IV}(t)}}{\sqrt{p_{III}(t)}} & [E_{p2A}+E_{p2B}+\frac{q^2}{d_{2A-2B}}+\hbar\frac{d}{dt}\Theta_{IV}(t)] \\
\end{pmatrix} \nonumber \\
\times
\begin{pmatrix}
p_{I}(t) \\
p_{II}(t) \\
p_{III}(t) \\
p_{IV}(t) \\
\end{pmatrix}. 
\end{eqnarray*}
\normalsize
\section{Mapping quantum tight-binding model to stochastic finite state machine}
Let us start from equations of motion for 2 electrostatically coupled position based qubits as by \cite{2SEL} expressed by minimalistic tight-binding model and we have
\small
\begin{eqnarray}
\begin{pmatrix}
E_{p1A}+E_{p1B}+\frac{q^2}{d_{1A-1B}} & t_{s(2_B \rightarrow 1_B)} & t_{s(2_A \rightarrow 1_A)} & 0 \\
t_{s(1_B \rightarrow 2_B)}      & E_{p1A}+E_{p2B}+\frac{q^2}{d_{1A-2B}} & 0 & t_{s(2_A \rightarrow 1_A)} \\
t_{s(1_A \rightarrow 2_A)}      & 0 & E_{p2A}+E_{p1B}+\frac{q^2}{d_{2A-1B}} & t_{s(2_B \rightarrow 1_B)} \\
0      & t_{s(1_A \rightarrow 2_A)} & t_{s(1_B \rightarrow 2_B)} & E_{p2A}+E_{p2B}+\frac{q^2}{d_{2A-2B}} \\
\end{pmatrix}
\begin{pmatrix}
\sqrt{p_{I}(t)cos(\Theta_I(t))}+i\sqrt{p_{I}(t)sin(\Theta_I(t))} \\
\sqrt{p_{II}(t)cos(\Theta_{II}(t))}+i\sqrt{p_{II}(t)sin(\Theta_{II}(t))}  \\
\sqrt{p_{III}(t)cos(\Theta_{III}(t))}+i\sqrt{p_{III}(t)sin(\Theta_{III}(t))}  \\
\sqrt{p_{IV}(t)cos(\Theta_{IV}(t))}+i\sqrt{p_{IV}(t)sin(\Theta_{IV}(t))}  \\
\end{pmatrix}= \nonumber \\
=i \hbar \frac{d}{dt}
\begin{pmatrix}
\sqrt{p_{I}(t)cos(\Theta_I(t))}+i\sqrt{p_{I}(t)sin(\Theta_I(t))} \\
\sqrt{p_{II}(t)cos(\Theta_{II}(t))}+i\sqrt{p_{II}(t)sin(\Theta_{II}(t))}  \\
\sqrt{p_{III}(t)cos(\Theta_{III}(t))}+i\sqrt{p_{III}(t)sin(\Theta_{III}(t))}  \\
\sqrt{p_{IV}(t)cos(\Theta_{IV}(t))}+i\sqrt{p_{IV}(t)sin(\Theta_{IV}(t))}
\end{pmatrix}=
\begin{pmatrix}
i \hbar \frac{d}{dt}\sqrt{p_{I}(t)cos(\Theta_I(t))}- \hbar \frac{d}{dt}\sqrt{p_{I}(t)sin(\Theta_I(t))} \\
i \hbar \frac{d}{dt}\sqrt{p_{II}(t)cos(\Theta_{II}(t))}- \hbar \frac{d}{dt}\sqrt{p_{II}(t)sin(\Theta_{II}(t))}  \\
i \hbar \frac{d}{dt}\sqrt{p_{III}(t)cos(\Theta_{III}(t))}- \hbar \frac{d}{dt}\sqrt{p_{III}(t)sin(\Theta_{III}(t))}  \\
i \hbar \frac{d}{dt}\sqrt{p_{IV}(t)cos(\Theta_{IV}(t))}- \hbar \frac{d}{dt}\sqrt{p_{IV}(t)sin(\Theta_{IV}(t))}
\end{pmatrix}
\end{eqnarray}
and equivalently we can write
\begin{eqnarray}
(E_{p1A}+E_{p1B}+\frac{q^2}{d_{1A-1B}})\sqrt{p_{I}(t)cos(\Theta_I(t))} +( t_{sr(2_B \rightarrow 1_B)})\sqrt{p_{II}(t)cos(\Theta_{II}(t))}+ \nonumber \\
-( t_{sim(2_B \rightarrow 1_B)})\sqrt{p_{II}(t)sin(\Theta_{II}(t))}+( t_{sr(2_A \rightarrow 1_A)})\sqrt{p_{III}(t)cos(\Theta_{III}(t))}-( t_{sim(2_A \rightarrow 1_A)})\sqrt{p_{III}(t)sin(\Theta_{III}(t))}=-\hbar \frac{d}{dt}\sqrt{p_{I}(t)sin(\Theta_I(t))}
\nonumber \\
( t_{sim(2_B \rightarrow 1_B)})\sqrt{p_{II}(t)cos(\Theta_{II}(t))} + t_{sre(2_B \rightarrow 1_B)}\sqrt{p_{II}(t)sin(\Theta_{II}(t))}+( t_{sim(2_A \rightarrow 1_A)})\sqrt{p_{III}(t)cos(\Theta_{III}(t))} + t_{sre(2_A \rightarrow 1_A)}\sqrt{p_{III}(t)sin(\Theta_{III}(t))} , \nonumber \\
=\hbar \frac{d}{dt}\sqrt{p_{I}(t)cos(\Theta_I(t))} \nonumber \\
(E_{p1A}+E_{p2B}+\frac{q^2}{d_{1A-2B}})\sqrt{p_{II}(t)cos(\Theta_{II}(t))} +( t_{sr(1_B \rightarrow 2_B)})\sqrt{p_{I}(t)cos(\Theta_I(t))} - t_{sim(1_B \rightarrow 2_B)}\sqrt{p_{I}(t)sin(\Theta_I(t))}+ \nonumber \\(t_{sr(2_A \rightarrow 1_A)}\sqrt{p_{IV}(t)cos(\Theta_{IV}(t))}) - t_{sim(2_A \rightarrow 1_A)}\sqrt{p_{IV}(t)sin(\Theta_{IV}(t))}  = - \hbar \frac{d}{dt}\sqrt{p_{II}(t)sin(\Theta_{II}(t))} \\
( t_{sim(1_B \rightarrow 2_B)})\sqrt{p_{I}(t)cos(\Theta_I(t))} + t_{sre(1_B \rightarrow 2_B)}\sqrt{p_{I}(t)sin(\Theta_I(t))}+( t_{sim(2_A \rightarrow 1_A)})\sqrt{p_{IV}(t)cos(\Theta_{IV}(t))} + t_{sre(2_A \rightarrow 1_A)}\sqrt{p_{IV}(t)sin(\Theta_{IV}(t))} , \nonumber \\
=\hbar \frac{d}{dt}\sqrt{p_{II}(t)cos(\Theta_{II}(t))} \nonumber \\
(E_{p2A}+E_{p1B}+\frac{q^2}{d_{2A-1B}})\sqrt{p_{III}(t)cos(\Theta_{III}(t))} +( t_{sr(2_B \rightarrow 1_B)})\sqrt{p_{IV}(t)cos(\Theta_{IV}(t))} - t_{sim(2_B \rightarrow 1_B)}\sqrt{p_{IV}(t)sin(\Theta_{IV}(t))}+ \nonumber \\(t_{sr(1_A \rightarrow 2_A)}\sqrt{p_{I}(t)cos(\Theta_{I}(t))}) - t_{sim(1_A \rightarrow 2_A)}\sqrt{p_{I}(t)sin(\Theta_{I}(t))}  = - \hbar \frac{d}{dt}\sqrt{p_{III}(t)sin(\Theta_{III}(t))} \\
( t_{sim(1_A \rightarrow 2_A)})\sqrt{p_{I}(t)cos(\Theta_I(t))} + t_{sre(1_A \rightarrow 2_A)}\sqrt{p_{I}(t)sin(\Theta_{I}(t))}+( t_{sim(2_B \rightarrow 1_B)})\sqrt{p_{IV}(t)cos(\Theta_{IV}(t))} + t_{sre(2_B \rightarrow 1_B)}\sqrt{p_{IV}(t)sin(\Theta_{IV}(t))} , \nonumber \\
=\hbar \frac{d}{dt}\sqrt{p_{III}(t)cos(\Theta_{III}(t))} \nonumber \\
(E_{p2A}+E_{p2B}+\frac{q^2}{d_{2A-2B}})\sqrt{p_{IV}(t)cos(\Theta_{IV}(t))} +( t_{sr(1_B \rightarrow 2_B)})\sqrt{p_{III}(t)cos(\Theta_{III}(t))} - t_{sim(1_B \rightarrow 2_B)}\sqrt{p_{III}(t)sin(\Theta_{III}(t))}+ \nonumber \\(t_{sr(1_A \rightarrow 2_A)}\sqrt{p_{II}(t)cos(\Theta_{II}(t))}) - t_{sim(1_A \rightarrow 2_A)}\sqrt{p_{II}(t)sin(\Theta_{II}(t))}  = - \hbar \frac{d}{dt}\sqrt{p_{IV}(t)sin(\Theta_{IV}(t))} \\
( t_{sim(1_A \rightarrow 2_A)})\sqrt{p_{II}(t)cos(\Theta_{II}t))} + t_{sre(1_A \rightarrow 2_A)}\sqrt{p_{II}(t)sin(\Theta_{II}(t))}+( t_{sim(1_B \rightarrow 2_B)})\sqrt{p_{III}(t)cos(\Theta_{III}(t))} + t_{sre(1_B \rightarrow 2_B)}\sqrt{p_{III}(t)sin(\Theta_{III}(t))} , \nonumber \\
=\hbar \frac{d}{dt}\sqrt{p_{IV}(t)cos(\Theta_{IV}(t))} \nonumber \\
\end{eqnarray}
The established relation can be written in the compact form as
\begin{landscape}
\tiny
\begin{eqnarray}
\begin{pmatrix}
0 & (E_{p1A}+E_{p1B}+\frac{q^2}{d_{1A-1B}}) & ( t_{sim(2_B \rightarrow 1_B)}) & ( t_{sr(2_B \rightarrow 1_B)}) & ( t_{sim(2_A \rightarrow 1_A)}) & ( t_{sr(2_A \rightarrow 1_A)}) & 0 & 0 \\
(E_{p1A}+E_{p1B}+\frac{q^2}{d_{1A-1B}}) & 0 & ( t_{sr(2_B \rightarrow 1_B)}) & ( t_{im(2_B \rightarrow 1_B)}) & ( t_{sr(2_A \rightarrow 1_A)}) & ( t_{sim(2_A \rightarrow 1_A)}) & 0 & 0 \\
( t_{sim(1_B \rightarrow 2_B)}) & ( t_{sre(1_B \rightarrow 2_B)}) & 0 & (E_{p1A}+E_{p2B}+\frac{q^2}{d_{1A-2B}}) & 0 & 0 & ( t_{sim(2_A \rightarrow 1_A)}) & ( t_{sr(2_A \rightarrow 1_A)}) \\
( t_{sre(1_B \rightarrow 2_B)}) & ( t_{sim(1_B \rightarrow 2_B)}) & (E_{p1A}+E_{p2B}+\frac{q^2}{d_{1A-2B}}) & 0 & 0 & 0 & ( t_{sr(2_A \rightarrow 1_A)}) & ( t_{sim(2_A \rightarrow 1_A)}) \\
( t_{sim(1_A \rightarrow 2_A)}) & ( t_{sim(1_A \rightarrow 2_A)}) & 0 & 0 & 0 & (E_{p2A}+E_{p1B}+\frac{q^2}{d_{2A-1B}}) & ( t_{sim(2_B \rightarrow 1_B)}) & ( t_{sim(2_B \rightarrow 1_B)}) \\
( t_{sim(1_A \rightarrow 2_A)}) & ( t_{sim(1_A \rightarrow 2_A)}) & 0 & 0 & (E_{p2A}+E_{p1B}+\frac{q^2}{d_{2A-1B}}) & 0 & ( t_{sim(2_B \rightarrow 1_B)}) & ( t_{sr(2_B \rightarrow 1_B)}) \\
0 & 0 & ( t_{sr(1_A \rightarrow 2_A)}) & ( t_{im(1_A \rightarrow 2_A)}) & ( t_{sim(1_B \rightarrow 2_B)}) & ( t_{sr(1_B \rightarrow 2_B)}) & 0 & (E_{p2A}+E_{p2B}+\frac{q^2}{d_{2A-2B}}) \\
0 & 0 & ( t_{sim(1_A \rightarrow 2_A)}) & ( t_{sr(1_A \rightarrow 2_A)}) & ( t_{sim(1_B \rightarrow 2_B)}) & ( t_{sr(1_B \rightarrow 2_B)}) & (E_{p2A}+E_{p2B}+\frac{q^2}{d_{2A-2B}}) & 0 \\
\end{pmatrix} 
\begin{pmatrix}
\sqrt{p_{I}(t)cos(\Theta_{I}(t))} \\
\sqrt{p_{I}(t)sin(\Theta_{I}(t))} \\
\sqrt{p_{II}(t)cos(\Theta_{II}(t))} \\
\sqrt{p_{II}(t)sin(\Theta_{II}(t))} \\
\sqrt{p_{III}(t)cos(\Theta_{III}(t))} \\
\sqrt{p_{III}(t)sin(\Theta_{III}(t))} \\
\sqrt{p_{IV}(t)cos(\Theta_{IV}(t))} \\
\sqrt{p_{IV}(t)sin(\Theta_{IV}(t))} \\
\end{pmatrix}
=\hbar \frac{d}{dt}
\begin{pmatrix}
+\sqrt{p_{I}(t)cos(\Theta_{I}(t))} \\
-\sqrt{p_{I}(t)sin(\Theta_{I}(t))} \\
+\sqrt{p_{II}(t)cos(\Theta_{II}(t))} \\
-\sqrt{p_{II}(t)sin(\Theta_{II}(t))} \\
+\sqrt{p_{III}(t)cos(\Theta_{III}(t))} \\
-\sqrt{p_{III}(t)sin(\Theta_{III}(t))} \\
+\sqrt{p_{IV}(t)cos(\Theta_{IV}(t))} \\
-\sqrt{p_{IV}(t)sin(\Theta_{IV}(t))} \\
\end{pmatrix}
\end{eqnarray} \normalsize
and equivalently we have
\tiny
\begin{eqnarray}
2
\begin{pmatrix}
\sqrt{p_{I}(t)}cos(\Theta_{I}(t)) & 0 & 0 & 0 & 0 & 0 & 0 & 0 \\
0 & -\sqrt{p_{I}(t)}sin(\Theta_{I}(t)) & 0 & 0 & 0 & 0 & 0 & 0 \\
0 & 0 & +\sqrt{p_{II}(t)}cos(\Theta_{II}(t)) & 0 & 0 & 0 & 0 & 0 \\
0 & 0 & 0 & -\sqrt{p_{II}(t)}sin(\Theta_{II}(t)) & 0 & 0 & 0 & 0 \\
0 & 0 & 0 & 0 & +\sqrt{p_{III}(t)}cos(\Theta_{III}(t)) & 0 & 0 & 0 \\
0 & 0 & 0 & 0 & 0 & -\sqrt{p_{III}(t)}sin(\Theta_{III}(t)) & 0 & 0 \\
0 & 0 & 0 & 0 & 0 & 0 & +\sqrt{p_{IV}(t)}cos(\Theta_{IV}(t)) & 0 \\
0 & 0 & 0 & 0 & 0 & 0 & 0 & -\sqrt{p_{IV}(t)}sin(\Theta_{IV}(t)) \\
\end{pmatrix} \times \nonumber \\ \times
\begin{pmatrix}
0 & (E_{p1A}+E_{p1B}+\frac{q^2}{d_{1A-1B}}) & ( t_{sim(2_B \rightarrow 1_B)}) & ( t_{sr(2_B \rightarrow 1_B)}) & ( t_{sim(2_A \rightarrow 1_A)}) & ( t_{sr(2_A \rightarrow 1_A)}) & 0 & 0 \\
(E_{p1A}+E_{p1B}+\frac{q^2}{d_{1A-1B}}) & 0 & ( t_{sr(2_B \rightarrow 1_B)}) & ( t_{im(2_B \rightarrow 1_B)}) & ( t_{sr(2_A \rightarrow 1_A)}) & ( t_{sim(2_A \rightarrow 1_A)}) & 0 & 0 \\
( t_{sim(1_B \rightarrow 2_B)}) & ( t_{sre(1_B \rightarrow 2_B)}) & 0 & (E_{p1A}+E_{p2B}+\frac{q^2}{d_{1A-2B}}) & 0 & 0 & ( t_{sim(2_A \rightarrow 1_A)}) & ( t_{sr(2_A \rightarrow 1_A)}) \\
( t_{sre(1_B \rightarrow 2_B)}) & ( t_{sim(1_B \rightarrow 2_B)}) & (E_{p1A}+E_{p2B}+\frac{q^2}{d_{1A-2B}}) & 0 & 0 & 0 & ( t_{sr(2_A \rightarrow 1_A)}) & ( t_{sim(2_A \rightarrow 1_A)}) \\
( t_{sim(1_A \rightarrow 2_A)}) & ( t_{sim(1_A \rightarrow 2_A)}) & 0 & 0 & 0 & (E_{p2A}+E_{p1B}+\frac{q^2}{d_{2A-1B}}) & ( t_{sim(2_B \rightarrow 1_B)}) & ( t_{sim(2_B \rightarrow 1_B)}) \\
( t_{sim(1_A \rightarrow 2_A)}) & ( t_{sim(1_A \rightarrow 2_A)}) & 0 & 0 & (E_{p2A}+E_{p1B}+\frac{q^2}{d_{2A-1B}}) & 0 & ( t_{sim(2_B \rightarrow 1_B)}) & ( t_{sr(2_B \rightarrow 1_B)}) \\
0 & 0 & ( t_{sr(1_A \rightarrow 2_A)}) & ( t_{im(1_A \rightarrow 2_A)}) & ( t_{sim(1_B \rightarrow 2_B)}) & ( t_{sr(1_B \rightarrow 2_B)}) & 0 & (E_{p2A}+E_{p2B}+\frac{q^2}{d_{2A-2B}}) \\
0 & 0 & ( t_{sim(1_A \rightarrow 2_A)}) & ( t_{sr(1_A \rightarrow 2_A)}) & ( t_{sim(1_B \rightarrow 2_B)}) & ( t_{sr(1_B \rightarrow 2_B)}) & (E_{p2A}+E_{p2B}+\frac{q^2}{d_{2A-2B}}) & 0 \\
\end{pmatrix} \times \nonumber \\
\begin{pmatrix}
\frac{1}{\sqrt{p_{I}(t)}cos(\Theta_{I}(t))} & 0 & 0 & 0 & 0 & 0 & 0 & 0 \\
0 & \frac{1}{\sqrt{p_{I}(t)}sin(\Theta_{I}(t))} & 0 & 0 & 0 & 0 & 0 & 0 \\
0 & 0 & \frac{1}{\sqrt{p_{II}(t)}cos(\Theta_{II}(t))} & 0 & 0 & 0 & 0 & 0 \\
0 & 0 & 0 & \frac{1}{\sqrt{p_{II}(t)}sin(\Theta_{II}(t))} & 0 & 0 & 0 & 0 \\
0 & 0 & 0 & 0 & \frac{1}{\sqrt{p_{III}(t)}cos(\Theta_{III}(t))} & 0 & 0 & 0 \\
0 & 0 & 0 & 0 & 0 & \frac{1}{\sqrt{p_{II}(t)}sin(\Theta_{II}(t))} & 0 & 0 \\
0 & 0 & 0 & 0 & 0 & 0 & \frac{1}{\sqrt{p_{IV}(t)}cos(\Theta_{IV}(t))} & 0 \\
0 & 0 & 0 & 0 & 0 & 0 & 0 & \frac{1}{\sqrt{p_{IV}(t)}cos(\Theta_{IV}(t))} \\
\end{pmatrix} \times \nonumber \\ \times
\begin{pmatrix}
\sqrt{p_{I}(t)}cos(\Theta_{I}(t)) & 0 & 0 & 0 & 0 & 0 & 0 & 0 \\
0 & \sqrt{p_{I}(t)}sin(\Theta_{I}(t)) & 0 & 0 & 0 & 0 & 0 & 0 \\
0 & 0 & \sqrt{p_{II}(t)}cos(\Theta_{II}(t)) & 0 & 0 & 0 & 0 & 0 \\
0 & 0 & 0 & \sqrt{p_{II}(t)}sin(\Theta_{II}(t)) & 0 & 0 & 0 & 0 \\
0 & 0 & 0 & 0 & \sqrt{p_{III}(t)}cos(\Theta_{III}(t)) & 0 & 0 & 0 \\
0 & 0 & 0 & 0 & 0 & \sqrt{p_{III}(t)}sin(\Theta_{III}(t)) & 0 & 0 \\
0 & 0 & 0 & 0 & 0 & 0 & \sqrt{p_{IV}(t)}cos(\Theta_{IV}(t)) & 0 \\
0 & 0 & 0 & 0 & 0 & 0 & 0 & \sqrt{p_{IV}(t)}sin(\Theta_{IV}(t)) \\
\end{pmatrix}
\begin{pmatrix}
\sqrt{p_{I}(t)}cos(\Theta_{I}) \\
\sqrt{p_{I}(t)}sin(\Theta_{I}) \\
\sqrt{p_{II}(t)}cos(\Theta_{II}) \\
\sqrt{p_{II}(t)}sin(\Theta_{II}) \\
\sqrt{p_{III}(t)}cos(\Theta_{III}) \\
\sqrt{p_{III}(t)}sin(\Theta_{III}) \\
\sqrt{p_{IV}(t)}cos(\Theta_{IV}) \\
\sqrt{p_{IV}(t)}sin(\Theta_{IV}) \\
\end{pmatrix}= \nonumber \\
=2
\begin{pmatrix}
\sqrt{p_{I}(t)}cos(\Theta_{I}(t)) & 0 & 0 & 0 & 0 & 0 & 0 & 0 \\
0 & -\sqrt{p_{I}(t)}sin(\Theta_{I}(t)) & 0 & 0 & 0 & 0 & 0 & 0 \\
0 & 0 & +\sqrt{p_{II}(t)}cos(\Theta_{II}(t)) & 0 & 0 & 0 & 0 & 0 \\
0 & 0 & 0 & -\sqrt{p_{II}(t)}sin(\Theta_{II}(t)) & 0 & 0 & 0 & 0 \\
0 & 0 & 0 & 0 & +\sqrt{p_{III}(t)}cos(\Theta_{III}(t)) & 0 & 0 & 0 \\
0 & 0 & 0 & 0 & 0 & -\sqrt{p_{III}(t)}sin(\Theta_{III}(t)) & 0 & 0 \\
0 & 0 & 0 & 0 & 0 & 0 & +\sqrt{p_{IV}(t)}cos(\Theta_{IV}(t)) & 0 \\
0 & 0 & 0 & 0 & 0 & 0 & 0 & -\sqrt{p_{IV}(t)}sin(\Theta_{IV}(t)) \\
\end{pmatrix} \times
\hbar \frac{d}{dt}
\begin{pmatrix}
+\sqrt{p_{I}(t)}cos(\Theta_{I}(t)) \\
-\sqrt{p_{I}(t)}sin(\Theta_{I}(t)) \\
+\sqrt{p_{II}(t)}cos(\Theta_{II}(t)) \\
-\sqrt{p_{II}(t)}sin(\Theta_{II}(t)) \\
+\sqrt{p_{III}(t)}cos(\Theta_{III}(t)) \\
-\sqrt{p_{III}(t)}sin(\Theta_{III}(t)) \\
+\sqrt{p_{IV}(t)}cos(\Theta_{IV}(t)) \\
-\sqrt{p_{IV}(t)}sin(\Theta_{IV}(t)) \\
\end{pmatrix}=
\hbar \frac{d}{dt}
\begin{pmatrix}
p_{I}(t)(cos(\Theta_{I}(t)))^2 \\
p_{I}(t)(sin(\Theta_{I}(t)))^2 \\
p_{II}(t)(cos(\Theta_{II}(t)))^2 \\
p_{II}(t)(sin(\Theta_{II}(t)))^2 \\
p_{III}(t)(cos(\Theta_{III}(t)))^2 \\
p_{III}(t)(sin(\Theta_{III}(t)))^2 \\
p_{IV}(t)(cos(\Theta_{IV}(t)))^2 \\
p_{IV}(t)(sin(\Theta_{IV}(t)))^2 \\
\end{pmatrix}=
\hbar \frac{d}{dt}
\begin{pmatrix}
p_1(t) \\
p_2(t) \\
p_3(t) \\
p_4(t) \\
p_5(t) \\
p_6(t) \\
p_7(t) \\
p_8(t) \\
\end{pmatrix}
\end{eqnarray}
\normalsize
The S matrix from classical epidemic model can be written as
\begin{eqnarray}
\hat{S}(t)
\begin{pmatrix}
p_1(t) \\
p_2(t) \\
p_3(t) \\
p_4(t) \\
p_5(t) \\
p_6(t) \\
p_7(t) \\
p_8(t) \\
\end{pmatrix} =
\begin{pmatrix}
s_{1,1}(t) & s_{1,2}(t) & s_{1,3}(t) & s_{1,4}(t) & s_{1,5}(t) & s_{1,6}(t) & s_{1,7}(t) & s_{1,8}(t) \\
s_{2,1}(t) & s_{2,2}(t) & s_{2,3}(t) & s_{2,4}(t) & s_{2,5}(t) & s_{2,6}(t) & s_{2,7}(t) & s_{2,8}(t) \\
s_{3,1}(t) & s_{3,2}(t) & s_{3,3}(t) & s_{3,4}(t) & s_{3,5}(t) & s_{3,6}(t) & s_{3,7}(t) & s_{3,8}(t) \\
s_{4,1}(t) & s_{4,2}(t) & s_{4,3}(t) & s_{4,4}(t) & s_{4,5}(t) & s_{4,6}(t) & s_{4,7}(t) & s_{4,8}(t) \\
s_{5,1}(t) & s_{5,2}(t) & s_{5,3}(t) & s_{5,4}(t) & s_{5,5}(t) & s_{5,6}(t) & s_{5,7}(t) & s_{5,8}(t) \\
s_{6,1}(t) & s_{6,2}(t) & s_{6,3}(t) & s_{6,4}(t) & s_{6,5}(t) & s_{6,6}(t) & s_{6,7}(t) & s_{6,8}(t) \\
s_{7,1}(t) & s_{7,2}(t) & s_{7,3}(t) & s_{7,4}(t) & s_{7,5}(t) & s_{7,6}(t) & s_{7,7}(t) & s_{7,8}(t) \\
s_{8,1}(t) & s_{8,2}(t) & s_{8,3}(t) & s_{8,4}(t) & s_{8,5}(t) & s_{8,6}(t) & s_{8,7}(t) & s_{8,8}(t) \\
\end{pmatrix}
\begin{pmatrix}
p_1(t) \\
p_2(t) \\
p_3(t) \\
p_4(t) \\
p_5(t) \\
p_6(t) \\
p_7(t) \\
p_8(t) \\
\end{pmatrix}
=\frac{d}{dt}
\begin{pmatrix}
p_1(t) \\
p_2(t) \\
p_3(t) \\
p_4(t) \\
p_5(t) \\
p_6(t) \\
p_7(t) \\
p_8(t) \\
\end{pmatrix}.
\end{eqnarray}
and can be written in relation to 2 coupled single electron devices as
\tiny
\begin{eqnarray}
\hat{S}(t)=
\begin{pmatrix}
s_{1,1}(t) & s_{1,2}(t) & s_{1,3}(t) & s_{1,4}(t) & s_{1,5}(t) & s_{1,6}(t) & s_{1,7}(t) & s_{1,8}(t) \\
s_{2,1}(t) & s_{2,2}(t) & s_{2,3}(t) & s_{2,4}(t) & s_{2,5}(t) & s_{2,6}(t) & s_{2,7}(t) & s_{2,8}(t) \\
s_{3,1}(t) & s_{3,2}(t) & s_{3,3}(t) & s_{3,4}(t) & s_{3,5}(t) & s_{3,6}(t) & s_{3,7}(t) & s_{3,8}(t) \\
s_{4,1}(t) & s_{4,2}(t) & s_{4,3}(t) & s_{4,4}(t) & s_{4,5}(t) & s_{4,6}(t) & s_{4,7}(t) & s_{4,8}(t) \\
s_{5,1}(t) & s_{5,2}(t) & s_{5,3}(t) & s_{5,4}(t) & s_{5,5}(t) & s_{5,6}(t) & s_{5,7}(t) & s_{5,8}(t) \\
s_{6,1}(t) & s_{6,2}(t) & s_{6,3}(t) & s_{6,4}(t) & s_{6,5}(t) & s_{6,6}(t) & s_{6,7}(t) & s_{6,8}(t) \\
s_{7,1}(t) & s_{7,2}(t) & s_{7,3}(t) & s_{7,4}(t) & s_{7,5}(t) & s_{7,6}(t) & s_{7,7}(t) & s_{7,8}(t) \\
s_{8,1}(t) & s_{8,2}(t) & s_{8,3}(t) & s_{8,4}(t) & s_{8,5}(t) & s_{8,6}(t) & s_{8,7}(t) & s_{8,8}(t) \\
\end{pmatrix}= \nonumber \\
=\frac{2}{\hbar}
\begin{pmatrix}
\sqrt{p_{I}(t)}cos(\Theta_{I}(t)) & 0 & 0 & 0 & 0 & 0 & 0 & 0 \\
0 & -\sqrt{p_{I}(t)}sin(\Theta_{I}(t)) & 0 & 0 & 0 & 0 & 0 & 0 \\
0 & 0 & +\sqrt{p_{II}(t)}cos(\Theta_{II}(t)) & 0 & 0 & 0 & 0 & 0 \\
0 & 0 & 0 & -\sqrt{p_{II}(t)}sin(\Theta_{II}(t)) & 0 & 0 & 0 & 0 \\
0 & 0 & 0 & 0 & +\sqrt{p_{III}(t)}cos(\Theta_{III}(t)) & 0 & 0 & 0 \\
0 & 0 & 0 & 0 & 0 & -\sqrt{p_{III}(t)}sin(\Theta_{III}(t)) & 0 & 0 \\
0 & 0 & 0 & 0 & 0 & 0 & +\sqrt{p_{IV}(t)}cos(\Theta_{IV}(t)) & 0 \\
0 & 0 & 0 & 0 & 0 & 0 & 0 & -\sqrt{p_{IV}(t)}sin(\Theta_{IV}(t)) \\
\end{pmatrix} \times \nonumber \\ \times
\begin{pmatrix}
Im(E_{p1A}+E_{p1B} & +Re(E_{p1A}+E_{p1B}+\frac{q^2}{d_{1A-1B}}) & ( t_{sim(2_B \rightarrow 1_B)}) & +( t_{sr(2_B \rightarrow 1_B)}) & ( t_{sim(2_A \rightarrow 1_A)}) & ( t_{sr(2_A \rightarrow 1_A)}) & 0 & 0 \\
-Re(E_{p1A}+E_{p1B}+\frac{q^2}{d_{1A-1B}}) & Im(E_{p1A}+E_{p1B} & -( t_{sr(2_B \rightarrow 1_B)}) & ( t_{sim(2_B \rightarrow 1_B)}) & ( -t_{sr(2_A \rightarrow 1_A)}) & ( t_{sim(2_A \rightarrow 1_A)}) & 0 & 0 \\
( -t_{sim(1_B \rightarrow 2_B)}) & ( t_{sre(1_B \rightarrow 2_B)}) & Im(E_{p1A}+E_{p2B}) & +Re(E_{p1A}+E_{p2B}+\frac{q^2}{d_{1A-2B}}) & 0 & 0 & ( t_{sim(2_A \rightarrow 1_A)}) & ( t_{sr(2_A \rightarrow 1_A)}) \\
( -t_{sr(1_B \rightarrow 2_B)}) & ( -t_{sim(1_B \rightarrow 2_B)}) & -Re(E_{p1A}+E_{p2B}+\frac{q^2}{d_{1A-2B}}) & Im(E_{p1A}+E_{p2B}) & 0 & 0 & ( -t_{sr(2_A \rightarrow 1_A)}) & ( t_{sim(2_A \rightarrow 1_A)}) \\
( -t_{sim(1_A \rightarrow 2_A)}) & ( t_{sr(1_A \rightarrow 2_A)}) & 0 & 0 & Im(E_{p2A}+E_{p1B}) & +Re(E_{p2A}+E_{p1B}+\frac{q^2}{d_{2A-1B}}) & ( t_{sim(2_B \rightarrow 1_B)}) & ( t_{sr(2_B \rightarrow 1_B)}) \\
( -t_{sr(1_A \rightarrow 2_A)}) & ( -t_{sim(1_A \rightarrow 2_A)}) & 0 & 0 & -Re(E_{p2A}+E_{p1B}+\frac{q^2}{d_{2A-1B}}) & Im(E_{p2A}+E_{p1B}) & ( -t_{sr(2_B \rightarrow 1_B)}) & ( t_{sim(2_B \rightarrow 1_B)}) \\
0 & 0 & ( -t_{sim(1_A \rightarrow 2_A)}) & ( t_{sr(1_A \rightarrow 2_A)}) & ( -t_{im(1_B \rightarrow 2_B)}) & ( t_{sr(1_B \rightarrow 2_B)}) & Im(E_{p2A}+E_{p2B}) & +Re(E_{p2A}+E_{p2B}+\frac{q^2}{d_{2A-2B}}) \\
0 & 0 & ( -t_{sr(1_A \rightarrow 2_A)}) & ( -t_{sim(1_A \rightarrow 2_A)}) & ( -t_{sr(1_B \rightarrow 2_B)}) & ( -t_{im(1_B \rightarrow 2_B)}) & -Re(E_{p2A}+E_{p2B}+\frac{q^2}{d_{2A-2B}}) & Im(E_{p2A}+E_{p2B}) \\
\end{pmatrix} \times \nonumber \\
\begin{pmatrix}
\frac{1}{\sqrt{p_{I}(t)}cos(\Theta_{I}(t))} & 0 & 0 & 0 & 0 & 0 & 0 & 0 \\
0 & \frac{1}{\sqrt{p_{I}(t)}sin(\Theta_{I}(t))} & 0 & 0 & 0 & 0 & 0 & 0 \\
0 & 0 & \frac{1}{\sqrt{p_{II}(t)}cos(\Theta_{II}(t))} & 0 & 0 & 0 & 0 & 0 \\
0 & 0 & 0 & \frac{1}{\sqrt{p_{II}(t)}sin(\Theta_{II}(t))} & 0 & 0 & 0 & 0 \\
0 & 0 & 0 & 0 & \frac{1}{\sqrt{p_{III}(t)}cos(\Theta_{III}(t))} & 0 & 0 & 0 \\
0 & 0 & 0 & 0 & 0 & \frac{1}{\sqrt{p_{II}(t)}sin(\Theta_{II}(t))} & 0 & 0 \\
0 & 0 & 0 & 0 & 0 & 0 & \frac{1}{\sqrt{p_{IV}(t)}cos(\Theta_{IV}(t))} & 0 \\
0 & 0 & 0 & 0 & 0 & 0 & 0 & \frac{1}{\sqrt{p_{IV}(t)}cos(\Theta_{IV}(t))} \\
\end{pmatrix}. 
\end{eqnarray}
\normalsize%
\normalsize
The obtained results can be summarized by four matrices builiding $\hat{S}$ matrix as
\small
\begin{eqnarray}
\hat{S}_{1,1}= \nonumber \\
\begin{pmatrix}
+Im(E_{p1A}+E_{p1B}) & [Re(E_{p1A}+E_{p1B})+\frac{q^2}{d_{1A-1B}}]\frac{\sqrt{Re(p_{I}(t))}}{\sqrt{Im(p_{I}(t))}} & t_{sim_{2B \rightarrow 1B}}\frac{\sqrt{Re(p_{I}(t))}}{\sqrt{Re(p_{II}(t))}} & t_{sr_{2B \rightarrow 1B}}\frac{\sqrt{Re(p_{I}(t))}}{\sqrt{Im(p_{II}(t))}} \\
-[Re(E_{p1A}+E_{p1B})+\frac{q^2}{d_{1A-1B}}]\frac{\sqrt{Im(p_{I}(t))}}{\sqrt{Re(p_{I}(t))}} & Im(E_{p1A}+E_{p1B}) & -t_{sr_{2B \rightarrow 1B}}\frac{\sqrt{Im(p_{I}(t))}}{\sqrt{Re(p_{II}(t))}} & t_{sim_{2B \rightarrow 1B}}\frac{\sqrt{Im(p_{I}(t))}}{\sqrt{Im(p_{II}(t))}} \\
-t_{sim_{1B \rightarrow 2B}}\frac{\sqrt{Re(p_{II}(t))}}{\sqrt{Re(p_{I}(t))}} & t_{sr_{1B \rightarrow 2B}}\frac{\sqrt{Re(p_{II}(t))}}{\sqrt{Im(p_{I}(t))}} & Im(E_{p1A}+E_{p2B}) & [Re(E_{p1A}+E_{p2B})+\frac{q^2}{d_{1A-2B}}]\frac{\sqrt{Re(p_{II}(t))}}{\sqrt{Im(p_{II}(t))}} \\
-t_{sr_{1B \rightarrow 2B}}\frac{\sqrt{Im(p_{II}(t))}}{\sqrt{Re(p_{I}(t))}} & -t_{sim_{1B \rightarrow 2B}}\frac{\sqrt{Im(p_{II}(t))}}{\sqrt{Im(p_{I}(t))}} & -Re(E_{p1A}+E_{p2B})+\frac{q^2}{d_{1A-2B}}]\frac{\sqrt{Im(p_{II}(t))}}{\sqrt{Re(p_{II}(t))}} & Im(E_{p1A}+E_{p2B}) \\
  \end{pmatrix}, \nonumber \\
\hat{S}_{1,2}=
\begin{pmatrix}
+t_{sim_{2A \rightarrow 1A}}\frac{\sqrt{Re(p_{I}(t))}}{\sqrt{Re(p_{III}(t))}} & t_{sr_{2A \rightarrow 1A}}\frac{\sqrt{Re(p_{I}(t))}}{\sqrt{Im(p_{III}(t))}}  & 0 & 0 \\
-t_{sr_{2A \rightarrow 1A}}\frac{\sqrt{Im(p_{I}(t))}}{\sqrt{Re(p_{III}(t))}}  & t_{sim_{2A \rightarrow 1A}}\frac{\sqrt{Im(p_{I}(t))}}{\sqrt{Im(p_{III}(t))}} & 0 & 0 \\
0 & 0 & t_{sim_{2A \rightarrow 1A}}\frac{\sqrt{Re(p_{II}(t))}}{\sqrt{Re(p_{IV}(t))}} & t_{sr_{2A \rightarrow 1A}}\frac{\sqrt{Re(p_{II}(t))}}{\sqrt{Im(p_{IV}(t))}} \\
0 & 0 & -t_{sr_{2A \rightarrow 1A}}\frac{\sqrt{Im(p_{II}(t))}}{\sqrt{Re(p_{IV}(t))}} & t_{sim_{2A \rightarrow 1A}}\frac{\sqrt{Im(p_{II}(t))}}{\sqrt{Im(p_{IV}(t))}} \\
  \end{pmatrix}, \nonumber \\
\hat{S}_{2,1}=
\begin{pmatrix}
-t_{sim_{1A \rightarrow 2A}}\frac{\sqrt{Re(p_{III}(t))}}{\sqrt{Re(p_{I}(t))}} & +t_{sr_{1A \rightarrow 2A}}\frac{\sqrt{Re(p_{III}(t))}}{\sqrt{Im(p_{I}(t))}} & 0 & 0 \\
-t_{sr_{1A \rightarrow 2A}}\frac{\sqrt{Im(p_{III}(t))}}{\sqrt{Re(p_{I}(t))}} & -t_{sim_{1A \rightarrow 2A}}\frac{\sqrt{Im(p_{III}(t))}}{\sqrt{Im(p_{I}(t))}} & 0 & 0 \\
0 & 0 & -t_{sim_{1A \rightarrow 2A}}\frac{\sqrt{Re(p_{IV}(t))}}{\sqrt{Re(p_{II}(t))}} & +t_{sr_{1A \rightarrow 2A}}\frac{\sqrt{Re(p_{IV}(t))}}{\sqrt{Im(p_{II}(t))}} \\
0 & 0 & -t_{sr_{1A \rightarrow 2A}}\frac{\sqrt{Im(p_{IV}(t))}}{\sqrt{Re(p_{II}(t))}} & -t_{sim_{1A \rightarrow 2A}}\frac{\sqrt{Re(p_{IV}(t))}}{\sqrt{Im(p_{II}(t))}} \\
  \end{pmatrix}, \nonumber \\
\hat{S}_{2,2}=
\begin{pmatrix}
Im(E_{p2A}+E_{p1B}) & [Re(E_{p2A}+E_{p1B})+\frac{q^2}{d_{2A-2B}}]\frac{\sqrt{Re(p_{III}(t))}}{\sqrt{Im(p_{III}(t))}} & t_{sre(2_B \rightarrow 1_B)}\frac{\sqrt{Re(p_{III}(t))}}{\sqrt{Re(p_{IV}(t))}} & t_{sim(2_B \rightarrow 1_B)}\frac{\sqrt{Re(p_{III}(t))}}{\sqrt{Im(p_{IV}(t))}} \\
[Re(E_{p2A}+E_{p1B})+\frac{q^2}{d_{2A-2B}}]\frac{\sqrt{Im(p_{III}(t))}}{\sqrt{Re(p_{III}(t))}} & [Im(E_{p2A}+E_{p1B})] & t_{sim(2_B \rightarrow 1_B)}\frac{\sqrt{Im(p_{III}(t))}}{\sqrt{Re(p_{IV}(t))}} & t_{sre(2_B \rightarrow 1_B)}\frac{\sqrt{Im(p_{III}(t))}}{\sqrt{Im(p_{IV}(t))}} \\
-t_{sim_{1B \rightarrow 2B}}\frac{\sqrt{Re(p_{IV}(t))}}{\sqrt{Re(p_{III}(t))}} & +t_{sre_{1B \rightarrow 2B}}\frac{\sqrt{Re(p_{IV}(t))}}{\sqrt{Im(p_{III}(t))}} & Im(E_{p2A}+E_{p2B}) & [Re(E_{p2A}+E_{p2B})+\frac{q^2}{d_{2A-2B}}]\frac{\sqrt{Re(p_{IV}(t))}}{\sqrt{Im(p_{IV}(t))}} \\
-t_{sre_{1B \rightarrow 2B}}\frac{\sqrt{Im(p_{IV}(t))}}{\sqrt{Re(p_{III}(t))}} & -t_{sim_{1B \rightarrow 2B}}\frac{\sqrt{Im(p_{IV}(t))}}{\sqrt{Im(p_{III}(t))}} & [Re(E_{p2A}+E_{p2B})+\frac{q^2}{d_{2A-2B}}]\frac{\sqrt{Im(p_{IV}(t))}}{\sqrt{Re(p_{IV}(t))}} & Im(E_{p2A}+E_{p2B}) \\
  \end{pmatrix},
\end{eqnarray}
\normalsize
\begin{eqnarray}
\hat{S}=
\begin{pmatrix}
\hat{S}_{11} & \hat{S}_{12}  \\
\hat{S}_{21} & \hat{S}_{22}  \\
  \end{pmatrix}. \nonumber \\
\end{eqnarray}
and with
\begin{eqnarray}
\frac{2}{\hbar}\hat{S} 
\begin{pmatrix}
Re(p_{I}(t))\\
Im(p_{I}(t))\\
Re(p_{II}(t))\\
Im(p_{II}(t))\\
Re(p_{III}(t))\\
Im(p_{III}(t))\\
Re(p_{IV}(t))\\
Im(p_{IV}(t))\\
\end{pmatrix}=
\frac{d}{dt}
\begin{pmatrix}
Re(p_{I}(t))\\
Im(p_{I}(t))\\
Re(p_{II}(t))\\
Im(p_{II}(t))\\
Re(p_{III}(t))\\
Im(p_{III}(t))\\
Re(p_{IV}(t))\\
Im(p_{IV}(t))\\
\end{pmatrix},
\begin{pmatrix}
Re(p_{I}(t))\\
Im(p_{I}(t))\\
Re(p_{II}(t))\\
Im(p_{II}(t))\\
Re(p_{III}(t))\\
Im(p_{III}(t))\\
Re(p_{IV}(t))\\
Im(p_{IV}(t))\\
\end{pmatrix}=
e^{\int_{t_0}^{t}\frac{2}{\hbar}\hat{S}(t')dt'} 
\begin{pmatrix}
Re(p_{I}(t_0))\\
Im(p_{I}(t_0))\\
Re(p_{II}(t_0))\\
Im(p_{II}(t_0))\\
Re(p_{III}(t_0))\\
Im(p_{III}(t_0)))\\
Re(p_{IV}(t_0))\\
Im(p_{IV}(t_0))\\
\end{pmatrix}=
e^{\frac{2}{\hbar} \begin{pmatrix}
\int_{t_0}^{t} \hat{S}_{11}(t')dt' & \int_{t_0}^{t} \hat{S}_{12}(t')dt'  \\
\int_{t_0}^{t} \hat{S}_{21}(t')dt' & \int_{t_0}^{t} \hat{S}_{22}(t')dt'  \\
  \end{pmatrix}} 
\begin{pmatrix}
Re(p_{I}(t_0))\\
Im(p_{I}(t_0))\\
Re(p_{II}(t_0))\\
Im(p_{II}(t_0))\\
Re(p_{III}(t_0))\\
Im(p_{III}(t_0)))\\
Re(p_{IV}(t_0))\\
Im(p_{IV}(t_0))\\
\end{pmatrix}
\end{eqnarray}
\end{landscape}
where all quantum information is mapped to classical stochastic machine in the way as
\begin{eqnarray}
Re(p_{I}(t))=\cos(\Theta_I(t))^2p_{I}(t),  \\
Im(p_{I}(t))=\sin(\Theta_I(t))^2p_{I}(t),  \\
Re(p_{II}(t))=\cos(\Theta_{II}(t))^2p_{II}(t), \\
Im(p_{II}(t))=\sin(\Theta_{II}(t))^2p_{II}(t), \\
Re(p_{III}(t))=\cos(\Theta_{III}(t))^2p_{III}(t), \\
Im(p_{III}(t))=\sin(\Theta_{III}(t))^2p_{III}(t), \\
Re(p_{IV}(t))=\cos(\Theta_{IV}(t))^2p_{IV}(t), \\
Im(p_{IV}(t))=\sin(\Theta_{IV}(t))^2p_{IV}(t).  \\
\end{eqnarray}
After conducted considerations it becomes explictly visible that usage of magnetic Aharonov-Bohm effect will have impact on the occupancy of nodes $(1_A,1_B)$,$(1_A,2_B)$,$(2_A,1_B)$,$(2_A,2_B)$. Let us assume that two interacting qubits are symmetric and aligned along x axes. In such case we need to renormalize phases dynamics with
time so
\begin{eqnarray}
\Theta_{I}(t) \rightarrow \Theta_{I}(t)+(A_x(1_A,t)+A_x(1_B,t))\Delta L\frac{e}{\hbar}, \nonumber \\
\Theta_{II}(t) \rightarrow \Theta_{II}(t)+(A_x(1_A,t)+A_x(2_B,t))\Delta L\frac{e}{\hbar}, \nonumber \\
\Theta_{III}(t) \rightarrow \Theta_{III}(t)+(A_x(2_A,t)+A_x(1_B,t))\Delta L\frac{e}{\hbar}, \nonumber \\
\Theta_{IV}(t) \rightarrow \Theta_{IV}(t)+(A_x(2_A,t)+A_x(2_B,t))\Delta L \frac{e}{\hbar},
\end{eqnarray}

where $\Delta L$ is the diameter each of 4 quantum dots and e is elementary charge of electron. Such situation was described by means of Schroedinger formalism in \cite{Spie}. Furthermore the noninvasive detection of charge movement by single electron devices
 \cite{Noninvasive} can also be also mapped to epidemic model. Various analytical solutions of tight-binding model that can be used in epidemic model can be found in \cite{Decoherence}, \cite{Photonic}, \cite{Nbodies}, \cite{Cryogenics}.

\normalsize
Here we have assumed that there is possible escape of electron(s) from the system of quantum dots to outside environment by tunneling from the system of 2 coupled quantum dots what can be reflected in imaginary values of $E_{p1A}$, $E_{p2A}$, $E_{p1B}$, $E_{p2B}$. Furthermore we can also assume that electron is being injected to the structure of
2 coupled quantum dots what also corresponds to non-zero imaginary value of $E_{p1A}$, $E_{p2A}$, $E_{p1B}$, $E_{p2B}$ and is denoted as $Im(E_{p1A})$, $Im(E_{p2A})$, $Im(E_{p1B})$, $Im(E_{p2B})$. Presence of electrons at positions (1A,1B),(1A,2B), (2A,1B), (2A,2B) is reflected by probabilities $p_1(t)+p_2(t)$, $p_3(t)+p_4(t)$, $p_5(t)+p_6(t)$, $p_7(t)+p_8(t)$. Dependence of evolution of quantum phases $\Theta_{I}(t)$,$\Theta_{II}(t)$,$\Theta_{III}(t)$,$\Theta_{IV}(t)$ is encoded by square of tangents of phases given by $tan(\Theta_{I}(t))^2=\frac{p_2(t)}{p_1(t)}$, $tan(\Theta_{II}(t))^2=\frac{p_4(t)}{p_3(t)}$, $tan(\Theta_{III}(t))^2=\frac{p_6(t)}{p_5(t)}$, $tan(\Theta_{IV}(t))^2=\frac{p_8(t)}{p_7(t)}$. If tight-binding model is time-independent the evolution of phase can be tracked in unique way by classical epidemic model.
Furthermore matrix $\hat{S}$(t) of classical epidemic model can be determined in unique way, since all its eigenstates and eigenvectors can be determined analytically.
This is not the case of bigger matrices than 4 by 4. Furthermore one can incorporate the noise in case of 2 coupled position based qubits with use of delta functions and such situation can also be mapped to classical epidemic model. We notice that for the case of coupled position based qubits system expressed by matrix N by N and N possible states we have mapping to classical epidemic model with obtained matrix $\hat{S}$ that has dimension 2N by 2N and it described evolution of 2N classical states in the framework of stochastic finite state machine.
\section{Concept of complex value time \label{complextime}}
In real situations we can have the extended concept of time to be complex value time as given by \cite{Kopnin}.
Simply we have $t_{ext}=te^{i\Theta(t)}$ that can be translated to be $\frac{dt_{ext}}{dt}=e^{i\Theta(t)}(1+it\frac{d}{dt}\Theta(t))$ and equivalently can be written as
$i\hbar\frac{1}{e^{i\Theta(t)}(1+it\frac{d}{dt}\Theta(t))}\frac{d}{dt}=i\hbar\frac{d}{dt_{ext}}$. In such case delivering or moving away the energy or mass can be accounted
by $\Theta(t)$ function that takes into account the fact of broken normalization of wavefunction.
However this dissipative process shall be different for different eigenenergies and eigenstates due to different shape of their eigenfunctions.
In such case one shall postulate
\begin{equation}
 i\hbar\frac{1}{e^{i\Theta_k(t)}(1+it\frac{d}{dt}\Theta_k(t))}\frac{d}{dt}=i\hbar\frac{d}{dt_{ext(k)}}=E_k\ket{E_k}\bra{E_k}.
\end{equation}
Finally we obtain
\begin{equation}
\sum_{k=0}^{N}E_k\ket{E_k}\bra{E_k}(1+it\frac{d}{dt}\Theta_k(t))E_ke^{i\Theta_k(t)}=i \hbar \frac{d}{dt}
\end{equation}
Therefore we have introduced here N+1 dimensions of time that can describe quantum system properties in most general case.

%
\section{Equivalence of Schr\"{o}dinger and Wannier formalism in one dimension}
So far we have shown the mathematical equivalence of finite stochastic machine and tight-binding quantum model. However from perspective of quantum mechanics development
the most fundamental picture is given by Schroedinger formalism. It is relatively to show the equivalence of Schroedinger formalism and tight-binding approach basing on specially defined Wannier functions. Let us consider the system of two coupled quantum dots as depicted in Fig.\ref{fig:WannierQ} and let us assign the occupancy of left quantum dot by electron as wavepacket presence in $x \in (-\infty,0)$ and wavepacket occupancy of right quantum dot as wavepacket presence in $x \in (0,+\infty)$. We can assume that Wannier wavefunctions are linear transformation of system eigenenergy wavefunctions.

We propose maximum localized orthonormal Wannier functions of the form
\begin{eqnarray}
w_L(x)=(+\alpha \psi_{E1}(x)+\beta \psi_{E2}(x))=w_{1,1}\psi_{E1}(x)+w_{1,2}\psi_{E2}(x), \nonumber \\
w_R(x)=(-\beta \psi_{E1}(x)+\alpha\psi_{E2}(x))=w_{2,1}\psi_{E1}(x)+w_{2,2}\psi_{E2}(x), \nonumber \\
\end{eqnarray}
and formally we have
\begin{eqnarray}
\begin{pmatrix}
w_L(x) \\
w_R(x)
\end{pmatrix}= \hat{W}
\begin{pmatrix}
\psi_{E1}(x) \\
\psi_{E2}(x)
\end{pmatrix}
=
\begin{pmatrix}
w_{1,1} & w_{1,2} \\
w_{2,1} & w_{2,2}
\end{pmatrix}
\begin{pmatrix}
\psi_{E1}(x) \\
\psi_{E2}(x)
\end{pmatrix},
\end{eqnarray}

We have 4 conditions to be fulfilled:
\begin{eqnarray} \label{eqnset}
1=\int_{-\infty}^{+\infty} dx w_L^{*}(x)w_L(x)=\int dx (+\alpha^{\dag} \psi_{E1}(x)^{\dag}+\beta^{\dag} \psi_{E2}(x)^{\dag})(\alpha\psi_{E1}(x)+\beta \psi_{E2}(x)), \nonumber \\
1=\int_{-\infty}^{+\infty} dx w_R^{*}(x)w_R(x)=\int dx (-\beta^{\dag} \psi_{E1}(x)+\alpha^{\dag}\psi_{E2}(x)^{\dag})(-\beta\psi_{E1}(x)+\alpha \psi_{E2}(x)), \nonumber \\
0=\int_{-\infty}^{+\infty} dx w_R^{*}(x)w_L(x)=\int dx (-\beta^{\dag} \psi_{E1}(x)^{\dag}+\alpha^{\dag} \psi_{E2}(x)^{\dag})(\alpha\psi_{E1}(x)+\beta \psi_{E2}(x)), \nonumber \\
0=\int_{-\infty}^{+\infty} dx w_L^{*}(x)w_R(x)=\int dx (\alpha^{\dag}\psi_{E1}(x)^{\dag}+\beta^{\dag} \psi_{E2}(x)^{\dag})(-\beta \psi_{E1}(x)+\alpha \psi_{E2}(x)). 
\end{eqnarray}
Due to orthogonality of 2 wavefunctions $\psi_{E1}$ and $\psi_{E2}$ we have from first two equations $|\alpha|^2+|\beta|^2=1$, so $|\beta|^2=1-|\alpha|^2$ and hence $|\alpha|=cos(\gamma)$ and $|\beta|=sin(\gamma)$.
From 3rd and 4th equation we have $\alpha^{\dag}\beta=\alpha\beta^{\dag}$ that is fulfilled when $\alpha=|\alpha|e^{i\delta}, \beta=|\beta|e^{+i\delta}=\sqrt{1-|\alpha|^2}e^{+i\delta}$.


We propose maximum localized orthonormal Wannier functions of the form
\begin{eqnarray}
w_L(x)=(+|\alpha|e^{i\delta} \psi_{E1}(x)+\sqrt{1-|\alpha|^2}e^{+i\delta} \psi_{E2}(x)), \nonumber \\
w_R(x)=(-\sqrt{1-|\alpha|^2}e^{+i\delta} \psi_{E1}(x)+|\alpha|e^{i\delta}\psi_{E2}(x)), \nonumber \\
\end{eqnarray}

The last criteria to be matched is that $w_L(x)$ is maximum localized on the left quantum dot that geometric position is given by $x \in (-\infty,0)$ and that $w_R(x)$ is maximum localized on the right quantum dot that is denoted by $x \in (0,+\infty)$. Formally we can define
\begin{eqnarray}
S_L(\alpha)[\psi_{E1}(x),\psi_{E2}(x)]=S_L(\gamma)[\psi_{E1}(x),\psi_{E2}(x)]=\int_{-\infty}^{0}w_L(x)^{*}w_L(x)dx= \nonumber \\
=\int_{-\infty}^{0}dx[+|\alpha| \psi_{E1}^{\dag}(x)+\sqrt{1-|\alpha|^2}\psi_{E2}^{\dag}(x))][+|\alpha| \psi_{E1}(x)+\sqrt{1-|\alpha|^2} \psi_{E2}(x))]= \nonumber \\
=\int_{-\infty}^{0}dx [ (1-|\alpha|^2)|\psi_{E2}|^2+|\alpha|^2|\psi_{E1}(x)|^2+|\alpha|\sqrt{1-|\alpha|^2}(\psi_{E1}\psi_{E2}^{\dag}(x)+\psi_{E1}^{\dag}\psi_{E2}(x)) ] = \nonumber \\
=\int_{-\infty}^{0}dx [ (1-cos(\gamma)^2)|\psi_{E2}(x)|^2+cos(\gamma)^2|\psi_{E1}(x)|^2+sin(\gamma)cos(\gamma)(\psi_{E1}(x)\psi_{E2}^{\dag}(x)+\psi_{E1}(x)^{\dag}\psi_{E2}(x)) ]. 
\end{eqnarray}
Since $S_L(\gamma)[\psi_{E1}(x),\psi_{E2}(x)]$ reaches maximum with respect to $\gamma$ what implies $\frac{d}{d \gamma}S_L(\gamma)[\psi_{E1}(x),\psi_{E2}(x)]=0$.
\begin{eqnarray}
0=\int_{-\infty}^{0}dx [-2 sin(\gamma) cos(\gamma)(|\psi_{E1}(x)|^2-|\psi_{E2}(x)|^2)+(cos(\gamma)^2-sin(\gamma)^2)(\psi_{E1}(x)\psi_{E2}^{\dag}(x)+\psi_{E1}(x)^{\dag}\psi_{E2}(x)) ].
\end{eqnarray}
that can be summarized as
\begin{eqnarray}
0=\int_{-\infty}^{0}dx [-sin(2\gamma)(|\psi_{E1}(x)|^2-|\psi_{E2}(x)|^2)+(cos(2\gamma))(\psi_{E1}(x)\psi_{E2}^{\dag}(x)+\psi_{E1}(x)^{\dag}\psi_{E2}(x)) ].
\end{eqnarray}
and finally we have
\begin{eqnarray}
\gamma=\frac{1}{2}ArcTan [ \frac{\int_{-\infty}^{0}dx (\psi_{E1}(x)\psi_{E2}^{\dag}(x)+\psi_{E1}(x)^{\dag}\psi_{E2}(x))}{\int_{-\infty}^{0}dx (|\psi_{E1}(x)|^2-|\psi_{E2}(x)|^2)}]=\frac{1}{2}ArcTan [r],
\label{GammaForumula}
\end{eqnarray}
where
\begin{eqnarray}
r=\frac{\int_{-\infty}^{0}dx (\psi_{E1}(x)\psi_{E2}^{\dag}(x)+\psi_{E1}(x)^{\dag}\psi_{E2}(x))}{\int_{-\infty}^{0}dx (|\psi_{E1}(x)|^2-|\psi_{E2}(x)|^2)}.
\label{special}
\end{eqnarray}

Consequently we have
\begin{eqnarray}
|\alpha|=
cos(\frac{1}{2} ArcTan [ \frac{\int_{-\infty}^{0}dx (\psi_{E1}(x)\psi_{E2}^{\dag}(x)+\psi_{E1}(x)^{\dag}\psi_{E2}(x))}{\int_{-\infty}^{0}dx (|\psi_{E1}(x)|^2-|\psi_{E2}(x)|^2)}]), \nonumber \\
|\beta|=
sin(\frac{1}{2} ArcTan [ \frac{\int_{-\infty}^{0}dx (\psi_{E1}(x)\psi_{E2}^{\dag}(x)+\psi_{E1}(x)^{\dag}\psi_{E2}(x))}{\int_{-\infty}^{0}dx (|\psi_{E1}(x)|^2-|\psi_{E2}(x)|^2)}]), \nonumber
\end{eqnarray}
Finally one can write

\begin{eqnarray}
\begin{pmatrix}
w_L(x) \\
w_R(x)
\end{pmatrix}
=
\begin{pmatrix}
+cos(\frac{1}{2}ArcTan [\frac{\int_{-\infty}^{0}dx (\psi_{E1}(x)\psi_{E2}^{\dag}(x)+\psi_{E1}(x)^{\dag}\psi_{E2}(x))}{\int_{-\infty}^{0}dx (|\psi_{E1}(x)|^2-|\psi_{E2}(x)|^2)}]) & sin(\frac{1}{2}ArcTan [\frac{\int_{-\infty}^{0}dx (\psi_{E1}(x)\psi_{E2}^{\dag}(x)+\psi_{E1}(x)^{\dag}\psi_{E2}(x))}{\int_{-\infty}^{0}dx (|\psi_{E1}(x)|^2-|\psi_{E2}(x)|^2)}]) \\
-sin(\frac{1}{2}ArcTan [ \frac{\int_{-\infty}^{0}dx (\psi_{E1}(x)\psi_{E2}^{\dag}(x)+\psi_{E1}(x)^{\dag}\psi_{E2}(x))}{\int_{-\infty}^{0}dx (|\psi_{E1}(x)|^2-|\psi_{E2}(x)|^2)}]) & cos(\frac{1}{2}ArcTan [ \frac{\int_{-\infty}^{0}dx (\psi_{E1}(x)\psi_{E2}^{\dag}(x)+\psi_{E1}(x)^{\dag}\psi_{E2}(x))}{\int_{-\infty}^{0}dx (|\psi_{E1}(x)|^2-|\psi_{E2}(x)|^2)}])
\end{pmatrix}
\begin{pmatrix}
\psi_{E1}(x) \\
\psi_{E2}(x)
\end{pmatrix}. \nonumber \\
\end{eqnarray}
 Such reasoning can be conducted for any 2 different energy levels as well as for N different energetic levels.
 If quantum state is given as
 \begin{eqnarray}
 \ket{\psi}=e^{\frac{E_1 (t-t_0)}{i \hbar}}e^{i\gamma_{E1}}\sqrt{p_{E1}}|E_1>+e^{\frac{E_2 (t-t_0)}{i \hbar}}e^{i\gamma_{E2}}\sqrt{p_{E2}}|E_2>
 \end{eqnarray}
 then

\begin{eqnarray}
\begin{pmatrix}
\alpha(t) w_L(x) \\
\beta(t) w_R(x)
\end{pmatrix}
=
\begin{pmatrix}
+cos(\frac{1}{2}ArcTan(r)) & +sin(\frac{1}{2}ArcTan(r)) \\
-sin(\frac{1}{2}ArcTan(r)) & +cos(\frac{1}{2}ArcTan(r))
\end{pmatrix}
\begin{pmatrix}
e^{\frac{E_1 (t-t_0)}{i \hbar}}e^{i\gamma_{E1}}\sqrt{p_{E1}} \psi_{E1}(x) \\
e^{\frac{E_2 (t-t_0)}{i \hbar}}e^{i\gamma_{E2}}\sqrt{p_{E2}} \psi_{E2}(x)
\end{pmatrix},
\end{eqnarray}
The last implies that
\begin{eqnarray}
\alpha_c(t)= \int_{-\infty}^{+\infty}dx
\begin{pmatrix}
[cos(\frac{1}{2}ArcTan(r))\psi_{E1}^{\dag}(x), & sin(\frac{1}{2}ArcTan(r))\psi_{E2}^{\dag}(x)
\end{pmatrix} \times \nonumber \\
\begin{pmatrix}
+cos(\frac{1}{2}ArcTan(r)), & +sin(\frac{1}{2}ArcTan(r)) \\
-sin(\frac{1}{2}ArcTan(r)), & +cos(\frac{1}{2}ArcTan(r))
\end{pmatrix}
\begin{pmatrix}
e^{\frac{E_1 (t-t_0)}{i \hbar}}e^{i\gamma_{E1}}\sqrt{p_{E1}} \psi_{E1}(x) \\
e^{\frac{E_2 (t-t_0)}{i \hbar}}e^{i\gamma_{E2}}\sqrt{p_{E2}} \psi_{E2}(x)
\end{pmatrix}= \nonumber \\
=\int_{-\infty}^{+\infty}dx
\begin{pmatrix}
w_{1,1}^{*}\psi_{E1}^{\dag}(x), & w_{1,2}^{*}\psi_{E2}^{\dag}(x)
\end{pmatrix} \times 
\begin{pmatrix}
w_{1,1} & w_{1,2} \\
w_{2,1} & w_{2,2}
\end{pmatrix}
\begin{pmatrix}
e^{\frac{E_1 (t-t_0)}{i \hbar}}e^{i\gamma_{E1}}\sqrt{p_{E1}} \psi_{E1}(x) \\
e^{\frac{E_2 (t-t_0)}{i \hbar}}e^{i\gamma_{E2}}\sqrt{p_{E2}} \psi_{E2}(x)
\end{pmatrix}= \nonumber \\
=\int_{-\infty}^{+\infty}dx
\begin{pmatrix}
w_{1,1}^{*}\psi_{E1}^{\dag}(x), & w_{1,2}^{*}\psi_{E2}^{\dag}(x)
\end{pmatrix}
\begin{pmatrix}
w_{1,1}e^{\frac{E_1 (t-t_0)}{i \hbar}}e^{i\gamma_{E1}}\sqrt{p_{E1}} \psi_{E1}(x)+w_{1,2}e^{\frac{E_2 (t-t_0)}{i \hbar}}e^{i\gamma_{E2}}\sqrt{p_{E2}} \psi_{E2}(x) \\
w_{2,1}e^{\frac{E_1 (t-t_0)}{i \hbar}}e^{i\gamma_{E1}}\sqrt{p_{E1}} \psi_{E1}(x)+w_{2,2}e^{\frac{E_2 (t-t_0)}{i \hbar}}e^{i\gamma_{E2}}\sqrt{p_{E2}} \psi_{E2}(x)
\end{pmatrix}= \nonumber \\
=w_{1,1}^{*}w_{1,1}e^{\frac{E_1 (t-t_0)}{i \hbar}}e^{i\gamma_{E1}}\sqrt{p_{E1}} +w_{1,2}^{*}w_{2,2}e^{\frac{E_2 (t-t_0)}{i \hbar}}e^{i\gamma_{E2}}\sqrt{p_{E2}}= \nonumber \\
=cos(\frac{1}{2}ArcTan(r))^2e^{\frac{E_1 (t-t_0)}{i \hbar}}e^{i\gamma_{E1}}\sqrt{p_{E1}} + 
sin(\frac{1}{2}ArcTan(r))cos(\frac{1}{2}ArcTan(r)) e^{\frac{E_2 (t-t_0)}{i \hbar}}e^{i\gamma_{E2}}\sqrt{p_{E2}}= \nonumber \\
=cos(\frac{1}{2}ArcTan(r)) [ cos(2ArcTan(r)) e^{\frac{E_1 (t-t_0)}{i \hbar}}e^{i\gamma_{E1}}\sqrt{p_{E1}}+ sin(\frac{1}{2}ArcTan(r)) e^{\frac{E_2 (t-t_0)}{i \hbar}}e^{i\gamma_{E2}}\sqrt{p_{E2}} ]
= \alpha(t). \nonumber \\
\end{eqnarray}
and
 \begin{eqnarray}
\beta_c(t)= \int_{-\infty}^{+\infty}dx
\begin{pmatrix}
 -sin(\frac{1}{2}ArcTan(r))\psi_{E1}^{\dag}(x), & cos(\frac{1}{2}ArcTan(r))\psi_{E2}^{\dag}(x)
\end{pmatrix} \times \nonumber \\
\begin{pmatrix}
+cos(\frac{1}{2}ArcTan(r)), & +sin(\frac{1}{2}ArcTan(r)) \\
-sin(\frac{1}{2}ArcTan(r)), & +cos(\frac{1}{2}ArcTan(r))
\end{pmatrix}
\begin{pmatrix}
e^{\frac{E_1 (t-t_0)}{i \hbar}}e^{i\gamma_{E1}}\sqrt{p_{E1}} \psi_{E1}(x) \\
e^{\frac{E_2 (t-t_0)}{i \hbar}}e^{i\gamma_{E2}}\sqrt{p_{E2}} \psi_{E2}(x)
\end{pmatrix}= \nonumber \\
=\int_{-\infty}^{+\infty}dx
\begin{pmatrix}
w_{2,1}^{*}\psi_{E1}^{\dag}(x), & w_{2,2}^{*}\psi_{E2}^{\dag}(x)
\end{pmatrix} \times 
\begin{pmatrix}
w_{1,1} & w_{1,2} \\
w_{2,1} & w_{2,2}
\end{pmatrix}
\begin{pmatrix}
e^{\frac{E_1 (t-t_0)}{i \hbar}}e^{i\gamma_{E1}}\sqrt{p_{E1}} \psi_{E1}(x) \\
e^{\frac{E_2 (t-t_0)}{i \hbar}}e^{i\gamma_{E2}}\sqrt{p_{E2}} \psi_{E2}(x)
\end{pmatrix}= \nonumber \\
=\int_{-\infty}^{+\infty}dx
\begin{pmatrix}
w_{2,1}^{*}\psi_{E1}^{\dag}(x), & w_{2,2}^{*}\psi_{E2}^{\dag}(x)
\end{pmatrix}
\begin{pmatrix}
w_{1,1}e^{\frac{E_1 (t-t_0)}{i \hbar}}e^{i\gamma_{E1}}\sqrt{p_{E1}} \psi_{E1}(x)+w_{1,2}e^{\frac{E_2 (t-t_0)}{i \hbar}}e^{i\gamma_{E2}}\sqrt{p_{E2}} \psi_{E2}(x) \\
w_{2,1}e^{\frac{E_1 (t-t_0)}{i \hbar}}e^{i\gamma_{E1}}\sqrt{p_{E1}} \psi_{E1}(x)+w_{2,2}e^{\frac{E_2 (t-t_0)}{i \hbar}}e^{i\gamma_{E2}}\sqrt{p_{E2}} \psi_{E2}(x)
\end{pmatrix}= \nonumber \\
=w_{2,1}^{*}w_{1,1}e^{\frac{E_1 (t-t_0)}{i \hbar}}e^{i\gamma_{E1}}\sqrt{p_{E1}} +w_{2,2}^{*}w_{2,2}e^{\frac{E_2 (t-t_0)}{i \hbar}}e^{i\gamma_{E2}}\sqrt{p_{E2}}= \nonumber \\
=-sin(\frac{1}{2}ArcTan(r))cos(\frac{1}{2}ArcTan(r))e^{\frac{E_1 (t-t_0)}{i \hbar}}e^{i\gamma_{E1}}\sqrt{p_{E1}} 
+cos(\frac{1}{2}ArcTan(r))^2e^{\frac{E_2 (t-t_0)}{i \hbar}}e^{i\gamma_{E2}}\sqrt{p_{E2}}= \nonumber \\
=cos(\frac{1}{2}ArcTan(r))[-sin(\frac{1}{2}ArcTan(r))e^{\frac{E_1 (t-t_0)}{i \hbar}}e^{i\gamma_{E1}}\sqrt{p_{E1}} 
+cos(\frac{1}{2}ArcTan(r))e^{\frac{E_2 (t-t_0)}{i \hbar}}e^{i\gamma_{E2}}\sqrt{p_{E2}}]=\beta(t)
\end{eqnarray}

\normalsize
In such way it was shown how to convert the quantum information represented by eigenergy qubits (as mostly used with formula $|\psi>=\sqrt{p_{E1}}e^{i\gamma_{E1}}|\psi>_{E1}+\sqrt{p_{E2}}e^{i\gamma_{E2}}|\psi>_{E2}$) in position based format $|\psi>=\alpha(t)|w>_{1}+\beta(t)|w>_{2}$ (Wannier qubit format).
We notice that
\begin{eqnarray}
\frac{\alpha_c(t)}{\beta_c(t)}=\frac{[ cos(\frac{1}{2}ArcTan(r)) e^{\frac{E_1 (t-t_0)}{i \hbar}}e^{i\gamma_{E1}}\sqrt{p_{E1}}+ sin(\frac{1}{2}ArcTan(r)) e^{\frac{E_2 (t-t_0)}{i \hbar}}e^{i\gamma_{E2}}\sqrt{p_{E2}} ]}{[-sin(\frac{1}{2}ArcTan(r))e^{\frac{E_1 (t-t_0)}{i \hbar}}e^{i\gamma_{E1}}\sqrt{p_{E1}} 
+cos(\frac{1}{2}ArcTan(r))e^{\frac{E_2 (t-t_0)}{i \hbar}}e^{i\gamma_{E2}}\sqrt{p_{E2}}]}= \nonumber \\
=\frac{[ cos(\frac{1}{2}ArcTan(r)) \sqrt{p_{E1}}+ sin(\frac{1}{2}ArcTan(r)) e^{\frac{(E_2-E_1)(t-t_0)}{i \hbar}}e^{i(\gamma_{E2}-\gamma_{E1})}\sqrt{p_{E2}} ]}{[-sin(\frac{1}{2}ArcTan(r))\sqrt{p_{E1}} 
+cos(\frac{1}{2}ArcTan(r))e^{\frac{(E_2-E_1) (t-t_0)}{i \hbar}}e^{i(\gamma_{E2}-\gamma_{E1})}\sqrt{p_{E2}}]}= \nonumber \\
=\frac{[ \frac{\sqrt{p_{E1}}}{\sqrt{p_{E2}}}+ Tan(\frac{1}{2}ArcTan(r)) e^{-i\frac{(E_2-E_1)(t-t_0)}{\hbar}}e^{i(\gamma_{E2}-\gamma_{E1})} ]}{[-Tan(\frac{1}{2}ArcTan(r))\frac{\sqrt{p_{E1}}}{\sqrt{p_{E2}}}] 
+e^{-i\frac{(E_2-E_1) (t-t_0)}{\hbar}}e^{i(\gamma_{E2}-\gamma_{E1})}},
\end{eqnarray}
what implies that occupancy of full Bloch sphere is not achievable by single Wannier qubit in static electric and magnetic field. Still we can approach arbitrary close to South and North pole of Bloch sphere by regulating $r$ \newline ( achieved from different effective potential generated by biasing electrodes) and by setting arbitrary ratio $\frac{\sqrt{p_{E1}}}{\sqrt{p_{E2}}}$. Furthermore we immediately obtain
\begin{eqnarray} \label{formts11}
E_{p1}=\int_{-\infty}^{+\infty}dx[w_L^{*}(x)\hat{H}w_L(x)]=\int_{-\infty}^{+\infty}dx[(\alpha^{\dag} \psi_{E1}^{\dag}(x) +\beta^{\dag} \psi_{E2}^{\dag}(x)) 
 \hat{H}(\alpha \psi_{E1}(x)+\beta \psi_{E2}(x))]= \nonumber \\
=|\alpha|^2 E_1+|\beta|^2 E_2 
=(1-|\beta|^2) E_1+|\beta|^2 E_2=E_1+|\beta|^2(E_2-E_1)= \nonumber \\
=E_1+(E_2-E_1)|sin(\frac{1}{2}ArcTan [\frac{\int_{-\infty}^{0}dx (\psi_{E1}(x)\psi_{E2}^{\dag}(x)+\psi_{E1}(x)^{\dag}\psi_{E2}(x))}{\int_{-\infty}^{0}dx (|\psi_{E1}(x)|^2-|\psi_{E2}(x)|^2)}])|^2
, \nonumber \\
\end{eqnarray}
\begin{eqnarray} \label{formts22}
E_{p2}=\int_{-\infty}^{+\infty}dx[w_R^{*}(x)\hat{H}w_R(x)]=\int_{-\infty}^{+\infty}dx(-\beta^{\dag} \psi_{E1}^{\dag}(x)+\alpha^{\dag} \psi_{E2}^{\dag}(x) ) \times \nonumber \\
\times \hat{H}(-\beta\psi_{E1}(x)+\alpha \psi_{E2}(x))=|\beta|^2 E_1+|\alpha|^2 E_2=E_1+(E_2-E_1)|\alpha|^2= \nonumber \\
=E_1+(E_2-E_1)|cos(\frac{1}{2}ArcTan [\frac{\int_{-\infty}^{0}dx (\psi_{E1}(x)\psi_{E2}^{\dag}(x)+\psi_{E1}(x)^{\dag}\psi_{E2}(x))}{\int_{-\infty}^{0}dx (|\psi_{E1}(x)|^2-|\psi_{E2}(x)|^2)}])|^2,
\end{eqnarray}

\begin{eqnarray} \label{formts21}
t_{s,2 \rightarrow 1}=\int_{-\infty}^{+\infty}dx[w_R^{*}(x)\hat{H}w_L(x)]=\int_{-\infty}^{+\infty}dx(-\beta^{\dag} \psi_{E1}^{\dag}(x) +\alpha^{\dag} \psi_{E2}^{\dag}(x) )\times \nonumber \\
\times \hat{H}(\alpha\psi_{E1}(x)+\beta \psi_{E2}(x))=-\alpha\beta^{*} E_1+\alpha^{*}\beta E_2=(E_2-E_1)\alpha \beta = \nonumber \\
= \frac{1}{2}sin(2 \frac{\int_{-\infty}^{0}dx (\psi_{E1}(x)\psi_{E2}^{\dag}(x)+\psi_{E1}(x)^{\dag}\psi_{E2}(x))}{\int_{-\infty}^{0}dx (|\psi_{E1}(x)|^2-|\psi_{E2}(x)|^2)})(E_2-E_1)= \nonumber \\
=\frac{(E_2-E_1)}{2}\frac{\frac{\int_{-\infty}^{0}dx (\psi_{E1}(x)\psi_{E2}^{\dag}(x)+\psi_{E1}(x)^{\dag}\psi_{E2}(x))}{\int_{-\infty}^{0}dx (|\psi_{E1}(x)|^2-|\psi_{E2}(x)|^2)}}{\sqrt{1+[\frac{\int_{-\infty}^{0}dx (\psi_{E1}(x)\psi_{E2}^{\dag}(x)+\psi_{E1}(x)^{\dag}\psi_{E2}(x))}{\int_{-\infty}^{0}dx (|\psi_{E1}(x)|^2-|\psi_{E2}(x)|^2)}]^2}},
\end{eqnarray}

\begin{eqnarray} \label{formts12}
t_{s,1 \rightarrow 2}=\int_{-\infty}^{+\infty}dx[w_L^{*}(x)\hat{H}w_R(x)]=\int_{-\infty}^{+\infty}dx(\alpha^{\dag} \psi_{E1}^{\dag}(x) +\beta^{\dag} \psi_{E2}^{\dag}(x) )\times \nonumber \\
\times \hat{H}(-\beta\psi_{E1}(x)+\alpha \psi_{E2}(x))=-\alpha^{*}\beta E_1+\alpha\beta^{*} E_2=(E_2-E_1)\alpha \beta= \nonumber \\ =(E_2-E_1)\frac{1}{2}2cos(\frac{1}{2}ArcTan(r))sin(\frac{1}{2}ArcTan(r))=(E_2-E_1)\frac{1}{2}sin(ArcTan(r))= \nonumber \\
=(E_2-E_1)\frac{1}{2}sin(ArcTan(r))=\frac{(E_2-E_1)}{2}\frac{r}{\sqrt{1+r^2}}=\nonumber \\
=\frac{(E_2-E_1)}{2}\frac{\frac{\int_{-\infty}^{0}dx (\psi_{E1}(x)\psi_{E2}^{\dag}(x)+\psi_{E1}(x)^{\dag}\psi_{E2}(x))}{\int_{-\infty}^{0}dx (|\psi_{E1}(x)|^2-|\psi_{E2}(x)|^2)}}{\sqrt{1+[\frac{\int_{-\infty}^{0}dx (\psi_{E1}(x)\psi_{E2}^{\dag}(x)+\psi_{E1}(x)^{\dag}\psi_{E2}(x))}{\int_{-\infty}^{0}dx (|\psi_{E1}(x)|^2-|\psi_{E2}(x)|^2)}]^2}}
%
\end{eqnarray}
Therefore tight-binding model useful for description of Wannier qubits (position based qubits) was fundamentally derived from Schr\"{o}dinger formalism.
The obtained results can be summarized by tight-binding model being functional of eigenenergies of Schr\"{o}dinger Hamiltonian in the form as
\begin{eqnarray}
\label{MainFormula}
\begin{pmatrix}
E_{p1} & t_{s21}  \\
t_{s12} & E_{p2}
\end{pmatrix}=
\begin{pmatrix}
E_1+|sin(\frac{1}{2}ArcTan(r))|^2(E_2-E_1) & (E_2-E_1)\frac{1}{2}sin(ArcTan(r))  \\
(E_2-E_1)\frac{1}{2}sin(ArcTan(r)) & E_1+(E_2-E_1)|cos(\frac{1}{2}ArcTan(r))|^2
\end{pmatrix},
\end{eqnarray}
with r given by \ref{special} and use of formulas as \ref{formts11} ,\ref{formts22}, \ref{formts12}, \ref{formts21} basing on formula \ref{GammaForumula}. 
If we assume the possible escape of electron from 2 coupled quantum dot system the wavefunction is no longer normalized to one and we can replace real value eigenenergies with
complex value energies, so $E_1 \rightarrow E_{1r}+iE_{1i}$ and $E_2 \rightarrow E_{2r}+iE_{2i}$. In such case the effective tight-binding model corresponding to complex value eigenenergies can be expressed as
\begin{eqnarray}
\label{MainFormulaD}
\begin{pmatrix}
E_{p1D} & t_{s21D}  \\
t_{s12D} & E_{p2D}
\end{pmatrix}=
\begin{pmatrix}
E_{1r}+|sin(\frac{1}{2}ArcTan(r))|^2(E_{2r}-E_{1r}) & (E_{2r}-E_{1r})\frac{1}{2}sin(ArcTan(r))  \\
(E_{2r}-E_{1r})\frac{1}{2}sin(ArcTan(r)) & E_{1r}+(E_{2r}-E_{1r})|cos(\frac{1}{2}ArcTan(r))|^2
\end{pmatrix}+ \nonumber \\
+\sqrt{-1}
\begin{pmatrix}
E_{1i}+|sin(\frac{1}{2}ArcTan(r))|^2(E_{2i}-E_{1i}) & (E_{2i}-E_{1i})\frac{1}{2}sin(ArcTan(r))  \\
(E_{2i}-E_{1i})\frac{1}{2}sin(ArcTan(r)) & E_{1i}+(E_{2i}-E_{1i})|cos(\frac{1}{2}ArcTan(r))|^2
\end{pmatrix}
\end{eqnarray}
The dissipative version of tight-binding model accounting for electron escape from 2 quantum dot system due to tunneling is non-Hermitian, while non-dissipative version of tight-binding model is Hermitian.

\section{From 2-dimensional Schroedinger equation to Wannier functions and stochastic Finite State Machine}
\begin{figure}
\centering
\includegraphics[scale=0.9]{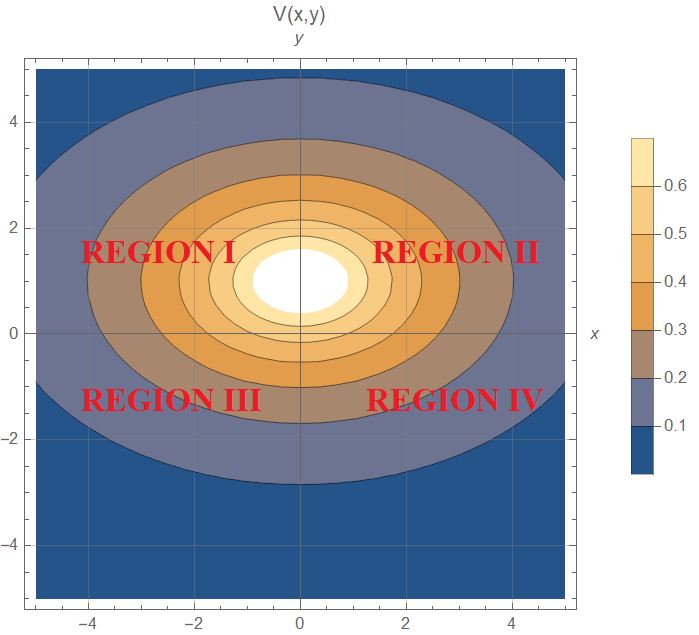}
\caption{Example of confining potential that can be specified to 4 regions: Region I with ($x \in (-\infty,0)$,$y \in (0,\infty)$ ), Region II with ($x \in (0,\infty,)$,$y \in (0,\infty)$ ), Region III with ($x \in (-\infty,0)$,$y \in (-\infty,0)$ ) and Region IV with ($x \in (0,+\infty,)$,$y \in (-\infty,0)$ ). }
\label{2dimBox}
\end{figure}

Let us consider the particle in two dimensional box described by Schroedinger equation.
The example of confining 2-dimensional potential is depicted in Fig.\ref{2dimBox} and one can always divide the space into 4 or N distinct geometric regions.
One can always associate maximum localized wavefunctions that will be called Wannier functions to those 4 or N distinct regions.
Region I will correspond to condition given by $x \in (-\infty,0)$ and $y \in (0,+\infty)$, Region II to condition $x \in (0,\infty,)$ and $y \in (0,+\infty)$,
Region III to condition $x \in (-\infty,0)$ and $y \in (-\infty,0)$ and Region IV will be specified by condition $x \in (0,\infty)$ and $y \in (-\infty,0)$.


We can write generalized Hamiltonian for 4 possible energy levels as
\begin{eqnarray*}
\hat{H}(t)_E=E_1(t)\ket{E_1}\bra{E_1}+E_2(t)\ket{E_2}\bra{E_2}+E_3(t)\ket{E_3}\bra{E_3}+E_4(t)\ket{E_4}\bra{E_4}+f_{E_1 \rightarrow E_2}(t)\ket{E_2}\bra{E_1}+f_{E_1 \rightarrow E_3}(t)\ket{E_3}\bra{E_1}+f_{E_1 \rightarrow E_4}(t)\ket{E_4}\bra{E_1}+ \nonumber \\
+f_{E_2 \rightarrow E_1}(t)\ket{E_1}\bra{E_2}+f_{E_2 \rightarrow E_3}(t)\ket{E_3}\bra{E_2}+f_{E_2 \rightarrow E_4}(t)\ket{E_4}\bra{E_2} 
+f_{E_3 \rightarrow E_1}(t)\ket{E_3}\bra{E_1}+f_{E_3 \rightarrow E_2}(t)\ket{E_2}\bra{E_1}+f_{E_3 \rightarrow E_4}(t)\ket{E_4}\bra{E_3} + \nonumber \\
+f_{E_1 \rightarrow E_4}(t)\ket{E_4}\bra{E_1}+f_{E_4 \rightarrow E_2}(t)\ket{E_2}\bra{E_4}+f_{E_4 \rightarrow E_3}(t)\ket{E_3}\bra{E_4}. \nonumber \\
\end{eqnarray*}
what can be expressed by Schroedinger equations of motion as
\begin{eqnarray}
\begin{pmatrix}
H_{11}=E_1(t) &  H_{12}=f_{E_2 \rightarrow E_1}(t) & H_{13}=f_{E_3 \rightarrow E_1}(t) & H_{14}=f_{E_4 \rightarrow E_1}(t) \nonumber \\
H_{21}=f_{E_1 \rightarrow E_2}(t) &  H_{22}=E_2(t) & H_{23}=f_{E_3 \rightarrow E_2}(t) & H_{24}=f_{E_4 \rightarrow E_2}(t) \nonumber \\
H_{31}=f_{E_1 \rightarrow E_3}(t) &  H_{32}=f_{E_2 \rightarrow E_3}(t) & H_{33}=E_3(t) & H_{34}=f_{E_4 \rightarrow E_3}(t) \nonumber \\
H_{41}=f_{E_1 \rightarrow E_4}(t) &  H_{42}=f_{E_2 \rightarrow E_4}(t) & H_{43}=f_{E_3 \rightarrow E_4}(t) & H_{44}=E_4(t) \nonumber \\
\end{pmatrix}_E
\begin{pmatrix}
\sqrt{p_{E1}(t)}e^{i \phi_1(t)}\psi_{E1}(x,y,t) , \nonumber \\
\sqrt{p_{E2}(t)}e^{i \phi_2(t)}\psi_{E2}(x,y,t) , \nonumber \\
\sqrt{p_{E3}(t)}e^{i \phi_3(t)}\psi_{E3}(x,y,t) , \nonumber \\
\sqrt{p_{E4}(t)}e^{i \phi_4(t)}\psi_{E4}(x,y,t) , \nonumber \\
\end{pmatrix}= \nonumber \\
=i \hbar \frac{d}{dt}
\begin{pmatrix}
\sqrt{p_{E1}(t)}e^{i \phi_1(t)}\psi_{E1}(x,y,t) , \nonumber \\
\sqrt{p_{E2}(t)}e^{i \phi_2(t)}\psi_{E2}(x,y,t) , \nonumber \\
\sqrt{p_{E3}(t)}e^{i \phi_3(t)}\psi_{E3}(x,y,t) , \nonumber \\
\sqrt{p_{E4}(t)}e^{i \phi_4(t)}\psi_{E4}(x,y,t) , \nonumber \\
\end{pmatrix} ,
\end{eqnarray}
where $\psi_{Ek}(x,y,t)$ are eigenenergies with $k=1..4$.

Now we introduce identity matrix $[.]_I$ into equations of motion what results in

\begin{eqnarray}
\begin{pmatrix}
H_{11} &  H_{12} & H_{13} & H_{14} \nonumber \\
H_{21} &  H_{22} & H_{23} & H_{24} \nonumber \\
H_{31} &  H_{32} & H_{33} & H_{34} \nonumber \\
H_{41} &  H_{42} & H_{43} & H_{44} \nonumber \\
\end{pmatrix}_E
\Bigg[
\begin{pmatrix}
+\alpha_1 & +\alpha_2 \\
-\alpha_2 & +\alpha_1 \\
\end{pmatrix}^{-1} \times
\begin{pmatrix}
+\beta_1 & +\beta_2 \\
-\beta_2 & +\beta_1 \\
\end{pmatrix}^{-1}
\begin{pmatrix}
+\alpha_1 & +\alpha_2 \\
-\alpha_2 & +\alpha_1 \\
\end{pmatrix} \times
\begin{pmatrix}
+\beta_1 & +\beta_2 \\
-\beta_2 & +\beta_1 \\
\end{pmatrix}
\Bigg]_{I}
\begin{pmatrix}
\sqrt{p_{E1}(t)}e^{i \phi_1(t)}\psi_{E1}(x,y,t)  \nonumber \\
\sqrt{p_{E2}(t)}e^{i \phi_2(t)}\psi_{E2}(x,y,t)  \nonumber \\
\sqrt{p_{E3}(t)}e^{i \phi_3(t)}\psi_{E3}(x,y,t)  \nonumber \\
\sqrt{p_{E4}(t)}e^{i \phi_4(t)}\psi_{E4}(x,y,t)  \nonumber \\
\end{pmatrix}= \nonumber \\
i \hbar \frac{d}{dt}
\begin{pmatrix}
\sqrt{p_{E1}(t)}e^{i \phi_1(t)}\psi_{E1}(x,y,t)  \nonumber \\
\sqrt{p_{E2}(t)}e^{i \phi_2(t)}\psi_{E2}(x,y,t)  \nonumber \\
\sqrt{p_{E3}(t)}e^{i \phi_3(t)}\psi_{E3}(x,y,t)  \nonumber \\
\sqrt{p_{E4}(t)}e^{i \phi_4(t)}\psi_{E4}(x,y,t)  \nonumber \\
\end{pmatrix}=i \hbar
\begin{pmatrix}
[i\frac{d\phi_1(t)}{dt}\sqrt{p_{E1}(t)}+\frac{d p_{E1}(t)}{dt}\frac{1}{2\sqrt{p_{E1}(t)}}]e^{i \phi_1(t)}]\psi_{E1}(x,y,t)+e^{i \phi_1(t)}\sqrt{p_{E1}(t)}\frac{d}{dt}\psi_{E1}(x,y,t)  \nonumber \\
i\frac{d\phi_2(t)}{dt}\sqrt{p_{E2}(t)}+\frac{d p_{E2}(t)}{dt}\frac{1}{2\sqrt{p_{E2}(t)}}]e^{i \phi_2(t)}]\psi_{E2}(x,y,t)+e^{i \phi_2(t)}\sqrt{p_{E2}(t)}\frac{d}{dt}\psi_{E2}(x,y,t)  \nonumber \\
i\frac{d\phi_3(t)}{dt}\sqrt{p_{E3}(t)}+\frac{d p_{E3}(t)}{dt}\frac{1}{2\sqrt{p_{E3}(t)}}]e^{i \phi_3(t)}]\psi_{E3}(x,y,t)+e^{i \phi_3(t)}\sqrt{p_{E3}(t)}\frac{d}{dt}\psi_{E3}(x,y,t)  \nonumber \\
i\frac{d\phi_4(t)}{dt}\sqrt{p_{E4}(t)}+\frac{d p_{E4}(t)}{dt}\frac{1}{2\sqrt{p_{E4}(t)}}]e^{i \phi_1(t)}]\psi_{E4}(x,y,t)+e^{i \phi_4(t)}\sqrt{p_{E4}(t)}\frac{d}{dt}\psi_{E4}(x,y,t)  \nonumber \\
\end{pmatrix}
,
\end{eqnarray}
what finally brings equations of motion into Wannier form
\begin{eqnarray}
\Bigg[
\begin{pmatrix}
+\alpha_1 & +\alpha_2 \\
-\alpha_2 & +\alpha_1 \\
\end{pmatrix} \times
\begin{pmatrix}
+\beta_1 & +\beta_2 \\
-\beta_2 & +\beta_1 \\
\end{pmatrix}
\Bigg] \times
  \nonumber \\
\begin{pmatrix}
H_{11} &  H_{12} & H_{13} & H_{14} \nonumber \\
H_{21} &  H_{22} & H_{23} & H_{24} \nonumber \\
H_{31} &  H_{32} & H_{33} & H_{34} \nonumber \\
H_{41} &  H_{42} & H_{43} & H_{44} \nonumber \\
\end{pmatrix}_E
\Bigg[
\begin{pmatrix}
+\alpha_1 & +\alpha_2 \\
-\alpha_2 & +\alpha_1 \\
\end{pmatrix}^{-1} \times
\begin{pmatrix}
+\beta_1 & +\beta_2 \\
-\beta_2 & +\beta_1 \\
\end{pmatrix}^{-1}
\begin{pmatrix}
+\alpha_1 & +\alpha_2 \\
-\alpha_2 & +\alpha_1 \\
\end{pmatrix} \times
\begin{pmatrix}
+\beta_1 & +\beta_2 \\
-\beta_2 & +\beta_1 \\
\end{pmatrix}
\Bigg]
\begin{pmatrix}
\sqrt{p_{E1}(t)}e^{i \phi_1(t)}\psi_{E1}(x,y,t) \nonumber \\
\sqrt{p_{E2}(t)}e^{i \phi_2(t)}\psi_{E2}(x,y,t) \nonumber \\
\sqrt{p_{E3}(t)}e^{i \phi_3(t)}\psi_{E3}(x,y,t) \nonumber \\
\sqrt{p_{E4}(t)}e^{i \phi_4(t)}\psi_{E4}(x,y,t) \nonumber \\
\end{pmatrix}= \nonumber \\
\begin{pmatrix}
+\alpha_1 & +\alpha_2 \\
-\alpha_2 & +\alpha_1 \\
\end{pmatrix} \times
\begin{pmatrix}
+\beta_1 & +\beta_2 \\
-\beta_2 & +\beta_1 \\
\end{pmatrix}
i \hbar \frac{d}{dt}
\begin{pmatrix}
\sqrt{p_{E1}(t)}e^{i \phi_1(t)}\psi_{E1}(x,y,t) \nonumber \\
\sqrt{p_{E2}(t)}e^{i \phi_2(t)}\psi_{E2}(x,y,t) \nonumber \\
\sqrt{p_{E3}(t)}e^{i \phi_3(t)}\psi_{E3}(x,y,t) \nonumber \\
\sqrt{p_{E4}(t)}e^{i \phi_4(t)}\psi_{E4}(x,y,t) \nonumber \\
\end{pmatrix}= \nonumber \\
=
i \hbar \frac{d}{dt}
\Bigg[
\begin{pmatrix}
+\alpha_1 & +\alpha_2 \\
-\alpha_2 & +\alpha_1 \\
\end{pmatrix} \times
\begin{pmatrix}
+\beta_1 & +\beta_2 \\
-\beta_2 & +\beta_1 \\
\end{pmatrix}
\begin{pmatrix}
\sqrt{p_{E1}(t)}e^{i \phi_1(t)}\psi_{E1}(x,y,t) \nonumber \\
\sqrt{p_{E2}(t)}e^{i \phi_2(t)}\psi_{E2}(x,y,t) \nonumber \\
\sqrt{p_{E3}(t)}e^{i \phi_3(t)}\psi_{E3}(x,y,t) \nonumber \\
\sqrt{p_{E4}(t)}e^{i \phi_4(t)}\psi_{E4}(x,y,t) \nonumber \\
\end{pmatrix}
\Bigg] + \nonumber \\
-
\Bigg[ i \hbar \frac{d}{dt}
\Big[
\begin{pmatrix}
+\alpha_1 & +\alpha_2 \\
-\alpha_2 & +\alpha_1 \\
\end{pmatrix} \times
\begin{pmatrix}
+\beta_1 & +\beta_2 \\
-\beta_2 & +\beta_1 \\
\end{pmatrix}
\Big]
\begin{pmatrix}
\sqrt{p_{E1}(t)}e^{i \phi_1(t)}\psi_{E1}(x,y,t) \nonumber \\
\sqrt{p_{E2}(t)}e^{i \phi_2(t)}\psi_{E2}(x,y,t) \nonumber \\
\sqrt{p_{E3}(t)}e^{i \phi_3(t)}\psi_{E3}(x,y,t) \nonumber \\
\sqrt{p_{E4}(t)}e^{i \phi_4(t)}\psi_{E4}(x,y,t) \nonumber \\
\end{pmatrix}
\Bigg] = \nonumber \\
=
i \hbar \frac{d}{dt}
\Bigg[
\begin{pmatrix}
+\alpha_1 & +\alpha_2 \\
-\alpha_2 & +\alpha_1 \\
\end{pmatrix} \times
\begin{pmatrix}
+\beta_1 & +\beta_2 \\
-\beta_2 & +\beta_1 \\
\end{pmatrix}
\begin{pmatrix}
\sqrt{p_{E1}(t)}e^{i \phi_1(t)}\psi_{E1}(x,y,t) \nonumber \\
\sqrt{p_{E2}(t)}e^{i \phi_2(t)}\psi_{E2}(x,y,t) \nonumber \\
\sqrt{p_{E3}(t)}e^{i \phi_3(t)}\psi_{E3}(x,y,t) \nonumber \\
\sqrt{p_{E4}(t)}e^{i \phi_4(t)}\psi_{E4}(x,y,t) \nonumber \\
\end{pmatrix}
\Bigg] + \nonumber \\
-
\Bigg[
\Big[ i \hbar \frac{d}{dt}
\begin{pmatrix}
+\alpha_1 & +\alpha_2 \\
-\alpha_2 & +\alpha_1 \\
\end{pmatrix} \times
\begin{pmatrix}
+\beta_1 & +\beta_2 \\
-\beta_2 & +\beta_1 \\
\end{pmatrix}
\Big]
\Big[
\begin{pmatrix}
+\alpha_1 & +\alpha_2 \\
-\alpha_2 & +\alpha_1 \\
\end{pmatrix} \times
\begin{pmatrix}
+\beta_1 & +\beta_2 \\
-\beta_2 & +\beta_1 \\
\end{pmatrix}
\Big]^{-1}
\times \nonumber \\
\times
\Big[
\begin{pmatrix}
+\alpha_1 & +\alpha_2 \\
-\alpha_2 & +\alpha_1 \\
\end{pmatrix} \times
\begin{pmatrix}
+\beta_1 & +\beta_2 \\
-\beta_2 & +\beta_1 \\
\end{pmatrix}
\begin{pmatrix}
\sqrt{p_{E1}(t)}e^{i \phi_1(t)}\psi_{E1}(x,y,t) \nonumber \\
\sqrt{p_{E2}(t)}e^{i \phi_2(t)}\psi_{E2}(x,y,t) \nonumber \\
\sqrt{p_{E3}(t)}e^{i \phi_3(t)}\psi_{E3}(x,y,t) \nonumber \\
\sqrt{p_{E4}(t)}e^{i \phi_4(t)}\psi_{E4}(x,y,t) \nonumber \\
\end{pmatrix}
\Big]
\Bigg] = \nonumber \\
=i \hbar \frac{d}{dt}
\begin{pmatrix}
w_{1}(x,y,t) \nonumber \\
w_{2}(x,y,t)  \nonumber \\
w_{3}(x,y,t)  \nonumber \\
w_{4}(x,y,t)  \nonumber \\
\end{pmatrix} -
\Big[ i \hbar \frac{d}{dt}
\begin{pmatrix}
+\alpha_1 & +\alpha_2 \\
-\alpha_2 & +\alpha_1 \\
\end{pmatrix} \times
\begin{pmatrix}
+\beta_1 & +\beta_2 \\
-\beta_2 & +\beta_1 \\
\end{pmatrix}
\Big]
\Big[
\begin{pmatrix}
+\alpha_1 & +\alpha_2 \\
-\alpha_2 & +\alpha_1 \\
\end{pmatrix} \times
\begin{pmatrix}
+\beta_1 & +\beta_2 \\
-\beta_2 & +\beta_1 \\
\end{pmatrix}
\Big]^{-1}
\begin{pmatrix}
w_{1}(x,y,t) \nonumber \\
w_{2}(x,y,t)  \nonumber \\
w_{3}(x,y,t)  \nonumber \\
w_{4}(x,y,t)  \nonumber \\
\end{pmatrix} \\
\end{eqnarray}
Having Hamiltonian in time independent case we have
$
\Big[ i \hbar \frac{d}{dt}
\begin{pmatrix}
+\alpha_1 & +\alpha_2 \\
-\alpha_2 & +\alpha_1 \\
\end{pmatrix} \times
\begin{pmatrix}
+\beta_1 & +\beta_2 \\
-\beta_2 & +\beta_1 \\
\end{pmatrix}
\Big]
\Big[
\begin{pmatrix}
+\alpha_1 & +\alpha_2 \\
-\alpha_2 & +\alpha_1 \\
\end{pmatrix} \times
\begin{pmatrix}
+\beta_1 & +\beta_2 \\
-\beta_2 & +\beta_1 \\
\end{pmatrix}
\Big]^{-1} = \hat{0}
$ what greatly simplifies the equation of motion. One obtains tight-binding model that is Schroedinger equivalent and encounters the
appearance of Wannier functions $w_1(x,y)$, .. , $w_4(x,y)$  given by following equation of motion

\begin{eqnarray}
\Bigg[
\begin{pmatrix}
+\alpha_1 & +\alpha_2 \\
-\alpha_2 & +\alpha_1 \\
\end{pmatrix} \times
\begin{pmatrix}
+\beta_1 & +\beta_2 \\
-\beta_2 & +\beta_1 \\
\end{pmatrix} 
\begin{pmatrix}
H_{11} &  H_{12} & H_{13} & H_{14} \nonumber \\
H_{21} &  H_{22} & H_{23} & H_{24} \nonumber \\
H_{31} &  H_{32} & H_{33} & H_{34} \nonumber \\
H_{41} &  H_{42} & H_{43} & H_{44} \nonumber \\
\end{pmatrix}_E
\begin{pmatrix}
+\alpha_1 & -\alpha_2 \\
+\alpha_2 & +\alpha_1 \\
\end{pmatrix} \times
\begin{pmatrix}
+\beta_1 & -\beta_2 \\
+\beta_2 & +\beta_1 \\
\end{pmatrix}
\Bigg]
\begin{pmatrix}
\sqrt{\eta_1(t)} e^{i \xi_1(t)}w_1(x,y) , \nonumber \\
\sqrt{\eta_2(t)} e^{i \xi_2(t)}w_2(x,y) , \nonumber \\
\sqrt{\eta_3(t)} e^{i \xi_3(t)}w_3(x,y) , \nonumber \\
\sqrt{\eta_4(t)} e^{i \xi_4(t)}w_4(x,y) , \nonumber \\
\end{pmatrix}
= \nonumber \\
i \hbar \frac{d}{dt}
\begin{pmatrix}
\sqrt{\eta_1(t)} e^{i \xi_1(t)}w_1(x,y) , \nonumber \\
\sqrt{\eta_2(t)} e^{i \xi_2(t)}w_2(x,y) , \nonumber \\
\sqrt{\eta_3(t)} e^{i \xi_3(t)}w_3(x,y) , \nonumber \\
\sqrt{\eta_4(t)} e^{i \xi_4(t)}w_4(x,y) , \nonumber \\
\end{pmatrix}
\end{eqnarray}
and with Hamiltonian in Wannier form given by

\begin{eqnarray}
\hat{H}_w=
\Bigg[
\begin{pmatrix}
+\alpha_1 & +\alpha_2 \\
-\alpha_2 & +\alpha_1 \\
\end{pmatrix} \times
\begin{pmatrix}
+\beta_1 & +\beta_2 \\
-\beta_2 & +\beta_1 \\
\end{pmatrix} 
\begin{pmatrix}
H_{11} &  H_{12} & H_{13} & H_{14} \nonumber \\
H_{21} &  H_{22} & H_{23} & H_{24} \nonumber \\
H_{31} &  H_{32} & H_{33} & H_{34} \nonumber \\
H_{41} &  H_{42} & H_{43} & H_{44} \nonumber \\
\end{pmatrix}_E
\begin{pmatrix}
+\alpha_1 & -\alpha_2 \\
+\alpha_2 & +\alpha_1 \\
\end{pmatrix} \times
\begin{pmatrix}
+\beta_1 & -\beta_2 \\
+\beta_2 & +\beta_1 \\
\end{pmatrix}
\Bigg]= \nonumber \\
=
\Bigg[
\begin{pmatrix}
+\alpha_1\beta_1 & +\alpha_1\beta_2 & +\alpha_2\beta_1 & +\alpha_2\beta_2  \\
-\alpha_1\beta_2 & +\alpha_1\beta_1 & -\alpha_2\beta_2 & + \alpha_2 \beta_1 \\
-\alpha_2\beta_1 & -\alpha_2\beta_2 & +\alpha_1\beta_1 & +\alpha_1\beta_2  \\
\alpha_2\beta_2 & -\alpha_2\beta_1 & -\alpha_1\beta_2 & +\alpha_1\beta_1  \\
\end{pmatrix}
\begin{pmatrix}
H_{11} &  H_{12} & H_{13} & H_{14} \nonumber \\
H_{21} &  H_{22} & H_{23} & H_{24} \nonumber \\
H_{31} &  H_{32} & H_{33} & H_{34} \nonumber \\
H_{41} &  H_{42} & H_{43} & H_{44} \nonumber \\
\end{pmatrix}_E
\begin{pmatrix}
+\alpha_1\beta_1 & -\alpha_1\beta_2 & -\alpha_2\beta_1 & +\alpha_2\beta_2 \\
+\alpha_1\beta_2 & +\alpha_1\beta_1 & -\alpha_2\beta_2 & -\alpha_2\beta_2 \\
+\alpha_2\beta_1 & -\alpha_2\beta_2 & +\alpha_1\beta_1 & -\alpha_1\beta_2 \\
+\alpha_2\beta_2 & +\alpha_2\beta_1 & +\alpha_1\beta_2 & +\alpha_1\beta_1 \\
\end{pmatrix}= \nonumber \\
=
\begin{pmatrix}
H_{w11} &  H_{w12} & H_{w13} & H_{w14} \nonumber \\
H_{w21} &  H_{w22} & H_{w23} & H_{w24} \nonumber \\
H_{w31} &  H_{w32} & H_{w33} & H_{w34} \nonumber \\
H_{w41} &  H_{w42} & H_{w43} & H_{w44} \nonumber \\
\end{pmatrix}_w
\end{eqnarray}
that also can be written in symbolic way as

\begin{eqnarray}
\hat{H}_w=H_{w11}\ket{w_1}\bra{w_1}+H_{w22}\ket{w_2}\bra{w_2}+H_{w33}\ket{w_3}\bra{w_3}+H_{w44}\ket{w_4}\bra{w_4} + 
+H_{w21}\ket{w_2}\bra{w_1}+H_{w23}\ket{w_3}\bra{w_2}+H_{w24}\ket{w_4}\bra{w_2}+ \nonumber \\
+H_{w31}\ket{w_3}\bra{w_1}+H_{w32}\ket{w_2}\bra{w_3}+H_{w34}\ket{w_4}\bra{w_3}+H_{w41}\ket{w_1}\bra{w_4}+H_{w42}\ket{w_2}\bra{w_4}+H_{w43}\ket{w_3}\bra{w_4},
\end{eqnarray}
with quantum state given in symbolic way as
\begin{eqnarray}
\ket{\psi(t)}_w=\sqrt{\eta_1(t)} e^{i \xi_1(t)}\ket{w_1}+\sqrt{\eta_2(t)} e^{i \xi_2(t)}\ket{w_2}+\sqrt{\eta_3(t)} e^{i \xi_3(t)}\ket{w_3}+\sqrt{\eta_4(t)} e^{i \xi_4(t)}\ket{w_4}
\end{eqnarray}
that implies probabilities of presence in regions 1, .. 4 by $\eta_1(t)$,.., $\eta_4(t)$ .

We have shown that Hamiltonian matrix in Wannier representation is particularly simplified if we have time-independent Hamiltonian given as
\begin{eqnarray*}
\hat{H}=E_1\ket{E_1}\bra{E_1}+E_2\ket{E_2}\bra{E_2}+E_3\ket{E_3}\bra{E_3}+E_4\ket{E_4}\bra{E_4}.
\end{eqnarray*}
In general case of time-dependent Hamiltonian one obtains
\begin{eqnarray}
\Bigg[
\begin{pmatrix}
+\alpha_1 & +\alpha_2 \\
-\alpha_2 & +\alpha_1 \\
\end{pmatrix}_t \times
\begin{pmatrix}
+\beta_1 & +\beta_2 \\
-\beta_2 & +\beta_1 \\
\end{pmatrix} 
\begin{pmatrix}
H_{11} &  H_{12} & H_{13} & H_{14} \nonumber \\
H_{21} &  H_{22} & H_{23} & H_{24} \nonumber \\
H_{31} &  H_{32} & H_{33} & H_{34} \nonumber \\
H_{41} &  H_{42} & H_{43} & H_{44} \nonumber \\
\end{pmatrix}_E
\begin{pmatrix}
+\alpha_1 & -\alpha_2 \\
+\alpha_2 & +\alpha_1 \\
\end{pmatrix} \times
\begin{pmatrix}
+\beta_1 & -\beta_2 \\
+\beta_2 & +\beta_1 \\
\end{pmatrix}
\begin{pmatrix}
\sqrt{\eta_1(t)} e^{i \xi_1(t)}w_1(x,y,t) , \nonumber \\
\sqrt{\eta_2(t)} e^{i \xi_2(t)}w_2(x,y,t) , \nonumber \\
\sqrt{\eta_3(t)} e^{i \xi_3(t)}w_3(x,y,t) , \nonumber \\
\sqrt{\eta_4(t)} e^{i \xi_4(t)}w_4(x,y,t) , \nonumber \\
\end{pmatrix}= \nonumber \\
=i \hbar \frac{d}{dt}
\begin{pmatrix}
w_{1}(x,y,t) \nonumber \\
w_{2}(x,y,t)  \nonumber \\
w_{3}(x,y,t)  \nonumber \\
w_{4}(x,y,t)  \nonumber \\
\end{pmatrix} - i \hbar
\Big[  \frac{d}{dt} \Big[
\begin{pmatrix}
+\alpha_1 & +\alpha_2 \\
-\alpha_2 & +\alpha_1 \\
\end{pmatrix} \times
\begin{pmatrix}
+\beta_1 & +\beta_2 \\
-\beta_2 & +\beta_1 \\
\end{pmatrix}
\Big]
\Big]
\Big[
\begin{pmatrix}
+\alpha_1 & -\alpha_2 \\
+\alpha_2 & +\alpha_1 \\
\end{pmatrix} \times
\begin{pmatrix}
+\beta_1 & -\beta_2 \\
+\beta_2 & +\beta_1 \\
\end{pmatrix}
\Big]
\begin{pmatrix}
w_{1}(x,y,t) \nonumber \\
w_{2}(x,y,t)  \nonumber \\
w_{3}(x,y,t)  \nonumber \\
w_{4}(x,y,t)  \nonumber \\
\end{pmatrix} \\
\end{eqnarray}
.
Let us now apply the operator
\begin{eqnarray}
\begin{pmatrix}
\int_{-\infty}^{+\infty}dx\int_{-\infty}^{+\infty}dy w_1^{*}(x,y,t) & 0 & 0 & 0 \\
0 & \int_{-\infty}^{+\infty}dx\int_{-\infty}^{+\infty}dy w_2^{*}(x,y,t) & 0 & 0 \\
0 & 0 & \int_{-\infty}^{+\infty}dx\int_{-\infty}^{+\infty}dy w_1^{*}(x,y,t) & 0  \\
0 & 0 & 0 & \int_{-\infty}^{+\infty}dx\int_{-\infty}^{+\infty}dy w_2^{*}(x,y,t) \\
\end{pmatrix}
\end{eqnarray}

\begin{landscape}
to both sides of previous equation and we obtain

\begin{eqnarray}
\Bigg[
\begin{pmatrix}
+\alpha_1 & +\alpha_2 \\
-\alpha_2 & +\alpha_1 \\
\end{pmatrix}_t \times
\begin{pmatrix}
+\beta_1 & +\beta_2 \\
-\beta_2 & +\beta_1 \\
\end{pmatrix} 
\begin{pmatrix}
H_{11} &  H_{12} & H_{13} & H_{14} \nonumber \\
H_{21} &  H_{22} & H_{23} & H_{24} \nonumber \\
H_{31} &  H_{32} & H_{33} & H_{34} \nonumber \\
H_{41} &  H_{42} & H_{43} & H_{44} \nonumber \\
\end{pmatrix}_E
\begin{pmatrix}
+\alpha_1 & -\alpha_2 \\
+\alpha_2 & +\alpha_1 \\
\end{pmatrix} \times
\begin{pmatrix}
+\beta_1 & -\beta_2 \\
+\beta_2 & +\beta_1 \\
\end{pmatrix}
\begin{pmatrix}
\sqrt{\eta_1(t)} e^{i \xi_1(t)} \nonumber \\
\sqrt{\eta_2(t)} e^{i \xi_2(t)} \nonumber \\
\sqrt{\eta_3(t)} e^{i \xi_3(t)} \nonumber \\
\sqrt{\eta_4(t)} e^{i \xi_4(t)}  \nonumber \\
\end{pmatrix}= \nonumber \\
=
i \hbar \frac{d}{dt}
\begin{pmatrix}
\sqrt{\eta_1(t)} e^{i \xi_1(t)} \nonumber \\
\sqrt{\eta_2(t)} e^{i \xi_2(t)} \nonumber \\
\sqrt{\eta_3(t)} e^{i \xi_3(t)} \nonumber \\
\sqrt{\eta_4(t)} e^{i \xi_4(t)}  \nonumber \\
\end{pmatrix}
- i \hbar
\Big[  \frac{d}{dt} \Big[
\begin{pmatrix}
+\alpha_1 & +\alpha_2 \\
-\alpha_2 & +\alpha_1 \\
\end{pmatrix} \times
\begin{pmatrix}
+\beta_1 & +\beta_2 \\
-\beta_2 & +\beta_1 \\
\end{pmatrix}
\Big]
\Big]
\Big[
\begin{pmatrix}
+\alpha_1 & -\alpha_2 \\
+\alpha_2 & +\alpha_1 \\
\end{pmatrix} \times
\begin{pmatrix}
+\beta_1 & -\beta_2 \\
+\beta_2 & +\beta_1 \\
\end{pmatrix}
\Big]
\begin{pmatrix}
\sqrt{\eta_1(t)} e^{i \xi_1(t)} \nonumber \\
\sqrt{\eta_2(t)} e^{i \xi_2(t)} \nonumber \\
\sqrt{\eta_3(t)} e^{i \xi_3(t)}  \nonumber \\
\sqrt{\eta_4(t)} e^{i \xi_4(t)}   \nonumber \\
\end{pmatrix}. \\
\end{eqnarray}
The last equation can be rewritten to be the most general form of disspative Wannier equation of motion that is fully equivalent to non-dissipative or dissipative Schroedinger equation of motion under time-dependent Hamiltonian and is of the form

\begin{eqnarray}
\Bigg[
\begin{pmatrix}
+\alpha_1\beta_1 & +\alpha_1\beta_2 & +\alpha_2\beta_1 & +\alpha_2\beta_2  \\
-\alpha_1\beta_2 & +\alpha_1\beta_1 & -\alpha_2\beta_2 & + \alpha_2 \beta_1 \\
-\alpha_2\beta_1 & -\alpha_2\beta_2 & +\alpha_1\beta_1 & +\alpha_1\beta_2  \\
\alpha_2\beta_2 & -\alpha_2\beta_1 & -\alpha_1\beta_2 & +\alpha_1\beta_1  \\
\end{pmatrix}_t
\begin{pmatrix}
H_{11}(t) &  H_{12}(t) & H_{13}(t) & H_{14}(t) \nonumber \\
H_{21}(t) &  H_{22}(t) & H_{23}(t) & H_{24}(t) \nonumber \\
H_{31}(t) &  H_{32}(t) & H_{33}(t) & H_{34}(t) \nonumber \\
H_{41}(t) &  H_{42}(t) & H_{43}(t) & H_{44}(t) \nonumber \\
\end{pmatrix}_{E(t)}
\begin{pmatrix}
+\alpha_1\beta_1 & -\alpha_1\beta_2 & -\alpha_2\beta_1 & +\alpha_2\beta_2 \\
+\alpha_1\beta_2 & +\alpha_1\beta_1 & -\alpha_2\beta_2 & -\alpha_2\beta_2 \\
+\alpha_2\beta_1 & -\alpha_2\beta_2 & +\alpha_1\beta_1 & -\alpha_1\beta_2 \\
+\alpha_2\beta_2 & +\alpha_2\beta_1 & +\alpha_1\beta_2 & +\alpha_1\beta_1 \\
\end{pmatrix}_t
\Bigg]_t
\begin{pmatrix}
\sqrt{\eta_1(t)} e^{i \xi_1(t)} \nonumber \\
\sqrt{\eta_2(t)} e^{i \xi_2(t)} \nonumber \\
\sqrt{\eta_3(t)} e^{i \xi_3(t)} \nonumber \\
\sqrt{\eta_4(t)} e^{i \xi_4(t)}  \nonumber \\
\end{pmatrix}= \nonumber \\
=
i \hbar \frac{d}{dt}
\begin{pmatrix}
\sqrt{\eta_1(t)} e^{i \xi_1(t)} \nonumber \\
\sqrt{\eta_2(t)} e^{i \xi_2(t)} \nonumber \\
\sqrt{\eta_3(t)} e^{i \xi_3(t)} \nonumber \\
\sqrt{\eta_4(t)} e^{i \xi_4(t)}  \nonumber \\
\end{pmatrix}
- i \hbar
\Bigg[
\begin{pmatrix}
+\frac{d}{dt}(\alpha_1\beta_1) & +\frac{d}{dt}(\alpha_1\beta_2) & +\frac{d}{dt}(\alpha_2\beta_1) & +\frac{d}{dt}(\alpha_2\beta_2) \\
-\frac{d}{dt}(\alpha_1\beta_2) & +\frac{d}{dt}(\alpha_1\beta_1) & -\frac{d}{dt}(\alpha_2\beta_2) & +\frac{d}{dt}(\alpha_2\beta_1) \\
-\frac{d}{dt}(\alpha_2\beta_1) & -\frac{d}{dt}(\alpha_2\beta_2) & +\frac{d}{dt}(\alpha_1\beta_1) & +\frac{d}{dt}(\alpha_1\beta_2) \\
+\frac{d}{dt}(\alpha_2\beta_2) & -\frac{d}{dt}(\alpha_2\beta_1) & -\frac{d}{dt}(\alpha_1\beta_2) & +\frac{d}{dt}(\alpha_1\beta_1) \\
\end{pmatrix}
\begin{pmatrix}
+\alpha_1\beta_1 & -\alpha_1\beta_2 & -\alpha_2\beta_1 & +\alpha_2\beta_2 \\
+\alpha_1\beta_2 & +\alpha_1\beta_1 & -\alpha_2\beta_2 & -\alpha_2\beta_1 \\
+\alpha_2\beta_1 & -\alpha_2\beta_2 & +\alpha_1\beta_1 & -\alpha_1\beta_2 \\
+\alpha_2\beta_2 & +\alpha_2\beta_1 & +\alpha_1\beta_2 & +\alpha_1\beta_1 \\
\end{pmatrix}_t
\Bigg]
\begin{pmatrix}
\sqrt{\eta_1(t)} e^{i \xi_1(t)} \nonumber \\
\sqrt{\eta_2(t)} e^{i \xi_2(t)} \nonumber \\
\sqrt{\eta_3(t)} e^{i \xi_3(t)}  \nonumber \\
\sqrt{\eta_4(t)} e^{i \xi_4(t)}   \nonumber \\
\end{pmatrix}. \\
\label{eqncentral}
\end{eqnarray}
Here matrix
$- i \hbar
\Bigg[
\begin{pmatrix}
+\frac{d}{dt}(\alpha_1\beta_1) & +\frac{d}{dt}(\alpha_1\beta_2) & +\frac{d}{dt}(\alpha_2\beta_1) & +\frac{d}{dt}(\alpha_2\beta_2) \\
-\frac{d}{dt}(\alpha_1\beta_2) & +\frac{d}{dt}(\alpha_1\beta_1) & -\frac{d}{dt}(\alpha_2\beta_2) & +\frac{d}{dt}(\alpha_2\beta_1) \\
-\frac{d}{dt}(\alpha_2\beta_1) & -\frac{d}{dt}(\alpha_2\beta_2) & +\frac{d}{dt}(\alpha_1\beta_1) & +\frac{d}{dt}(\alpha_1\beta_2) \\
+\frac{d}{dt}(\alpha_2\beta_2) & -\frac{d}{dt}(\alpha_2\beta_1) & -\frac{d}{dt}(\alpha_1\beta_2) & +\frac{d}{dt}(\alpha_1\beta_1) \\
\end{pmatrix}
\begin{pmatrix}
+\alpha_1\beta_1 & -\alpha_1\beta_2 & -\alpha_2\beta_1 & +\alpha_2\beta_2 \\
+\alpha_1\beta_2 & +\alpha_1\beta_1 & -\alpha_2\beta_2 & -\alpha_2\beta_1 \\
+\alpha_2\beta_1 & -\alpha_2\beta_2 & +\alpha_1\beta_1 & -\alpha_1\beta_2 \\
+\alpha_2\beta_2 & +\alpha_2\beta_1 & +\alpha_1\beta_2 & +\alpha_1\beta_1 \\
\end{pmatrix}_t
\Bigg]$ plays the role of imaginary value potential that can be treated as dissipative term. One can map equation of motion specified by \ref{eqncentral}
into complex value time that was presented in the Section \ref{complextime}. However it is the subject of future work.

Considering time-independent Hamiltonian in eigenergy base one obtains the following Wannier Hamiltonian
\normalsize
\begin{eqnarray}
\hat{H}_w=
\begin{pmatrix}
\alpha_1^2 (\beta_1^2 E_1 + \beta_2^2 E_2) +
 \alpha_2^2 (\beta_1^2 E_3 + \beta_2^2 E_4) &  \beta_1 \beta_2 (\alpha_1^2 (-E_1 + E_2) + \alpha_2^2 (-E_3 + E_4)) & \alpha_1 \alpha_2 (\beta_1^2 (-E_1 + E_3) + \beta_2^2 (-E_2 + E_4)) & \alpha_1 \alpha_2 \beta_1 \beta_2 (E_1 - E_2 - E_3 + E_4) \nonumber \\
\beta_1 \beta_2 (\alpha_1^2 (-E_1 + E_2) + \alpha_2^2 (-E_3 + E_4)) &  \alpha_1^2 (\beta_2^2 E_1 + \beta_1^2 E_2) +
 \alpha_2^2 (\beta_2^2 E_3 + \beta_1^2 E_4) & \alpha_1 \alpha_2 \beta_1 \beta_2 (E_1 - E_2 - E_3 + E_4) & \alpha_1 \alpha_2 (\beta_2^2 (-E_1 + E_3) + \beta_1^2 (-E_2 + E_4)) \nonumber \\
\alpha_1 \alpha_2 (\beta_1^2 (-E_1 + E_3) + \beta_2^2 (-E_2 + E_4)) &  \alpha_1 \alpha_2 \beta_1 \beta_2 (E_1 - E_2 - E_3 + E_4) & \alpha_2^2 (\beta_1^2 E_1 + \beta2^2 E_2) + \alpha_1^2 (\beta_1^2 E_3 + \beta_2^2 E_4 & \beta_1 \beta_2 (\alpha_2^2 (-E_1 + E_2) + \alpha_1^2 (-E_3 + E_4)) \nonumber \\
\alpha_1 \alpha_2 \beta_1 \beta_2 (E_1 - E_2 - E_3 + E_4) &  \alpha_1 \alpha_2 (\beta_2^2 (-E_1 + E_3) + \beta_1^2 (-E_2 + E_4)) & \beta_1 \beta_2 (\alpha_2^2 (-E_1 + E_2) + \alpha_1^2 (-E_3 + E_4)) & \alpha_2^2 (\beta_2^2 E_1 + \beta_1^2 E_2) +
 \alpha_1^2 (\beta2^2 E_3 + \beta_1^2 E_4) \nonumber \\
\end{pmatrix}
\end{eqnarray}
\normalsize
The equations of motion with Hamiltonian matrix in Wannier representation can be simplified to be in the form as
\small
\begin{eqnarray}
\begin{pmatrix}
\int_{-\infty}^{+\infty}dx\int_{-\infty}^{+\infty}dy w_1^{*}(x,y) & 0 & 0 & 0 \\
0 & \int_{-\infty}^{+\infty}dx\int_{-\infty}^{+\infty}dy w_2^{*}(x,y) & 0 & 0 \\
0 & 0 & \int_{-\infty}^{+\infty}dx\int_{-\infty}^{+\infty}dy w_1^{*}(x,y) & 0  \\
0 & 0 & 0 & \int_{-\infty}^{+\infty}dx\int_{-\infty}^{+\infty}dy w_2^{*}(x,y) \\
\end{pmatrix}
i\hbar \frac{d}{dt}
\begin{pmatrix}
\sqrt{\eta_1(t)} e^{i \xi_1(t)}w_1(x,y) \nonumber \\
\sqrt{\eta_2(t)} e^{i \xi_2(t)}w_2(x,y) \nonumber \\
\sqrt{\eta_3(t)} e^{i \xi_3(t)}w_3(x,y) \nonumber \\
\sqrt{\eta_4(t)} e^{i \xi_4(t)}w_4(x,y) \nonumber \\
\end{pmatrix}
= \nonumber \\
=
\begin{pmatrix}
\int_{-\infty}^{+\infty}dx\int_{-\infty}^{+\infty}dy w_1^{*}(x,y) & 0 & 0 & 0 \\
0 & \int_{-\infty}^{+\infty}dx\int_{-\infty}^{+\infty}dy w_2^{*}(x,y) & 0 & 0 \\
0 & 0 & \int_{-\infty}^{+\infty}dx\int_{-\infty}^{+\infty}dy w_1^{*}(x,y) & 0  \\
0 & 0 & 0 & \int_{-\infty}^{+\infty}dx\int_{-\infty}^{+\infty}dy w_2^{*}(x,y) \\
\end{pmatrix}
\times
\nonumber \\
\begin{pmatrix}
\alpha_1^2 (\beta_1^2 E_1 + \beta_2^2 E_2) +
 \alpha_2^2 (\beta_1^2 E_3 + \beta_2^2 E_4) &  \beta_1 \beta_2 (\alpha_1^2 (-E_1 + E_2) + \alpha_2^2 (-E_3 + E_4)) & \alpha_1 \alpha_2 (\beta_1^2 (-E_1 + E_3) + \beta_2^2 (-E_2 + E_4)) & \alpha_1 \alpha_2 \beta_1 \beta_2 (E_1 - E_2 - E_3 + E_4) \nonumber \\
\beta_1 \beta_2 (\alpha_1^2 (-E_1 + E_2) + \alpha_2^2 (-E_3 + E_4)) &  \alpha_1^2 (\beta_2^2 E_1 + \beta_1^2 E_2) +
 \alpha_2^2 (\beta_2^2 E_3 + \beta_1^2 E_4) & \alpha_1 \alpha_2 \beta_1 \beta_2 (E_1 - E_2 - E_3 + E_4) & \alpha_1 \alpha_2 (\beta_2^2 (-E_1 + E_3) + \beta_1^2 (-E_2 + E_4)) \nonumber \\
\alpha_1 \alpha_2 (\beta_1^2 (-E_1 + E_3) + \beta_2^2 (-E_2 + E_4)) &  \alpha_1 \alpha_2 \beta_1 \beta_2 (E_1 - E_2 - E_3 + E_4) & \alpha_2^2 (\beta_1^2 E_1 + \beta2^2 E_2) + \alpha_1^2 (\beta_1^2 E_3 + \beta_2^2 E_4 & \beta_1 \beta_2 (\alpha_2^2 (-E_1 + E_2) + \alpha_1^2 (-E_3 + E_4)) \nonumber \\
\alpha_1 \alpha_2 \beta_1 \beta_2 (E_1 - E_2 - E_3 + E_4) &  \alpha_1 \alpha_2 (\beta_2^2 (-E_1 + E_3) + \beta_1^2 (-E_2 + E_4)) & \beta_1 \beta_2 (\alpha_2^2 (-E_1 + E_2) + \alpha_1^2 (-E_3 + E_4)) & \alpha_2^2 (\beta_2^2 E_1 + \beta_1^2 E_2) +
 \alpha_1^2 (\beta2^2 E_3 + \beta_1^2 E_4) \nonumber \\
\end{pmatrix}
\begin{pmatrix}
\sqrt{\eta_1(t)} e^{i \xi_1(t)}w_1(x,y) \nonumber \\
\sqrt{\eta_2(t)} e^{i \xi_2(t)}w_2(x,y) \nonumber \\
\sqrt{\eta_3(t)} e^{i \xi_3(t)}w_3(x,y) \nonumber \\
\sqrt{\eta_4(t)} e^{i \xi_4(t)}w_4(x,y) \nonumber \\
\end{pmatrix}
\end{eqnarray}
\normalsize
 that yields in no-dependence of those equations on position and gives

\begin{eqnarray}
i\hbar \frac{d}{dt}
\begin{pmatrix}
\sqrt{\eta_1(t)} e^{i \xi_1(t)} \nonumber \\
\sqrt{\eta_2(t)} e^{i \xi_2(t)} \nonumber \\
\sqrt{\eta_3(t)} e^{i \xi_3(t)} \nonumber \\
\sqrt{\eta_4(t)} e^{i \xi_4(t)} \nonumber \\
\end{pmatrix}
= \nonumber \\
\begin{pmatrix}
\alpha_1^2 (\beta_1^2 E_1 + \beta_2^2 E_2) +
 \alpha_2^2 (\beta_1^2 E_3 + \beta_2^2 E_4) &  \beta_1 \beta_2 (\alpha_1^2 (-E_1 + E_2) + \alpha_2^2 (-E_3 + E_4)) & \alpha_1 \alpha_2 (\beta_1^2 (-E_1 + E_3) + \beta_2^2 (-E_2 + E_4)) & \alpha_1 \alpha_2 \beta_1 \beta_2 (E_1 - E_2 - E_3 + E_4) \nonumber \\
\beta_1 \beta_2 (\alpha_1^2 (-E_1 + E_2) + \alpha_2^2 (-E_3 + E_4)) &  \alpha_1^2 (\beta_2^2 E_1 + \beta_1^2 E_2) +
 \alpha_2^2 (\beta_2^2 E_3 + \beta_1^2 E_4) & \alpha_1 \alpha_2 \beta_1 \beta_2 (E_1 - E_2 - E_3 + E_4) & \alpha_1 \alpha_2 (\beta_2^2 (-E_1 + E_3) + \beta_1^2 (-E_2 + E_4)) \nonumber \\
\alpha_1 \alpha_2 (\beta_1^2 (-E_1 + E_3) + \beta_2^2 (-E_2 + E_4)) &  \alpha_1 \alpha_2 \beta_1 \beta_2 (E_1 - E_2 - E_3 + E_4) & \alpha_2^2 (\beta_1^2 E_1 + \beta2^2 E_2) + \alpha_1^2 (\beta_1^2 E_3 + \beta_2^2 E_4 & \beta_1 \beta_2 (\alpha_2^2 (-E_1 + E_2) + \alpha_1^2 (-E_3 + E_4)) \nonumber \\
\alpha_1 \alpha_2 \beta_1 \beta_2 (E_1 - E_2 - E_3 + E_4) &  \alpha_1 \alpha_2 (\beta_2^2 (-E_1 + E_3) + \beta_1^2 (-E_2 + E_4)) & \beta_1 \beta_2 (\alpha_2^2 (-E_1 + E_2) + \alpha_1^2 (-E_3 + E_4)) & \alpha_2^2 (\beta_2^2 E_1 + \beta_1^2 E_2) +
 \alpha_1^2 (\beta2^2 E_3 + \beta_1^2 E_4) \nonumber \\
\end{pmatrix} \nonumber \\
\begin{pmatrix}
\sqrt{\eta_1(t)} e^{i \xi_1(t)} \nonumber \\
\sqrt{\eta_2(t)} e^{i \xi_2(t)} \nonumber \\
\sqrt{\eta_3(t)} e^{i \xi_3(t)} \nonumber \\
\sqrt{\eta_4(t)} e^{i \xi_4(t)} \nonumber \\
\end{pmatrix}
\end{eqnarray}
 \normalsize
 and hence
 \normalsize
\begin{eqnarray}
\begin{pmatrix}
\sqrt{\eta_1(t)} e^{i \xi_1(t)} \nonumber \\
\sqrt{\eta_2(t)} e^{i \xi_2(t)} \nonumber \\
\sqrt{\eta_3(t)} e^{i \xi_3(t)} \nonumber \\
\sqrt{\eta_4(t)} e^{i \xi_4(t)} \nonumber \\
\end{pmatrix}
=
\nonumber \\
e^{\Bigg[
\frac{-i}{\hbar}
\begin{pmatrix}
\alpha_1^2 (\beta_1^2 E_1 + \beta_2^2 E_2) +
 \alpha_2^2 (\beta_1^2 E_3 + \beta_2^2 E_4) &  \beta_1 \beta_2 (\alpha_1^2 (-E_1 + E_2) + \alpha_2^2 (-E_3 + E_4)) & \alpha_1 \alpha_2 (\beta_1^2 (-E_1 + E_3) + \beta_2^2 (-E_2 + E_4)) & \alpha_1 \alpha_2 \beta_1 \beta_2 (E_1 - E_2 - E_3 + E_4) \nonumber \\
\beta_1 \beta_2 (\alpha_1^2 (-E_1 + E_2) + \alpha_2^2 (-E_3 + E_4)) &  \alpha_1^2 (\beta_2^2 E_1 + \beta_1^2 E_2) +
 \alpha_2^2 (\beta_2^2 E_3 + \beta_1^2 E_4) & \alpha_1 \alpha_2 \beta_1 \beta_2 (E_1 - E_2 - E_3 + E_4) & \alpha_1 \alpha_2 (\beta_2^2 (-E_1 + E_3) + \beta_1^2 (-E_2 + E_4)) \nonumber \\
\alpha_1 \alpha_2 (\beta_1^2 (-E_1 + E_3) + \beta_2^2 (-E_2 + E_4)) &  \alpha_1 \alpha_2 \beta_1 \beta_2 (E_1 - E_2 - E_3 + E_4) & \alpha_2^2 (\beta_1^2 E_1 + \beta2^2 E_2) + \alpha_1^2 (\beta_1^2 E_3 + \beta_2^2 E_4) & \beta_1 \beta_2 (\alpha_2^2 (-E_1 + E_2) + \alpha_1^2 (-E_3 + E_4)) \nonumber \\
\alpha_1 \alpha_2 \beta_1 \beta_2 (E_1 - E_2 - E_3 + E_4) &  \alpha_1 \alpha_2 (\beta_2^2 (-E_1 + E_3) + \beta_1^2 (-E_2 + E_4)) & \beta_1 \beta_2 (\alpha_2^2 (-E_1 + E_2) + \alpha_1^2 (-E_3 + E_4)) & \alpha_2^2 (\beta_2^2 E_1 + \beta_1^2 E_2) +
 \alpha_1^2 (\beta2^2 E_3 + \beta_1^2 E_4) \nonumber \\
\end{pmatrix}
(t-t_0) \Bigg] } \nonumber \\
\times
\begin{pmatrix}
\sqrt{\eta_1(t_0)} e^{i \xi_1(t_0)} \nonumber \\
\sqrt{\eta_2(t_0)} e^{i \xi_2(t_0)} \nonumber \\
\sqrt{\eta_3(t_0)} e^{i \xi_3(t_0)} \nonumber \\
\sqrt{\eta_4(t_0)} e^{i \xi_4(t_0)} \nonumber \\
\end{pmatrix}.
\end{eqnarray}
 \normalsize
 These equation of motion is particularly interesting if eigenergies $E_1$, .. $E_4$ have complex values, so dissipation in quantum system is allowed.
 Now we come back into equations of motion and we have

\begin{eqnarray}
\frac{d}{dt} \Bigg[
\begin{pmatrix}
\frac{1}{\sqrt{\eta_1(t)}} & 0 & 0 & 0 \\
0 & \frac{1}{\sqrt{\eta_2(t)}} & 0 & 0 \\
0 & 0 & \frac{1}{\sqrt{\eta_3(t)}} & 0 \\
0 & 0 & 0 & \frac{1}{\sqrt{\eta_4(t)}} \\
\end{pmatrix}
\begin{pmatrix}
\eta_1(t) e^{i \xi_1(t)} \nonumber \\
\eta_2(t) e^{i \xi_2(t)} \nonumber \\
\eta_3(t) e^{i \xi_3(t)} \nonumber \\
\eta_4(t) e^{i \xi_4(t)} \nonumber \\
\end{pmatrix}
\Bigg]
=
\begin{pmatrix}
\frac{1}{\sqrt{\eta_1(t)}} & 0 & 0 & 0 \\
0 & \frac{1}{\sqrt{\eta_2(t)}} & 0 & 0 \\
0 & 0 & \frac{1}{\sqrt{\eta_3(t)}} & 0 \\
0 & 0 & 0 & \frac{1}{\sqrt{\eta_4(t)}} \\
\end{pmatrix}
\frac{d}{dt}
\begin{pmatrix}
\eta_1(t) e^{i \xi_1(t)} \nonumber \\
\eta_2(t) e^{i \xi_2(t)} \nonumber \\
\eta_3(t) e^{i \xi_3(t)} \nonumber \\
\eta_4(t) e^{i \xi_4(t)} \nonumber \\
\end{pmatrix} + \nonumber \\
+
\begin{pmatrix}
-\frac{\frac{d}{dt}(\eta_1(t))}{2(\sqrt{\eta_1(t)})^2} & 0 & 0 & 0 \\
0 & -\frac{\frac{d}{dt}(\eta_2(t))}{2(\sqrt{\eta_2(t)})^2} & 0 & 0 \\
0 & 0 & -\frac{\frac{d}{dt}(\eta_3(t))}{2(\sqrt{\eta_3(t)})^2} & 0 \\
0 & 0 & 0 & -\frac{\frac{d}{dt}(\eta_4(t))}{2(\sqrt{\eta_4(t)})^2} \\
\end{pmatrix}
\begin{pmatrix}
\eta_1(t) e^{i \xi_1(t)} \nonumber \\
\eta_2(t) e^{i \xi_2(t)} \nonumber \\
\eta_3(t) e^{i \xi_3(t)} \nonumber \\
\eta_4(t) e^{i \xi_4(t)} \nonumber \\
\end{pmatrix}
=\nonumber \\
\Bigg[
-\frac{i}{\hbar}
\begin{pmatrix}
\alpha_1^2 (\beta_1^2 E_1 + \beta_2^2 E_2) +
 \alpha_2^2 (\beta_1^2 E_3 + \beta_2^2 E_4) &  \beta_1 \beta_2 (\alpha_1^2 (-E_1 + E_2) + \alpha_2^2 (-E_3 + E_4)) & \alpha_1 \alpha_2 (\beta_1^2 (-E_1 + E_3) + \beta_2^2 (-E_2 + E_4)) & \alpha_1 \alpha_2 \beta_1 \beta_2 (E_1 - E_2 - E_3 + E_4) \nonumber \\
\beta_1 \beta_2 (\alpha_1^2 (-E_1 + E_2) + \alpha_2^2 (-E_3 + E_4)) &  \alpha_1^2 (\beta_2^2 E_1 + \beta_1^2 E_2) +
 \alpha_2^2 (\beta_2^2 E_3 + \beta_1^2 E_4) & \alpha_1 \alpha_2 \beta_1 \beta_2 (E_1 - E_2 - E_3 + E_4) & \alpha_1 \alpha_2 (\beta_2^2 (-E_1 + E_3) + \beta_1^2 (-E_2 + E_4)) \nonumber \\
\alpha_1 \alpha_2 (\beta_1^2 (-E_1 + E_3) + \beta_2^2 (-E_2 + E_4)) &  \alpha_1 \alpha_2 \beta_1 \beta_2 (E_1 - E_2 - E_3 + E_4) & \alpha_2^2 (\beta_1^2 E_1 + \beta2^2 E_2) + \alpha_1^2 (\beta_1^2 E_3 + \beta_2^2 E_4 & \beta_1 \beta_2 (\alpha_2^2 (-E_1 + E_2) + \alpha_1^2 (-E_3 + E_4)) \nonumber \\
\alpha_1 \alpha_2 \beta_1 \beta_2 (E_1 - E_2 - E_3 + E_4) &  \alpha_1 \alpha_2 (\beta_2^2 (-E_1 + E_3) + \beta_1^2 (-E_2 + E_4)) & \beta_1 \beta_2 (\alpha_2^2 (-E_1 + E_2) + \alpha_1^2 (-E_3 + E_4)) & \alpha_2^2 (\beta_2^2 E_1 + \beta_1^2 E_2) +
 \alpha_1^2 (\beta2^2 E_3 + \beta_1^2 E_4) \nonumber \\
\end{pmatrix} \times \nonumber \\
\times
\begin{pmatrix}
\frac{1}{\sqrt{\eta_1(t)}} & 0 & 0 & 0 \\
0 & \frac{1}{\sqrt{\eta_2(t)}} & 0 & 0 \\
0 & 0 & \frac{1}{\sqrt{\eta_3(t)}} & 0 \\
0 & 0 & 0 & \frac{1}{\sqrt{\eta_4(t)}} \\
\end{pmatrix}
\Bigg]
\begin{pmatrix}
\eta_1(t) e^{i \xi_1(t)} \nonumber \\
\eta_2(t) e^{i \xi_2(t)} \nonumber \\
\eta_3(t) e^{i \xi_3(t)} \nonumber \\
\eta_4(t) e^{i \xi_4(t)} \nonumber \\
\end{pmatrix}
\end{eqnarray}
 \normalsize
 and we obtain

\begin{eqnarray}
\frac{d}{dt}
\begin{pmatrix}
\eta_1(t) e^{i \xi_1(t)} \nonumber \\
\eta_2(t) e^{i \xi_2(t)} \nonumber \\
\eta_3(t) e^{i \xi_3(t)} \nonumber \\
\eta_4(t) e^{i \xi_4(t)} \nonumber \\
\end{pmatrix}
=
\begin{pmatrix}
\frac{\frac{d}{dt}(\eta_1(t))}{2(\sqrt{\eta_1(t)})} & 0 & 0 & 0 \\
0 & \frac{\frac{d}{dt}(\eta_2(t))}{2(\sqrt{\eta_2(t)})} & 0 & 0 \\
0 & 0 & \frac{\frac{d}{dt}(\eta_3(t))}{2(\sqrt{\eta_3(t)})} & 0 \\
0 & 0 & 0 & \frac{\frac{d}{dt}(\eta_4(t))}{2(\sqrt{\eta_4(t)})} \\
\end{pmatrix}
+
(-\frac{i}{\hbar})
\begin{pmatrix}
H_{w11}  & H_{w12}\frac{\sqrt{\eta_1(t)}}{\sqrt{\eta_2(t)}} & H_{w13}\frac{\sqrt{\eta_1(t)}}{\sqrt{\eta_3(t)}} & H_{w14}\frac{\sqrt{\eta_1(t)}}{\sqrt{\eta_4(t)}} \\
H_{w21}\frac{\sqrt{\eta_2(t)}}{\sqrt{\eta_1(t)}} & H_{w22} & H_{w23}\frac{\sqrt{\eta_2(t)}}{\sqrt{\eta_3(t)}} & H_{w24}\frac{\sqrt{\eta_2(t)}}{\sqrt{\eta_4(t)}} \\
H_{w31}\frac{\sqrt{\eta_3(t)}}{\sqrt{\eta_1(t)}} & H_{w32}\frac{\sqrt{\eta_3(t)}}{\sqrt{\eta_2(t)}} & H_{w33} & H_{w34}\frac{\sqrt{\eta_3(t)}}{\sqrt{\eta_4(t)}} \\    
H_{w41}\frac{\sqrt{\eta_4(t)}}{\sqrt{\eta_1(t)}} & H_{w42}\frac{\sqrt{\eta_4(t)}}{\sqrt{\eta_2(t)}} & H_{w43}\frac{\sqrt{\eta_4(t)}}{\sqrt{\eta_3(t)}} & H_{w44} \\
\end{pmatrix}
\begin{pmatrix}
\eta_1(t) e^{i \xi_1(t)} \nonumber \\
\eta_2(t) e^{i \xi_2(t)} \nonumber \\
\eta_3(t) e^{i \xi_3(t)} \nonumber \\
\eta_4(t) e^{i \xi_4(t)} \nonumber \\
\end{pmatrix} .
\end{eqnarray}
\normalsize
Next step allows for identification of 8 probabilities embedded among 4 dimensional state vector in the way as given below

\normalsize

\begin{eqnarray}
\frac{d}{dt}
\begin{pmatrix}
\eta_1(t) e^{i \xi_1(t)}=\eta_1(t) cos(\xi_1(t))+ i\eta_1(t) sin(\xi_1(t))=P_{1re}(t)+iP_{1im}(t) \nonumber \\
\eta_2(t) e^{i \xi_2(t)}=\eta_2(t) cos(\xi_2(t))+ i\eta_2(t) sin(\xi_2(t))=P_{2re}(t)+iP_{2im}(t) \nonumber \\
\eta_3(t) e^{i \xi_3(t)}=\eta_3(t) cos(\xi_3(t))+ i\eta_3(t) sin(\xi_3(t))=P_{3re}(t)+iP_{3im}(t) \nonumber \\
\eta_4(t) e^{i \xi_4(t)}=\eta_4(t) cos(\xi_4(t))+ i\eta_4(t) sin(\xi_4(t))=P_{4re}(t)+iP_{4im}(t) \nonumber \\
\end{pmatrix}
=
(-\frac{i}{\hbar})
\begin{pmatrix}
H_{w11}+\hbar i \frac{\frac{d}{dt}(\eta_1(t))}{2(\sqrt{\eta_1(t)})}  & H_{w12}\frac{\sqrt{\eta_1(t)}}{\sqrt{\eta_2(t)}} & H_{w13}\frac{\sqrt{\eta_1(t)}}{\sqrt{\eta_3(t)}} & H_{w14}\frac{\sqrt{\eta_1(t)}}{\sqrt{\eta_4(t)}} \\
H_{w21}\frac{\sqrt{\eta_2(t)}}{\sqrt{\eta_1(t)}} & H_{w22}+\hbar i \frac{\frac{d}{dt}(\eta_2(t))}{2(\sqrt{\eta_2(t)})} & H_{w23}\frac{\sqrt{\eta_2(t)}}{\sqrt{\eta_3(t)}} & H_{w24}\frac{\sqrt{\eta_2(t)}}{\sqrt{\eta_4(t)}} \\
H_{w31}\frac{\sqrt{\eta_3(t)}}{\sqrt{\eta_1(t)}} & H_{w32}\frac{\sqrt{\eta_3(t)}}{\sqrt{\eta_2(t)}} & H_{w33}+\hbar i\frac{\frac{d}{dt}(\eta_3(t))}{2(\sqrt{\eta_3(t)})} & H_{w34}\frac{\sqrt{\eta_3(t)}}{\sqrt{\eta_4(t)}} \\    
H_{w41}\frac{\sqrt{\eta_4(t)}}{\sqrt{\eta_1(t)}} & H_{w42}\frac{\sqrt{\eta_4(t)}}{\sqrt{\eta_2(t)}} & H_{w43}\frac{\sqrt{\eta_4(t)}}{\sqrt{\eta_3(t)}} & H_{w44}+\hbar i \frac{\frac{d}{dt}(\eta_4(t))}{2(\sqrt{\eta_4(t)})} \\
\end{pmatrix}
\nonumber \\
\times
\begin{pmatrix}
\eta_1(t) e^{i \xi_1(t)}=\eta_1(t) cos(\xi_1(t))+ i\eta_1(t) sin(\xi_1(t))=P_{1re}(t)+iP_{1im}(t) , \nonumber \\
\eta_2(t) e^{i \xi_2(t)}=\eta_2(t) cos(\xi_2(t))+ i\eta_2(t) sin(\xi_2(t))=P_{2re}(t)+iP_{2im}(t), \nonumber \\
\eta_3(t) e^{i \xi_3(t)}=\eta_3(t) cos(\xi_3(t))+ i\eta_3(t) sin(\xi_3(t))=P_{3re}(t)+iP_{3im}(t), \nonumber \\
\eta_4(t) e^{i \xi_4(t)}=\eta_4(t) cos(\xi_4(t))+ i\eta_4(t) sin(\xi_4(t))=P_{4re}(t)+iP_{4im}(t), \nonumber \\
\end{pmatrix},
\end{eqnarray}
\normalsize
We can redefine the last equation system and equivalently obtain vector of quasi-probabilities as given by $P_{\textbf{k}re}(t)=\eta_{\textbf{k}}(t) cos(\xi_{\textbf{k}}(t))$, $P_{\textbf{k}im}(t)=\eta_{\textbf{k}}(t) sin(\xi_{\textbf{k}}(t))$ for $k=1..4$ so, it leads to the following equations of motion.
\begin{eqnarray}
\frac{d}{dt}
\begin{pmatrix}
P_{1re}(t)  \nonumber \\
P_{1im}(t)  \nonumber \\
P_{2re}(t)  \nonumber \\
P_{2im}(t)  \nonumber \\
P_{3re}(t)  \nonumber \\
P_{3im}(t)  \nonumber \\
P_{4re}(t)  \nonumber \\
P_{4im}(t)
\end{pmatrix}
=
\begin{pmatrix}
\frac{\frac{d}{dt}(\eta_1(t))}{2(\sqrt{\eta_1(t)})} & \frac{1}{\hbar}H_{w11} & 0 & \frac{1}{\hbar}H_{w12}\frac{\sqrt{\eta_1(t)}}{\sqrt{\eta_2(t)}} & 0 & \frac{1}{\hbar}H_{w13}\frac{\sqrt{\eta_1(t)}}{\sqrt{\eta_3(t)}} & 0 & \frac{1}{\hbar}H_{w14}\frac{\sqrt{\eta_1(t)}}{\sqrt{\eta_4(t)}} \\
-\frac{1}{\hbar}H_{w11} & -\frac{\frac{d}{dt}(\eta_1(t))}{2(\sqrt{\eta_1(t)})} & -\frac{1}{\hbar}H_{w12}\frac{\sqrt{\eta_1(t)}}{\sqrt{\eta_2(t)}} & 0 & -\frac{1}{\hbar}H_{w13}\frac{\sqrt{\eta_1(t)}}{\sqrt{\eta_3(t)}} & 0 & -\frac{1}{\hbar}H_{w14}\frac{\sqrt{\eta_1(t)}}{\sqrt{\eta_4(t)}} & 0 \\
0 & \frac{1}{\hbar}H_{w21}\frac{\sqrt{\eta_2(t)}}{\sqrt{\eta_1(t)}} & \frac{\frac{d}{dt}(\eta_2(t))}{2(\sqrt{\eta_2(t)})} & \frac{1}{\hbar}H_{w22} & 0 & \frac{1}{\hbar}H_{w23}\frac{\sqrt{\eta_2(t)}}{\sqrt{\eta_3(t)}} & 0 & \frac{1}{\hbar}H_{w24}\frac{\sqrt{\eta_2(t)}}{\sqrt{\eta_4(t)}} \\
-\frac{1}{\hbar}H_{w21}\frac{\sqrt{\eta_2(t)}}{\sqrt{\eta_1(t)}} & 0 & -\frac{1}{\hbar}H_{w22} & -\frac{\frac{d}{dt}(\eta_2(t))}{2(\sqrt{\eta_2(t)})} & -\frac{1}{\hbar}H_{w23}\frac{\sqrt{\eta_2(t)}}{\sqrt{\eta_3(t)}} & 0 & -\frac{1}{\hbar}H_{w24}\frac{\sqrt{\eta_2(t)}}{\sqrt{\eta_4(t)}} & 0 \\
0 & \frac{1}{\hbar}H_{w31}\frac{\sqrt{\eta_3(t)}}{\sqrt{\eta_1(t)}} & 0 & \frac{1}{\hbar}H_{w32}\frac{\sqrt{\eta_3(t)}}{\sqrt{\eta_2(t)}} & \frac{\frac{d}{dt}(\eta_3(t))}{2(\sqrt{\eta_3(t)})} & \frac{1}{\hbar}H_{w33} & 0 & \frac{1}{\hbar}H_{w34}\frac{\sqrt{\eta_3(t)}}{\sqrt{\eta_4(t)}} \\
-\frac{1}{\hbar}H_{w31}\frac{\sqrt{\eta_3(t)}}{\sqrt{\eta_1(t)}} & 0 & -\frac{1}{\hbar}H_{w32}\frac{\sqrt{\eta_3(t)}}{\sqrt{\eta_2(t)}} & 0 & -\frac{1}{\hbar}H_{w33} & -\frac{\frac{d}{dt}(\eta_3(t))}{2(\sqrt{\eta_3(t)})} & -\frac{1}{\hbar}H_{w34}\frac{\sqrt{\eta_3(t)}}{\sqrt{\eta_4(t)}} & 0 \\
0 & \frac{1}{\hbar}H_{w41}\frac{\sqrt{\eta_4(t)}}{\sqrt{\eta_1(t)}} & 0 & \frac{1}{\hbar}H_{w42}\frac{\sqrt{\eta_4(t)}}{\sqrt{\eta_2(t)}} & 0 & \frac{1}{\hbar}H_{w43}\frac{\sqrt{\eta_4(t)}}{\sqrt{\eta_3(t)}} & \frac{\frac{d}{dt}(\eta_4(t))}{2(\sqrt{\eta_4(t)})} & \frac{1}{\hbar}H_{w44} \\
-\frac{1}{\hbar}H_{w41}\frac{\sqrt{\eta_4(t)}}{\sqrt{\eta_1(t)}} & 0 & -\frac{1}{\hbar}H_{w42}\frac{\sqrt{\eta_4(t)}}{\sqrt{\eta_2(t)}} & 0 & -\frac{1}{\hbar}H_{w43}\frac{\sqrt{\eta_4(t)}}{\sqrt{\eta_3(t)}} & 0 & -\frac{1}{\hbar}H_{w44} & -\frac{\frac{d}{dt}(\eta_4(t))}{2(\sqrt{\eta_4(t)})} \\
\end{pmatrix}
\begin{pmatrix}
P_{1re}(t)  \nonumber \\
P_{1im}(t)  \nonumber \\
P_{2re}(t)  \nonumber \\
P_{2im}(t)  \nonumber \\
P_{3re}(t)  \nonumber \\
P_{3im}(t)  \nonumber \\
P_{4re}(t)  \nonumber \\
P_{4im}(t)  \\
\end{pmatrix}
\end{eqnarray}
Here we have used $\eta_{k}(t)=\sqrt{(P_{\textbf{k}re}(t))^2+(P_{\textbf{k}im}(t))^2}$ with $k=1..4$.
We can redefine $P_{kre}(t) \in (-1,+1)$ and $P_{kim}(t) \in (-1,+1)$, so $1+P_{kre}(t)=P_{krep}(t) \in (0,+1)$ and $1+P_{kim}(t)=P_{kimp}(t) \in (0,+1)$, so we have
\begin{eqnarray}
\frac{d}{dt}
\begin{pmatrix}
P_{1rep}(t)  \nonumber \\
P_{1imp}(t)  \nonumber \\
P_{2rep}(t)  \nonumber \\
P_{2imp}(t)  \nonumber \\
P_{3rep}(t)  \nonumber \\
P_{3imp}(t)  \nonumber \\
P_{4rep}(t)  \nonumber \\
P_{4imp}(t)  \nonumber \\
\end{pmatrix}
=
\begin{pmatrix}
\frac{\frac{d}{dt}(\eta_1(t))}{2(\sqrt{\eta_1(t)})} & \frac{1}{\hbar}H_{w11} & 0 & \frac{1}{\hbar}H_{w12}\frac{\sqrt{\eta_1(t)}}{\sqrt{\eta_2(t)}} & 0 & \frac{1}{\hbar}H_{w13}\frac{\sqrt{\eta_1(t)}}{\sqrt{\eta_3(t)}} & 0 & \frac{1}{\hbar}H_{w14}\frac{\sqrt{\eta_1(t)}}{\sqrt{\eta_4(t)}} \\
-\frac{1}{\hbar}H_{w11} & -\frac{\frac{d}{dt}(\eta_1(t))}{2(\sqrt{\eta_1(t)})} & -\frac{1}{\hbar}H_{w12}\frac{\sqrt{\eta_1(t)}}{\sqrt{\eta_2(t)}} & 0 & -\frac{1}{\hbar}H_{w13}\frac{\sqrt{\eta_1(t)}}{\sqrt{\eta_3(t)}} & 0 & -\frac{1}{\hbar}H_{w14}\frac{\sqrt{\eta_1(t)}}{\sqrt{\eta_4(t)}} & 0 \\
0 & \frac{1}{\hbar}H_{w21}\frac{\sqrt{\eta_2(t)}}{\sqrt{\eta_1(t)}} & \frac{\frac{d}{dt}(\eta_2(t))}{2(\sqrt{\eta_2(t)})} & \frac{1}{\hbar}H_{w22} & 0 & \frac{1}{\hbar}H_{w23}\frac{\sqrt{\eta_2(t)}}{\sqrt{\eta_3(t)}} & 0 & \frac{1}{\hbar}H_{w24}\frac{\sqrt{\eta_2(t)}}{\sqrt{\eta_4(t)}} \\
-\frac{1}{\hbar}H_{w21}\frac{\sqrt{\eta_2(t)}}{\sqrt{\eta_1(t)}} & 0 & -\frac{1}{\hbar}H_{w22} & -\frac{\frac{d}{dt}(\eta_2(t))}{2(\sqrt{\eta_2(t)})} & -\frac{1}{\hbar}H_{w23}\frac{\sqrt{\eta_2(t)}}{\sqrt{\eta_3(t)}} & 0 & -\frac{1}{\hbar}H_{w24}\frac{\sqrt{\eta_2(t)}}{\sqrt{\eta_4(t)}} & 0 \\
0 & \frac{1}{\hbar}H_{w31}\frac{\sqrt{\eta_3(t)}}{\sqrt{\eta_1(t)}} & 0 & \frac{1}{\hbar}H_{w32}\frac{\sqrt{\eta_3(t)}}{\sqrt{\eta_2(t)}} & \frac{\frac{d}{dt}(\eta_3(t))}{2(\sqrt{\eta_3(t)})} & \frac{1}{\hbar}H_{w33} & 0 & \frac{1}{\hbar}H_{w34}\frac{\sqrt{\eta_3(t)}}{\sqrt{\eta_4(t)}} \\
-\frac{1}{\hbar}H_{w31}\frac{\sqrt{\eta_3(t)}}{\sqrt{\eta_1(t)}} & 0 & -\frac{1}{\hbar}H_{w32}\frac{\sqrt{\eta_3(t)}}{\sqrt{\eta_2(t)}} & 0 & -\frac{1}{\hbar}H_{w33} & -\frac{\frac{d}{dt}(\eta_3(t))}{2(\sqrt{\eta_3(t)})} & -\frac{1}{\hbar}H_{w34}\frac{\sqrt{\eta_3(t)}}{\sqrt{\eta_4(t)}} & 0 \\
0 & \frac{1}{\hbar}H_{w41}\frac{\sqrt{\eta_4(t)}}{\sqrt{\eta_1(t)}} & 0 & \frac{1}{\hbar}H_{w42}\frac{\sqrt{\eta_4(t)}}{\sqrt{\eta_2(t)}} & 0 & \frac{1}{\hbar}H_{w43}\frac{\sqrt{\eta_4(t)}}{\sqrt{\eta_3(t)}} & \frac{\frac{d}{dt}(\eta_4(t))}{2(\sqrt{\eta_4(t)})} & \frac{1}{\hbar}H_{w44} \\
-\frac{1}{\hbar}H_{w41}\frac{\sqrt{\eta_4(t)}}{\sqrt{\eta_1(t)}} & 0 & -\frac{1}{\hbar}H_{w42}\frac{\sqrt{\eta_4(t)}}{\sqrt{\eta_2(t)}} & 0 & -\frac{1}{\hbar}H_{w43}\frac{\sqrt{\eta_4(t)}}{\sqrt{\eta_3(t)}} & 0 & -\frac{1}{\hbar}H_{w44} & -\frac{\frac{d}{dt}(\eta_4(t))}{2(\sqrt{\eta_4(t)})} \\
\end{pmatrix}
\begin{pmatrix}
P_{1rep}(t)-1  \nonumber \\
P_{1imp}(t)-1  \nonumber \\
P_{2rep}(t)-1  \nonumber \\
P_{2imp}(t)-1  \nonumber \\
P_{3rep}(t)-1  \nonumber \\
P_{3imp}(t)-1  \nonumber \\
P_{4rep}(t)-1  \nonumber \\
P_{4imp}(t)-1  \nonumber \\
\end{pmatrix}
\end{eqnarray}
Here we have used $\eta_{k}(t)=\sqrt{(P_{\textbf{k}rep}(t)-1)^2+(P_{\textbf{k}imp}(t)-1)^2}$ with $k=1..4$.
We recognize that the equation for stochastic Finite State Machine has the generalized form of epidemic model equation
\begin{eqnarray}
\frac{d}{dt}\vec{P}(t)=\hat{M}_t\vec{P}(t)+\vec{b}([\vec{P}]).
\end{eqnarray}
However here $ P_{\textbf{k}rep}(t)$ , $P_{\textbf{k}imp}(t)$ are positive-value and are not normalized, so one needs to renormalize it by
assigning the probability by $\frac{P_{\textbf{k}rep}(t)}{P_{\textbf{1}rep}(t)+P_{\textbf{1}imp}(t)+P_{\textbf{2}rep}(t)+P_{\textbf{2}imp}(t)+P_{\textbf{3}rep}(t)+P_{\textbf{3}imp}(t)+P_{\textbf{4}rep}(t)+P_{\textbf{4}imp}(t)+P_{\textbf{5}rep}(t)+P_{\textbf{5}imp}(t)+P_{\textbf{6}rep}(t)+P_{\textbf{6}imp}(t)+P_{\textbf{7}rep}(t)+P_{\textbf{7}imp}(t)+P_{\textbf{8}rep}(t)+P_{\textbf{8}imp}(t)}$.
 \end{landscape}
 \begin{landscape}
 Furthermore one can obtain alternative derivation of finite stochastic machine model from tight-binding model. We have
 \small
 \begin{eqnarray}
i\hbar \frac{d}{dt}
\begin{pmatrix}
\sqrt{\eta_1(t)} e^{i \xi_1(t)}, \nonumber \\
\sqrt{\eta_2(t)} e^{i \xi_2(t)}, \nonumber \\
\sqrt{\eta_3(t)} e^{i \xi_3(t)}, \nonumber \\
\sqrt{\eta_4(t)} e^{i \xi_4(t)}, \nonumber \\
\end{pmatrix}
=-i\hbar \frac{1}{2}
\begin{pmatrix}
\frac{(\frac{d}{dt}\eta_1(t))}{\sqrt{\eta_1(t)}} e^{i \xi_1(t)}, \nonumber \\
\frac{(\frac{d}{dt}\eta_2(t))}{\sqrt{\eta_2(t)}} e^{i \xi_2(t)}, \nonumber \\
\frac{(\frac{d}{dt}\eta_3(t))}{\sqrt{\eta_3(t)}} e^{i \xi_3(t)}, \nonumber \\
\frac{(\frac{d}{dt}\eta_4(t))}{\sqrt{\eta_4(t)}} e^{i \xi_4(t)}, \nonumber \\
\end{pmatrix}
+i\hbar
\begin{pmatrix}
i\sqrt{\eta_1(t)}e^{i \xi_1(t)}\frac{d}{dt}\xi_1(t), \nonumber \\
i\sqrt{\eta_2(t)}e^{i \xi_2(t)}\frac{d}{dt}\xi_2(t), \nonumber \\
i\sqrt{\eta_3(t)}e^{i \xi_3(t)}\frac{d}{dt}\xi_3(t), \nonumber \\
i\sqrt{\eta_4(t)}e^{i \xi_4(t)}\frac{d}{dt}\xi_4(t), \nonumber \\
\end{pmatrix}=
\nonumber \\
=
\begin{pmatrix}
\alpha_1^2 (\beta_1^2 E_1 + \beta_2^2 E_2) +
 \alpha_2^2 (\beta_1^2 E_3 + \beta_2^2 E_4) &  \beta_1 \beta_2 (\alpha_1^2 (-E_1 + E_2) + \alpha_2^2 (-E_3 + E_4)) & \alpha_1 \alpha_2 (\beta_1^2 (-E_1 + E_3) + \beta_2^2 (-E_2 + E_4)) & \alpha_1 \alpha_2 \beta_1 \beta_2 (E_1 - E_2 - E_3 + E_4) \nonumber \\
\beta_1 \beta_2 (\alpha_1^2 (-E_1 + E_2) + \alpha_2^2 (-E_3 + E_4)) &  \alpha_1^2 (\beta_2^2 E_1 + \beta_1^2 E_2) +
 \alpha_2^2 (\beta_2^2 E_3 + \beta_1^2 E_4) & \alpha_1 \alpha_2 \beta_1 \beta_2 (E_1 - E_2 - E_3 + E_4) & \alpha_1 \alpha_2 (\beta_2^2 (-E_1 + E_3) + \beta_1^2 (-E_2 + E_4)) \nonumber \\
\alpha_1 \alpha_2 (\beta_1^2 (-E_1 + E_3) + \beta_2^2 (-E_2 + E_4)) &  \alpha_1 \alpha_2 \beta_1 \beta_2 (E_1 - E_2 - E_3 + E_4) & \alpha_2^2 (\beta_1^2 E_1 + \beta2^2 E_2) + \alpha_1^2 (\beta_1^2 E_3 + \beta_2^2 E_4 & \beta_1 \beta_2 (\alpha_2^2 (-E_1 + E_2) + \alpha_1^2 (-E_3 + E_4)) \nonumber \\
\alpha_1 \alpha_2 \beta_1 \beta_2 (E_1 - E_2 - E_3 + E_4) &  \alpha_1 \alpha_2 (\beta_2^2 (-E_1 + E_3) + \beta_1^2 (-E_2 + E_4)) & \beta_1 \beta_2 (\alpha_2^2 (-E_1 + E_2) + \alpha_1^2 (-E_3 + E_4)) & \alpha_2^2 (\beta_2^2 E_1 + \beta_1^2 E_2) +
 \alpha_1^2 (\beta2^2 E_3 + \beta_1^2 E_4) \nonumber \\
\end{pmatrix} \times \nonumber \\
\times
\begin{pmatrix}
\frac{1}{\sqrt{\eta_1(t)}} e^{+i \xi_1(t)} & 0 & 0 & 0 \\
 0 & \frac{1}{\sqrt{\eta_2(t)}} e^{+i \xi_2(t)} & 0 & 0 \\
 0 & 0 & \frac{1}{\sqrt{\eta_3(t)}} e^{+i \xi_3(t)} & 0  \\
 0 & 0 & 0 & \frac{1}{\sqrt{\eta_4(t)}} e^{+i \xi_4(t)}   \\
\end{pmatrix}
\begin{pmatrix}
\sqrt{\eta_1(t)} e^{-i \xi_1(t)} & 0 & 0 & 0 \\
 0 & \sqrt{\eta_2(t)} e^{-i \xi_2(t)} & 0 & 0 \\
 0 & 0 & \sqrt{\eta_3(t)} e^{-i \xi_3(t)} & 0  \\
 0 & 0 & 0 & \sqrt{\eta_4(t)} e^{-i \xi_4(t)}   \\
\end{pmatrix}
\begin{pmatrix}
\sqrt{\eta_1(t)} e^{i \xi_1(t)}, \nonumber \\
\sqrt{\eta_2(t)} e^{i \xi_2(t)}, \nonumber \\
\sqrt{\eta_3(t)} e^{i \xi_3(t)}, \nonumber \\
\sqrt{\eta_4(t)} e^{i \xi_4(t)}, \nonumber \\
\end{pmatrix}
\end{eqnarray}
 \normalsize

 Further simplification gives
\small
 \begin{eqnarray}
\Bigg[ i\hbar \frac{1}{2}
\begin{pmatrix}
e^{i \xi_1(t)}\frac{1}{\sqrt{\eta_1(t)}} & 0 & 0 & 0 \\
0 & e^{i \xi_2(t)}\frac{1}{\sqrt{\eta_2(t)}} & 0 & 0 \\
0 & 0 & e^{i \xi_3(t)}\frac{1}{\sqrt{\eta_3(t)}} & 0 \\
0 & 0 & 0 & e^{i \xi_4(t)}\frac{1}{\sqrt{\eta_4(t)}} \\
\end{pmatrix}
\begin{pmatrix}
\frac{d}{dt}\eta_1(t) \nonumber \\
\frac{d}{dt}\eta_2(t) \nonumber \\
\frac{d}{dt}\eta_3(t) \nonumber \\
\frac{d}{dt}\eta_4(t) \nonumber \\
\end{pmatrix}
-\hbar
\begin{pmatrix}
e^{i \xi_1(t)}\frac{1}{\sqrt{\eta_1(t)}}\frac{d}{dt}\xi_1(t) & 0 & 0 & 0 \\
0 & e^{i \xi_2(t)}\frac{1}{\sqrt{\eta_2(t)}}\frac{d}{dt}\xi_2(t) & 0 & 0 \\
0 & 0 & e^{i \xi_3(t)}\frac{1}{\sqrt{\eta_3(t)}}\frac{d}{dt}\xi_3(t) & 0 \\
0 & 0 & 0 & e^{i \xi_4(t)}\frac{1}{\sqrt{\eta_4(t)}}\frac{d}{dt}\xi_4(t) \\
\end{pmatrix}
\begin{pmatrix}
 \eta_1(t) \nonumber \\
 \eta_2(t) \nonumber \\
 \eta_3(t) \nonumber \\
 \eta_4(t) \nonumber \\
\end{pmatrix}
\Bigg]
=
\nonumber \\
\Bigg[
\begin{pmatrix}
\alpha_1^2 (\beta_1^2 E_1 + \beta_2^2 E_2) +
 \alpha_2^2 (\beta_1^2 E_3 + \beta_2^2 E_4) &  \beta_1 \beta_2 (\alpha_1^2 (-E_1 + E_2) + \alpha_2^2 (-E_3 + E_4)) & \alpha_1 \alpha_2 (\beta_1^2 (-E_1 + E_3) + \beta_2^2 (-E_2 + E_4)) & \alpha_1 \alpha_2 \beta_1 \beta_2 (E_1 - E_2 - E_3 + E_4) \nonumber \\
\beta_1 \beta_2 (\alpha_1^2 (-E_1 + E_2) + \alpha_2^2 (-E_3 + E_4)) &  \alpha_1^2 (\beta_2^2 E_1 + \beta_1^2 E_2) +
 \alpha_2^2 (\beta_2^2 E_3 + \beta_1^2 E_4) & \alpha_1 \alpha_2 \beta_1 \beta_2 (E_1 - E_2 - E_3 + E_4) & \alpha_1 \alpha_2 (\beta_2^2 (-E_1 + E_3) + \beta_1^2 (-E_2 + E_4)) \nonumber \\
\alpha_1 \alpha_2 (\beta_1^2 (-E_1 + E_3) + \beta_2^2 (-E_2 + E_4)) &  \alpha_1 \alpha_2 \beta_1 \beta_2 (E_1 - E_2 - E_3 + E_4) & \alpha_2^2 (\beta_1^2 E_1 + \beta2^2 E_2) + \alpha_1^2 (\beta_1^2 E_3 + \beta_2^2 E_4 & \beta_1 \beta_2 (\alpha_2^2 (-E_1 + E_2) + \alpha_1^2 (-E_3 + E_4)) \nonumber \\
\alpha_1 \alpha_2 \beta_1 \beta_2 (E_1 - E_2 - E_3 + E_4) &  \alpha_1 \alpha_2 (\beta_2^2 (-E_1 + E_3) + \beta_1^2 (-E_2 + E_4)) & \beta_1 \beta_2 (\alpha_2^2 (-E_1 + E_2) + \alpha_1^2 (-E_3 + E_4)) & \alpha_2^2 (\beta_2^2 E_1 + \beta_1^2 E_2) +
 \alpha_1^2 (\beta2^2 E_3 + \beta_1^2 E_4) \nonumber \\
\end{pmatrix} \times \nonumber \\
\times
\begin{pmatrix}
\frac{1}{\sqrt{\eta_1(t)}} e^{+i \xi_1(t)} & 0 & 0 & 0 \\
 0 & \frac{1}{\sqrt{\eta_2(t)}} e^{+i \xi_2(t)} & 0 & 0 \\
 0 & 0 & \frac{1}{\sqrt{\eta_3(t)}} e^{+i \xi_3(t)} & 0  \\
 0 & 0 & 0 & \frac{1}{\sqrt{\eta_4(t)}} e^{+i \xi_4(t)}   \\
\end{pmatrix}
\Bigg]
\begin{pmatrix}
\eta_1(t)=P_1(t)  \nonumber \\
\eta_2(t)=P_2(t)  \nonumber \\
\eta_3(t)=P_3(t)  \nonumber \\
\eta_4(t)=P_4(t)  \nonumber \\
\end{pmatrix}.
\end{eqnarray}
 \normalsize

 and consequently we arrive to
 \tiny
 \begin{eqnarray}
\frac{d}{dt}
\begin{pmatrix}
\eta_1(t) \nonumber \\
\eta_2(t) \nonumber \\
\eta_3(t) \nonumber \\
\eta_4(t) \nonumber \\
\end{pmatrix}
=
-\frac{2i}{\hbar}
\begin{pmatrix}
[\alpha_1^2 (\beta_1^2 E_1 + \beta_2^2 E_2) +
 \alpha_2^2 (\beta_1^2 E_3 + \beta_2^2 E_4)+\hbar\frac{d}{dt}\xi_1(t)] &  \frac{\sqrt{\eta_1(t)}}{\sqrt{\eta_2(t)}} e^{+i (\xi_2(t)-\xi_1(t))}[\beta_1 \beta_2 (\alpha_1^2 (-E_1 + E_2) + \alpha_2^2 (-E_3 + E_4))] & \frac{\sqrt{\eta_1(t)}}{\sqrt{\eta_3(t)}} e^{+i (\xi_3(t)-\xi_1(t))}[\alpha_1 \alpha_2 (\beta_1^2 (-E_1 + E_3) + \beta_2^2 (-E_2 + E_4))] & \frac{\sqrt{\eta_1(t)}}{\sqrt{\eta_4(t)}} e^{+i (\xi_4(t)-\xi_1(t))}[\alpha_1 \alpha_2 \beta_1 \beta_2 (E_1 - E_2 - E_3 + E_4)] \nonumber \\
\frac{\sqrt{\eta_2(t)}}{\sqrt{\eta_1(t)}} e^{+i(\xi_1(t)-\xi_2(t))}[\beta_1 \beta_2 (\alpha_1^2 (-E_1 + E_2) + \alpha_2^2 (-E_3 + E_4))] &  [\alpha_1^2 (\beta_2^2 E_1 + \beta_1^2 E_2) +
 \alpha_2^2 (\beta_2^2 E_3 + \beta_1^2 E_4)+\hbar\frac{d}{dt}\xi_2(t)] & \frac{\sqrt{\eta_2(t)}}{\sqrt{\eta_3(t)}}e^{+i (\xi_3(t)-\xi_2(t))}[ \alpha_1 \alpha_2 \beta_1 \beta_2 (E_1 - E_2 - E_3 + E_4)] & \frac{\sqrt{\eta_2(t)}}{\sqrt{\eta_4(t)}} e^{+i (\xi_4(t)-\xi_2(t))}[\alpha_1 \alpha_2 (\beta_2^2 (-E_1 + E_3) + \beta_1^2 (-E_2 + E_4))] \nonumber \\
\frac{\sqrt{\eta_3(t)}}{\sqrt{\eta_1(t)}} e^{+i(\xi_1(t)-\xi_3(t))}[\alpha_1 \alpha_2 (\beta_1^2 (-E_1 + E_3) + \beta_2^2 (-E_2 + E_4))] &  \frac{\sqrt{\eta_3(t)}}{\sqrt{\eta_2(t)}} e^{+i(\xi_2(t)-\xi_3(t))}[\alpha_1 \alpha_2 \beta_1 \beta_2 (E_1 - E_2 - E_3 + E_4)] & [\alpha_2^2 (\beta_1^2 E_1 + \beta2^2 E_2) + \alpha_1^2 (\beta_1^2 E_3 + \beta_2^2 E_4+\hbar\frac{d}{dt}\xi_3(t)] & \frac{\sqrt{\eta_3(t)}}{\sqrt{\eta_4(t)}} e^{+i(\xi_4(t)-\xi_3(t))}[\beta_1 \beta_2 (\alpha_2^2 (-E_1 + E_2) + \alpha_1^2 (-E_3 + E_4))] \nonumber \\
\frac{\sqrt{\eta_4(t)}}{\sqrt{\eta_1(t)}} e^{+i (\xi_1(t)-\xi_4(t))}[\alpha_1 \alpha_2 \beta_1 \beta_2 (E_1 - E_2 - E_3 + E_4)] &  \frac{\sqrt{\eta_4(t)}}{\sqrt{\eta_2(t)}} e^{+i (\xi_2(t)-\xi_4(t))}[ \alpha_1 \alpha_2 (\beta_2^2 (-E_1 + E_3) + \beta_1^2 (-E_2 + E_4))] & \frac{\sqrt{\eta_4(t)}}{\sqrt{\eta_3(t)}} e^{+i (\xi_3(t)-\xi_4(t))}[\beta_1 \beta_2 (\alpha_2^2 (-E_1 + E_2) + \alpha_1^2 (-E_3 + E_4))] &
[\alpha_2^2 (\beta_2^2 E_1 + \beta_1^2 E_2) +
 \alpha_1^2 (\beta2^2 E_3 + \beta_1^2 E_4)+\hbar\frac{d}{dt}\xi_4(t)] \nonumber \\
\end{pmatrix} 
\begin{pmatrix}
\eta_1(t)  \nonumber \\
\eta_2(t)  \nonumber \\
\eta_3(t)  \nonumber \\
\eta_4(t)  \nonumber \\
\end{pmatrix} = \nonumber \\
\end{eqnarray}
\normalsize
\begin{eqnarray}
=(E_{val1}\ket{St1(t)}\bra{St1(t)}+E_{val2}\ket{St2(t)}\bra{St2(t)}+E_{val3}\ket{St3(t)}\bra{St3(t)}+E_{val4}\ket{St4(t)}\bra{St4(t)}) \nonumber \\
(p_{v1}e^{\int_{t_0}^{t}E_{val1}(t')dt'}\ket{St1(t)}+p_{v2}e^{\int_{t_0}^{t}E_{val2}(t')dt'}\ket{St2(t)}+p_{v3}e^{\int_{t_0}^{t}E_{val3}(t')dt'}\ket{St3(t)}+p_{v4}e^{\int_{t_0}^{t}E_{val4}(t')dt'}\ket{St4(t)})
= \nonumber \\
=(E_{val1}(t)p_{v1}[\eta_1(t_0),\eta_2(t_0),\eta_3(t_0),\eta_4(t_0)] ]e^{\int_{t_0}^{t}E_{val1}(t')dt'}\ket{St1(t)}+E_{val2}(t)p_{v2}[\eta_1(t_0),\eta_2(t_0),\eta_3(t_0),\eta_4(t_0)]e^{\int_{t_0}^{t}E_{val2}(t')dt'}\ket{St2(t)}+\nonumber \\ +E_{val3}(t)p_{v3}[\eta_1(t_0),\eta_2(t_0),\eta_3(t_0),\eta_4(t_0)]e^{\int_{t_0}^{t}E_{val3}(t')dt'}\ket{St3(t)}+E_{val4}(t)p_{v4}[\eta_1(t_0),\eta_2(t_0),\eta_3(t_0),\eta_4(t_0)]e^{\int_{t_0}^{t}E_{val4}(t')dt'}\ket{St4(t)}),
\end{eqnarray}
 where
\begin{eqnarray}
\ket{\psi(t)}_{class}=
\begin{pmatrix}
\eta_1(t)  \nonumber \\
\eta_2(t)  \nonumber \\
\eta_3(t)  \nonumber \\
\eta_4(t)  \nonumber \\
\end{pmatrix}=
(p_{v1}[\eta_1(t_0),\eta_2(t_0),\eta_3(t_0),\eta_4(t_0)]e^{\int_{t_0}^{t}E_{val1}(t')dt'}\ket{St1(t)}+p_{v2}[\eta_1(t_0),\eta_2(t_0),\eta_3(t_0),\eta_4(t_0)]e^{E_{val2}t}\ket{St2(t)}+ \nonumber \\
 +p_{v3}[\eta_1(t_0),\eta_2(t_0),\eta_3(t_0),\eta_4(t_0)]e^{\int_{t_0}^{t}E_{val3}(t')dt'}\ket{St3(t)}+p_{v4}[\eta_1(t_0),\eta_2(t_0),\eta_3(t_0),\eta_4(t_0)]e^{\int_{t_0}^{t}E_{val4}(t')dt'}\ket{St4(t)})
\end{eqnarray}
 and we have
\begin{eqnarray}
\bra{St1(t_0)}\ket{\psi(t_0)}_{class}=p_{v1}[\eta_1(t_0),\eta_2(t_0),\eta_3(t_0),\eta_4(t_0)],
\bra{St2(t_0)}\ket{\psi(t_0)}_{class}=p_{v2}[\eta_1(t_0),\eta_2(t_0),\eta_3(t_0),\eta_4(t_0)], \nonumber \\
\bra{St3(t_0)}\ket{\psi(t_0)}_{class}=p_{v3}[\eta_1(t_0),\eta_2(t_0),\eta_3(t_0),\eta_4(t_0)],
\bra{St4(t_0)}\ket{\psi(t_0)}_{class}=p_{v4}[\eta_1(t_0),\eta_2(t_0),\eta_3(t_0),\eta_4(t_0)], \nonumber \\
\end{eqnarray}

   \end{landscape}
We obtain explicit diagonal values of Hamiltonian in Wannier representation given as
\begin{eqnarray}
H_{w11}= \alpha_1^2 ( \beta_1^2 H_{11} + \beta_1 \beta_2 (H_{12} + H_{21}) + \beta_2^2 H_{22} ) + \alpha_1 \alpha_2 (\beta_1^2 (H_{1,3} + H_{3,1}) + \beta_1 \beta_2 (H_{1,4} + H_{2,3} + H_{3,2} + H_{4,1}) + \beta_2^2 (H_{2,4} + H_{4,2})) + \nonumber \\
 \alpha_2^2 (\beta_1^2 H_{3,3} + \beta_1 \beta_2 (H_{3,4} + H_{4,3}) + \beta_2^2 H_{4,4})
\end{eqnarray}

\begin{eqnarray}
H_{w22}=
\alpha_1^2 (\beta_2^2 H_{11} - \beta_1 \beta_2 (H_{12} + H_{21}) + \beta_1^2 H_{22}) +
 \alpha_1 \alpha_2 (\beta_2^2 (H_{13} + H_{31}) -
    \beta_1 \beta_2 (H_{14} + H_{23} + H_{32} + H_{41}) + \beta_1^2 (H_{24} + H_{42})) +  \nonumber \\
 \alpha_2^2 (\beta_2^2 H_{33} - \beta_1 \beta_2 (H_{34} + H_{43}) + \beta_1^2 H_{44})
\end{eqnarray}
\begin{eqnarray}
H_{w33}=\alpha2^2 (\beta_1^2 H_{11} + \beta_1 \beta_2 (H_{12} +
H21) + \beta_2^2 H_{22}) -
 \alpha_1 \alpha_2 (\beta_1^2 (H_{13} + H_{31}) + \nonumber \\
    \beta_1 \beta_2 (H_{14} + H_{23} + H_{32} + H_{41}) + \beta_2^2 (H_{24} + H_{42})) +
 \alpha_1^2 (\beta_1^2 H_{33} + \beta_1 \beta_2 (H_{34} + H_{43}) + \beta_2^2 H_{44})
\end{eqnarray}
\begin{eqnarray}
H_{w44}=\alpha2^2 (\beta_2^2 H_{11} - \beta_1 \beta_2 (H_{12} + H_{21}) + \beta_1^2 H_{22}) + \nonumber \\
 \alpha_1 \alpha_2 (-\beta_2^2 (H_{13} + H_{31}) +
    \beta_1 \beta_2 (H_{14} + H_{23} + H_{32} + H_{41}) - \beta_1^2 (H24 + H42)) +
 \alpha_1^2 (\beta_2^2 H_{33} - \beta_1 \beta_2 (H_{34} + H_{43}) + \beta_1^2 H_{44})
\end{eqnarray}
and values of diagonal given as
\begin{eqnarray}
H_{w12}= \alpha_1^2 ( \beta_1^2 H_{12} - \beta_2^2 H_{21} + \beta_1 \beta_2 (-H_{11} + H_{22})) +
 +\alpha_1 \alpha_2 (\beta_1^2 (H_{14} + H_{32}) - \beta_2^2 (H_{23} + H_{41}) + \nonumber \\
 \alpha_1 \alpha_2 (\beta_1^2 (H_{14} + H_{32}) - \beta_2^2 (H_{23} + H_{41}) +
    \beta_1 \beta_2 (-H_{13} + H_{24} - H_{31} + H_{42})) + \nonumber \\
 +\alpha_2^2 (-\beta_2^2 H_{34} + \beta_1^2 H_{43} + \beta_1 \beta2 (-H_{33} + H_{44}))
\end{eqnarray}

\begin{eqnarray}
H_{w21}=\alpha_1^2 (-\beta_2^2 H_{12} + \beta1^2 H_{21} + \beta_1 \beta_2 (-H_{11} + H_{22})) +
 \alpha_1 \alpha_2 (-\beta_2^2 (H_{14} + H_{32}) + \beta_1^2 (H_{23} + H_{41}) + \nonumber \\
    \beta_1 \beta_2 (-H_{13} + H_{24} - H_{31} + H_{42})) +
 \alpha_2^2 (-\beta_2^2 H_{34} + \beta_1^2 H_{43} + \beta_1 \beta_2 (-H_{33} + H_{44})).
\end{eqnarray}

Wannier transformation and its inverse for 4 energy level quantum system is expressed as tensor product of two matrices given as

\begin{eqnarray}
\hat{W}^{-1}=
\begin{pmatrix}
+\alpha_1 & -\alpha_2 \\
+\alpha_2 & +\alpha_1 \\
\end{pmatrix} \times
\begin{pmatrix}
+\beta_1 & -\beta_2 \\
+\beta_2 & +\beta_1 \\
\end{pmatrix}
= 
\begin{pmatrix}
+\alpha_1
\begin{pmatrix}
+\beta_1 & -\beta_2 \\
+\beta_2 & +\beta_1 \\
\end{pmatrix}  & -\alpha_2 \begin{pmatrix}
+\beta_1 & -\beta_2 \\
+\beta_2 & +\beta_1 \\
\end{pmatrix}  \\
+\alpha_2 \begin{pmatrix}
+\beta_1 & -\beta_2 \\
+\beta_2 & +\beta_1 \\
\end{pmatrix}  & +\alpha_1 \begin{pmatrix}
+\beta_1 & -\beta_2 \\
+\beta_2 & +\beta_1 \\
\end{pmatrix}  \\
\end{pmatrix} = \nonumber \\
\begin{pmatrix}
+\alpha_1\beta_1=+u_1 & -\alpha_1\beta_2=+u_2 & -\alpha_2\beta_1=+u_3 & +\alpha_2\beta_2=+u_4 \\
+\alpha_1\beta_2=-u_2 & +\alpha_1\beta_1=+u_1 & -\alpha_2\beta_2=-u_4 & -\alpha_2\beta_1=+u_3 \\
+\alpha_2\beta_1=-u_3 & -\alpha_2\beta_2=-u_4 & +\alpha_1\beta_1=+u_1 & -\alpha_1\beta_2=+u_2 \\
+\alpha_2\beta_2=+u_4 & +\alpha_2\beta_1=-u_3 & +\alpha_1\beta_2=-u_2 & +\alpha_1\beta_1=+u_1 \\
\end{pmatrix}=
\begin{pmatrix}
+u_1 & +u_2 & +u_3 & +u_4 \\
-u_2 & +u_1 & -u_4 & +u_3 \\
-u_3 & -u_4 & +u_1 & +u_2 \\
+u_4 & -u_3 & -u_2 & +u_1 \\
\end{pmatrix}, \nonumber \\
\hat{W}=
\begin{pmatrix}
+\alpha_1 & +\alpha_2 \\
-\alpha_2 & +\alpha_1 \\
\end{pmatrix} \times
\begin{pmatrix}
+\beta_1 & +\beta_2 \\
-\beta_2 & +\beta_1 \\
\end{pmatrix}
= 
\begin{pmatrix}
+\alpha_1
\begin{pmatrix}
+\beta_1 & +\beta_2 \\
-\beta_2 & +\beta_1 \\
\end{pmatrix}  & \alpha_2 \begin{pmatrix}
+\beta_1 & +\beta_2 \\
-\beta_2 & +\beta_1 \\
\end{pmatrix}  \\
-\alpha_2 \begin{pmatrix}
+\beta_1 & +\beta_2 \\
-\beta_2 & +\beta_1 \\
\end{pmatrix}  & +\alpha_1 \begin{pmatrix}
+\beta_1 & +\beta_2 \\
-\beta_2 & +\beta_1 \\
\end{pmatrix}  \\
\end{pmatrix} = \nonumber \\
\begin{pmatrix}
+\alpha_1\beta_1=+u_1 & +\alpha_1\beta_2=-u_2 & +\alpha_2\beta_1=-u_3 & +\alpha_2\beta_2=+u_4 \\
-\alpha_1\beta_2=+u_2 & +\alpha_1\beta_1=+u_1 & -\alpha_2\beta_2=-u_4 & +\alpha_2\beta_1=-u_3 \\
-\alpha_2\beta_1=+u_3 & -\alpha_2\beta_2=-u_4 & +\alpha_1\beta_1=+u_1 & +\alpha_1\beta_2=-u_2 \\
+\alpha_2\beta_2=+u_4 & -\alpha_2\beta_1=+u_3 & -\alpha_1\beta_2=-u_2 & +\alpha_1\beta_1=+u_1 \\
\end{pmatrix}=
\begin{pmatrix}
+u_1 & -u_2 & -u_3 & +u_4 \\
+u_2 & +u_1 & -u_4 & -u_3 \\
+u_3 & -u_4 & +u_1 & -u_2 \\
+u_4 & +u_3 & -u_2 & +u_1 \\
\end{pmatrix}, \nonumber \\
|\alpha_1|^2+|\alpha_2|^2=1, |\beta_1|^2+|\beta_2|^2=1,
|u_1|^2+|u_2|^2+|u_3|^2+|u_4|^2=1 \nonumber \\
\end{eqnarray}
The condition $|u_1|^2+|u_2|^2+|u_3|^2+|u_4|^2=1$ is equivalent to unit sphere in 4-dimensional Euclidean space.
Such reasoning for determination of Wannier transformation can be conduced for quantum system with $2^n$ energies.
For example one can obtain Wannier transformation for $2^3=8$ energy levels with predefined structure given as
\begin{eqnarray}
\hat{W}_{8}=
\begin{pmatrix}
+\alpha_1 & +\alpha_2 \\
-\alpha_2 & +\alpha_1 \\
\end{pmatrix} \times
\begin{pmatrix}
+\beta_1 & +\beta_2 \\
-\beta_2 & +\beta_1 \\
\end{pmatrix} \times
\begin{pmatrix}
+\gamma_1 & +\gamma_2 \\
-\gamma_2 & +\gamma_1 \\
\end{pmatrix}, \nonumber \\
\hat{W}_{8}^{-1}=
\begin{pmatrix}
+\alpha_1 & -\alpha_2 \\
+\alpha_2 & +\alpha_1 \\
\end{pmatrix} \times
\begin{pmatrix}
+\beta_1 & -\beta_2 \\
+\beta_2 & +\beta_1 \\
\end{pmatrix} \times
\begin{pmatrix}
+\gamma_1 & -\gamma_2 \\
+\gamma_2 & +\gamma_1 \\
\end{pmatrix}, \nonumber \\
|\alpha_1|^2+|\alpha_2|^2=1, |\beta_1|^2+|\beta_2|^2=1,
|\gamma_1|^2+|\gamma_2|^2=1,
\end{eqnarray}
Having quantum system with 4 eigenergies we are given normalization condition for its eigenergy wavefunctions that is formally expressed as
\begin{eqnarray}
 \int_{-\infty}^{+\infty}dx\int_{-\infty}^{+\infty}dy|\psi_{E1}(x,y)|^2 = 1, 
 \int_{-\infty}^{+\infty}dx\int_{-\infty}^{+\infty}dy|\psi_{E2}(x,y)|^2 = 1, \nonumber  \\
 \int_{-\infty}^{+\infty}dx\int_{-\infty}^{+\infty}dy|\psi_{E3}(x,y)|^2 = 1, 
 \int_{-\infty}^{+\infty}dx\int_{-\infty}^{+\infty}dy|\psi_{E4}(x,y)|^2 = 1.
\end{eqnarray}
what implies that 4 unique orthonormal Wannier functions referring to Fig.\ref{2dimBox}. can be identified as linear transformation of eigenergy wavefunctions, so
\begin{eqnarray}
w_{Upper,L}(x,y)=\alpha_1 \psi_{E1}(x,y)+\alpha_2 \psi_{E2}(x,y)+\alpha_3 \psi_{E3}(x,y)+\alpha_4 \psi_{E4}(x,y), \nonumber \\
w_{Upper,R}(x,y)=\beta_1 \psi_{E1}(x,y)+\beta_2 \psi_{E2}(x,y)+\beta_3 \psi_{E3}(x,y)+\beta_4 \psi_{E4}(x,y), \nonumber \\
w_{Lower,L}(x,y)=\gamma_1 \psi_{E1}(x,y)+\gamma_2 \psi_{E2}(x,y)+\gamma_3 \psi_{E3}(x,y)+\gamma_4 \psi_{E4}(x,y), \nonumber \\
w_{Lower,R}(x,y)=\eta_1 \psi_{E1}(x,y)+\eta_2 \psi_{E2}(x,y)+\eta_3 \psi_{E3}(x,y)+\eta_4 \psi_{E4}(x,y). \nonumber \\
\end{eqnarray}
In bra-ket notation intoduced by Dirac \cite{Dirac} we one can write $\ket{\vec{w}}=\hat{W}\ket{\vec{E_{k}}}$
  and we have 4 possible solutions for the system with 4 eigenergies

\begin{eqnarray}
w_{1}(x,y)=w_{Upper,Left}(x,y)= +\alpha_1\beta_1  \psi_{E1}(x,y) + \alpha_1\beta_2  \psi_{E2}(x,y)+\alpha_2\beta_1  \psi_{E3}(x,y)+\alpha_2\beta_2 \psi_{E4}(x,y), \nonumber \\
w_{2}(x,y)=w_{Upper,Right}(x,y)= -\alpha_1\beta_2  \psi_{E1}(x,y) + \alpha_1\beta_1 \psi_{E2}(x,y) -\alpha_2\beta_2 \psi_{E3}(x,y)+ \alpha_2 \beta_1  \psi_{E4}(x,y), \nonumber \\
w_{3}(x,y)=w_{Lower,Left}(x,y)=-\alpha_2\beta_1 \psi_{E1}(x,y)-\alpha_2\beta_2\psi_{E2}(x,y)+\alpha_1\beta_1 \psi_{E3}(x,y)+\alpha_1\beta_2 \psi_{E4}(x,y), \nonumber \\
w_{4}(x,y)=w_{Lower,Right}(x,y)=+\alpha_2\beta_2 \psi_{E1}(x,y)-\alpha_2\beta_1 \psi_{E2}(x,y)-\alpha_1\beta_2 \psi_{E3}(x,y)+\alpha_1\beta_1 \psi_{E4}(x,y), \nonumber \\
\end{eqnarray}
Alternatively we can write inverse Wannier transform in the form as $\hat{W}^{-1}\ket{\vec{w}}=\ket{\vec{E_{k}}}$
\begin{eqnarray}
\bra{x,y}
\hat{W}^{-1}\ket{\vec{w}}=
\Bigg[
\begin{pmatrix}
+\alpha_1 & -\alpha_2 \\
+\alpha_2 & +\alpha_1 \\
\end{pmatrix} \times
\begin{pmatrix}
+\beta_1 & -\beta_2 \\
+\beta_2 & +\beta_1 \\
\end{pmatrix}
\Bigg]
\begin{pmatrix}
\psi_{w1}(x,y) \nonumber \\
\psi_{w2}(x,y) \nonumber \\
\psi_{w3}(x,y) \nonumber \\
\psi_{w4}(x,y) \nonumber \\
\end{pmatrix}
=
\begin{pmatrix}
\psi_{E1}(x,y) \nonumber \\
\psi_{E2}(x,y) \nonumber \\
\psi_{E3}(x,y) \nonumber \\
\psi_{E4}(x,y) \nonumber \\
\end{pmatrix}=\bra{x,y} \ket{\vec{E_{k}}} .
\end{eqnarray}

Basing on normalization condition for Wannier functions we can set $\alpha_1=cos(\Theta_1)$,$\alpha_2=sin(\Theta_1)$, $\beta_1=cos(\Theta_2)$,$\beta_2=sin(\Theta_2)$.
Now we need to determine coefficients in Wannier transformation as by usage of fact that Wannier functions are maximum localized in certain geometrical regions what is formally expressed by the conditions
\begin{eqnarray}
\int_{-\infty}^{0}dx\int_{0}^{\infty}dy|w_{1}(x,y)|^2dxdy=f_1(\Theta_1,\Theta_2), \frac{d}{d\Theta_1}f_1(\Theta_1,\Theta_2)=0, \frac{d}{d\Theta_2}f_1(\Theta_1,\Theta_2)=0, \nonumber \\
\int_{0}^{+\infty}dx\int_{0}^{\infty}dy|w_{2}(x,y)|^2dxdy=f_2(\Theta_1,\Theta_2),  \frac{d}{d\Theta_1}f_2(\Theta_1,\Theta_2)=0, \frac{d}{d\Theta_2}f_2(\Theta_1,\Theta_2)=0 \nonumber \\
\int_{-\infty}^{0}dx\int_{-\infty}^{0}dy|w_{3}(x,y)|^2dxdy=f_3(\Theta_1,\Theta_2),  \frac{d}{d\Theta_1}f_3(\Theta_1,\Theta_2)=0, \frac{d}{d\Theta_2}f_3(\Theta_1,\Theta_2)=0 \nonumber \\
\int_{0}^{+\infty}dx\int_{-\infty}^{0}dy|w_{4}(x,y)|^2dxdy=f_4(\Theta_1,\Theta_2), \frac{d}{d\Theta_1}f_4(\Theta_1,\Theta_2)=0, \frac{d}{d\Theta_2}f_4(\Theta_1,\Theta_2)=0,\nonumber \\
0=\int_{-\infty}^{0}dx\int_{0}^{\infty}dy w_{1}^{*}(x,y)w_{2}(x,y)dxdy, 0=\int_{-\infty}^{0}dx\int_{0}^{\infty}dy w_{1}^{*}(x,y)w_{3}(x,y)dxdy, \nonumber \\
0=\int_{-\infty}^{0}dx\int_{0}^{\infty}dy w_{1}^{*}(x,y)w_{4}(x,y)dxdy, 0=\int_{-\infty}^{0}dx\int_{0}^{\infty}dy w_{3}^{*}(x,y)w_{2}(x,y)dxdy \nonumber \\
0=\int_{-\infty}^{0}dx\int_{0}^{\infty}dy w_{2}^{*}(x,y)w_{4}(x,y)dxdy, 0=\int_{-\infty}^{0}dx\int_{0}^{\infty}dy w_{3}^{*}(x,y)w_{4}(x,y)dxdy
\end{eqnarray}
indicating maxima with respect to $\Theta_1$ and $\Theta_2$ parameters, so consequently fractional derivatives shall be equal to zero .
Particular condition $\frac{1}{cos(\Theta_1)^2cos(\Theta_2)^2}\frac{d}{d\Theta_1}f_1(\Theta_1,\Theta_2)=0$ implies
 \begin{eqnarray}
 -2 tan(\Theta_1)\gamma_{1,1}-2tan(\Theta_1)tan(\Theta_2)\gamma_{1,2}+  \nonumber \\
 +(1-tan(\Theta_1)^2)\gamma_{1,3}+(1-tan(\Theta_1)^2)tan(\Theta_2)\gamma_{1,4} + \nonumber \\
 -2 tan(\Theta_1)tan(\Theta_2)\gamma_{2,1}-2 tan(\Theta_1)tan(\Theta_2)^2\gamma_{2,2}+  \nonumber \\
 +(1-tan(\Theta_1)^2)tan(\Theta_2)\gamma_{2,3}+(1-tan(\Theta_1)^2)tan(\Theta_2)^{2}\gamma_{2,4} + \nonumber \\
 +(1-tan(\Theta_1)^2)\gamma_{3,1}+(1-tan(\Theta_1)^2)tan(\Theta_2)\gamma_{3,2}  + \nonumber \\
 +(1-tan(\Theta_1)^2)\gamma_{3,3}+2tan(\Theta_1)tan(\Theta_2)\gamma_{3,4} + \nonumber \\
 +(1-tan(\Theta_1)^2)tan(\Theta_2)\gamma_{4,1}+(1-tan(\Theta_1)^2)tan(\Theta_2)^2\gamma_{4,2} + \nonumber \\
 +2tan(\Theta_1)tan(\Theta_2)\gamma_{4,3}+2tan(\Theta_1)^2tan(\Theta_2)^2\gamma_{4,4}
  \end{eqnarray}
  and condition
 $\frac{1}{cos(\Theta_1)^2cos(\Theta_2)^2}\frac{d}{d\Theta_2}f_1(\Theta_1,\Theta_2)=0$
 implies
 \begin{eqnarray}
 -2 tan(\Theta_2)\gamma_{1,1}+(1-tan(\Theta_2)^2)\gamma_{1,2}+  \nonumber \\
 -2tan(\Theta_1)tan(\Theta_2)\gamma_{1,3}+tan(\Theta_1)(1-tan(\Theta_2)^2)\gamma_{1,4} + \nonumber \\
+(1-tan(\Theta_2)^2)\gamma_{2,1}-2 tan(\Theta_2)\gamma_{2,2}+  \nonumber \\
 +tan(\Theta_1)(1-tan(\Theta_2)^2)\gamma_{2,3}+2tan(\Theta_1)tan(\Theta_2)\gamma_{2,4} + \nonumber \\
 -2tan(\Theta_1)tan(\Theta_2)\gamma_{3,1}+tan(\Theta_1)(1-(tan(\Theta_2))^2)\gamma_{3,2}  + \nonumber \\
 -2tan(\Theta_1)^2tan(\Theta_2)\gamma_{3,3}+tan(\Theta_1)^2(1-tan(\Theta_2)^2)\gamma_{3,4} + \nonumber \\
 +tan(\Theta_1)(1-tan(\Theta_2)^2)\gamma_{4,1}+2tan(\Theta_1)tan(\Theta_2)\gamma_{4,2} + \nonumber \\
 +tan(\Theta_1)^2(1-tan(\Theta_2)^2)\gamma_{4,3}+2tan(\Theta_1)^2tan(\Theta_2)\gamma_{4,4}
  \end{eqnarray}
We obtain two quadrature equations for $tan(\Theta_1)$ and $tan(\Theta_2)$ that have analytical solutions.

\section{Conclusions}
The obtained results shows that quantum mechanical phenomena might be almost entirely simulated by classical statistical model encoded in epidemic model or in more general form in stochastic Finite State Machine. It includes the quantum like entanglement and superposition of states. Therefore coupled epidemic models expressed by classical systems in terms of classical physics can be the base for possible incorporation of quantum technologies and in particular
for quantum like computation and quantum like communication. In the conduced computations Wolfram software was used \cite{Mathematica}. All work presented at \cite{QHSchannel},
\cite{Nbodies} can be expressed by classical epidemic model. It is expected that time crystals can be also described in the given framework \cite{Wilczek}, \cite{Sacha}.
It is open issue to what extent we can parameterize various condensed matter phenomena by stochastic finite state machine \cite{Spalek}.
The main achievements can be characterized as
\begin{itemize}
  \item  Position-based qubit with external source of vector potential can fully reproduce the behavior of linear and non-linear epidemic model.
         \newline
  \item  Entanglement in classical and quantum systems were identified. \\
  \item  Density matrix was formulated for classical and quantum system. \\
  \item  Entanglement entropy was formulated for classical and quantum system. \\
  \item  Dissipation was identified in classical and in quantum system as non-fundamental description of system dynamics.
\end{itemize}
$ $ \newline
and will be continued in next papers.
Various other measurements of classical entanglement can be proposed mainly basing on already existing measurements of quantum entropy \cite{Statistics1}.
Similar way of reasoning was also expressed by \cite{Statistics}. Various types of quantum gates \cite{Swap} can be implemented by usage of classical statistical model. One can also tackle with problem of entanglement of matter and radiation \cite{Jaynes},
\cite{EntanglementMR}, \cite{Birula}, \cite{QInternet} that is known in quantum mechanics
so it can be expressed by classical epidemic model.  One shall expect that phenomena as quantum time crystals \cite{Wilczek},\cite{Sacha} or movement of electron in crystal \cite{Spalek} can also be represented effectively by classical epidemic model or stochastic Finite State Machine.
Good example of application can be the analysis of financial data from stock market. Given price of stock can be categorized in N possible intervals. Then upon observation of data with time the $S$ matrix could be determined in stochastic Finite State Machine. This $S$ matrix can be later converted into tight-binding model that can be represented by single-electron devices in framework of position dependent qubit polarized with time-dependent voltages. In such case one can covert the classical information into quantum information that could be later processed by quantum algorithms. One of main results presented here is equivalence of Schroedinger and tight-binding formalism in any time-dependent Hamiltonianians as given by equation \ref{eqncentral}. This points mathematical equivalence of stochastic finite state machine and tight-binding (Wannier model) and Schroedinger model. Very last can have profound importance in quantum information processing technologies especially in reference to the semiconductor CMOS (and non-CMOS) single-electron devices that can have very high integration. This equivalence shall be not surprising as there is a lot of work on analogies of quantum mechanics and statistical physics as given by \cite{Mishin}, \cite{Statistics1} and by \cite{QMStat}.

\end{document}